\documentclass{ws-ijmpb}

\usepackage{color}

\begin{document}

\markboth{O. Derzhko, J. Richter, M. Maksymenko}
{Strongly correlated flat-band systems: The route from Heisenberg spins to Hubbard electrons}

%
\catchline{}{}{}{}{}
%

\title{STRONGLY CORRELATED FLAT-BAND SYSTEMS: THE ROUTE FROM HEISENBERG SPINS TO HUBBARD ELECTRONS}

\author{OLEG DERZHKO}

\address{Institute for Condensed Matter Physics, National Academy of Sciences of Ukraine, 
         Svientsitskii Street 1, 79011 L'viv, Ukraine}
\address{Institut f\"{u}r theoretische Physik, Otto-von-Guericke-Universit\"{a}t Magdeburg, 
         P.O. Box 4120, 39016 Magdeburg, Germany}
\address{Max-Planck-Institut f\"{u}r Physik komplexer Systeme, 
         N\"{o}thnitzer Stra\ss e 38, 01187 Dresden, Germany}
\address{Abdus Salam International Centre for Theoretical Physics,
         Strada Costiera 11, 34151 Trieste, Italy}
\address{Department for Metal Physics, Ivan Franko National University of L'viv,
         Kyryla \& Mephodiya Street 8, 79005 L'viv, Ukraine}

\author{JOHANNES RICHTER}

\address{Institut f\"{u}r theoretische Physik, Otto-von-Guericke-Universit\"{a}t Magdeburg, 
         P.O. Box 4120, 39016 Magdeburg, Germany}

\author{MYKOLA MAKSYMENKO\footnote{On leave from 
Institute for Condensed Matter Physics, National Academy of Sciences of Ukraine, 
Svientsitskii Street 1, 79011 L'viv, Ukraine.}}

\address{Department of Condensed Matter Physics, Weizmann Institute of Science, 
         76100 Rehovot, Israel}
\address{Max-Planck-Institut f\"{u}r Physik komplexer Systeme, 
         N\"{o}thnitzer Stra\ss e 38, 01187 Dresden, Germany}



\maketitle

\begin{history}
\received{Day Month Year}
\revised{Day Month Year}
\end{history}

\begin{abstract}
On a large class of lattices 
(such as  the sawtooth chain, the kagome and the pyrochlore lattices) 
the quantum Heisenberg and the repulsive Hubbard models 
may host a completely dispersionless (flat) energy band in the single-particle spectrum. 
The flat-band states can be viewed as completely localized within a finite volume (trap) of the lattice 
and allow for construction of many-particle states, roughly speaking, by occupying the traps with particles. 
If the flat band happens to be the lowest-energy one 
the manifold of such many-body states 
will often determine the ground-state and low-temperature physics of the models at hand 
even in the presence of strong interactions. 
The localized nature of these many-body states 
makes possible the mapping of this subset of eigenstates onto a corresponding classical hard-core system. 
As a result, 
the ground-state and low-temperature properties of the strongly correlated flat-band systems 
can be analyzed in detail using concepts and tools of classical statistical mechanics 
(e.g., classical lattice-gas approach or percolation approach),
in contrast to more challenging quantum many-body techniques 
usually necessary to examine strongly correlated quantum systems.

In this review 
we recapitulate the basic features  of the flat-band spin systems and briefly summarize earlier studies in the field.
Main emphasis is made on recent developments
which include results for both spin and electron flat-band models.
In particular,
for flat-band spin systems we highlight field-driven phase transitions 
for frustrated quantum Heisenberg antiferromagnets at low temperatures, 
chiral flat-band states, as well as
the effect of a slight dispersion of a previously strictly flat band due to nonideal lattice
geometry.
For electronic systems, we discuss 
the universal low-temperature behavior of several flat-band Hubbard models,
the emergence of ground-state ferromagnetism in the square-lattice Tasaki-Hubbard model 
and the related 
Pauli-correlated percolation problem,
as well as the dispersion-driven ground-state ferromagnetism in flat-band Hubbard
systems.
Closely related studies and
possible experimental realizations of the flat-band physics are also
described briefly.
\end{abstract}



\section{Introduction and Outline}
\label{sec01}

Flat-band physics attracts increasing attention nowadays.
A good evidence for this is a number of reviews which have been published recently,
mainly with a special focus on topological flat-band models.\cite{fqh-flat-review,bergholtz,liu-liu-wu}
However, the interest in strongly correlated flat-band systems  came up long ago.
For example, it was known since early 1990s that linear spin-wave theory for the antiferromagnetic 
Heisenberg model on the kagome lattice leads to flat modes.\cite{kagome_flat_1,kagome_flat_2}
This flat mode of the kagome antiferromagnet corresponds to a coherent motion of spins on a hexagon 
while the spins surrounding the hexagon are ordered ferromagnetically.\cite{kagome_flat_2}
However, these calculations were approximate (linear spin-wave theory, i.e.,
large-$s$ limit) and refer to the zero-field case.
Several years later new prominent features of quantum Heisenberg antiferromagnets 
on a class of highly frustrated lattices\footnote{Magnetic interactions are frustrated, 
if a spin cannot arrange its orientation such that it profits from the interaction with its neighbors 
as, for instance, in the case of antiferromagnetic interactions on the triangular lattice.
For a more in depth discussion we refer to Refs.~\refcite{Ramirez,Greedan}.}
were observed around the saturation field. 
These are the ground-state magnetization jump at the saturation field,\cite{lm1} 
a magnetic-field driven spin-Peierls instability,\cite{sp-Peierls}
a finite residual entropy at the saturation field,\cite{prb2004,mike-andreas,mike1}
and an unconventional low-temperature
thermodynamics.\cite{mike-andreas,mike1,epjb2006}
A crucial  feature leading to these effects is the fact 
that the lowest 
band of one-magnon
excitations above the fully polarized ferromagnetic state (which becomes the
ground state for magnetic fields
above the saturation field) is completely dispersionless (flat).
In many cases the flat-band states can be visualized as the states located within a small part of the lattice (trap)
and this localization in turn is a consequence of destructive quantum interference 
which occurs due to a special geometry of the lattice. 
The very existence of localized magnons as the lowest-energy one-particle states
opens an interesting perspective 
to construct and fully characterize many-magnon ground states of the considered frustrated quantum Heisenberg antiferromagnets. 
Moreover, the set of relevant low-energy many-magnon states of the quantum spin systems
can be mapped onto corresponding (highly non-trivial) classical systems 
and as a result one may use completely different methods from the toolbox of classical statistical mechanics 
to study the spin systems at hand. 
The striking difference to the previous linear spin-wave studies of flat-band antiferromagnets at zero field\cite{kagome_flat_1,kagome_flat_2} 
is that there are flat-band {\it many-body} states at high magnetic fields 
which are exact ground states for arbitrary spin quantum number $s$ 
and that the abovementioned features are true quantum  effects which disappear in the classical limit $s\to \infty$.

Later on it was recognized\cite{cmp2005} that the flat-band quantum spin systems 
are covertly related to another class of strongly correlated electron models, 
the so-called flat-band Hubbard ferromagnets, 
which were discussed even earlier.\cite{le1,le2,le3} 
While the seminal papers by A.~Mielke and H.~Tasaki are focused mainly on rigorous proofs of the ferromagnetic ground states
for the flat-band Hubbard model\cite{le1,le2,le3} and their stability,\cite{flband-ferro-stability}
similarities to localized-magnon systems lead to the discussion of the ground-state degeneracy and universal 
low-temperature behavior of flat-band Hubbard systems.\cite{fb-el1,fb-el2,fb-el3}
Furthermore, 
a percolation representation for the ground-state ferromagnetism
suggested by A.~Mielke and H.~Tasaki\cite{le3} 
has been recognized to present a new class of classical percolation problems 
-- the so-called Pauli-correlated percolation.\cite{fb-el3_4,fb-el4,mykola,fb-el5}

In the theory of frustrated magnetism\cite{lnp,lnp645,lacroix} the localized-magnon systems
represent a special class of models 
which in a special regime (high fields, low temperatures) admit a rather detailed study of their properties
which is based on the rigorous knowledge of the many-body localized-magnon states.
In the broad field of the theory of the Hubbard model,\cite{origin_FM,hubbard_nature}
and, in particular, of ferromagnetism of the Hubbard model,\cite{tasaki_rev1,tasaki_rev2,tasaki_rev3}
the Mielke-Tasaki flat-band ferromagnetism provides one specific root to ferromagnetism which
also allows rather rigorous analysis.
Both fields appear to be related: From the mathematical point of view the description of spin and electron systems has many
similarities,
since  
a key property  of both flat-band systems is that relevant quantum many-body
degrees of freedom can be mapped on classical ones.  

Earlier studies on localized magnons were reviewed some years ago,
see Refs.~\refcite{review1}, \refcite{review2} and \refcite{review3}.
In the present paper we intend to provide a snapshot of the state of art in the field
presenting new directions with larger focus on electron systems. 
Naturally, the main attention is paid to summarizing studies in which the authors are involved or closely related ones.
We also provide a detailed introduction to the basic concepts.

The outline of this review is as follows.
We begin with a short introduction to localized-magnon and localized-electron systems
focusing on lattices, Hamiltonians, concepts etc.
and recall the main results which were reviewed earlier 
(magnetization jump, lattice instability, residual entropy, low-temperature peak of specific heat),\cite{review1,review2,review3}
see Sec.~\ref{sec02}.
Concerning new results for spin systems,
we illustrate the emergence of Ising degrees of freedom in frustrated ladder and bilayer Heisenberg antiferromagnets\cite{ising_degrees} in Sec.~\ref{sec03},
chiral localized magnons (and localized electrons)\cite{chiral1,chiral2} in Sec.~\ref{sec04}, 
and discuss the effect of small deviations from the flat-band geometry\cite{deviations1,deviations2} in Sec.~\ref{sec05}. 
Next we discuss some studies stimulated by the localized-magnon picture (Sec.~\ref{sec06})
discussing in particular flat-band systems at densities slightly above the localized-magnon-crystal density\cite{huber-altman}
and
the magnetization plateaus in the spin-$\frac{1}{2}$ Heisenberg antiferromagnet
on the kagome lattice.\cite{nishimoto,capponi,Sakai}
Very recently it has been found that
localized-magnon states can be the relevant low-energy states even in zero magnetic field
for a sawtooth chain with ferro- and antiferromagnetic interactions.\cite{dmitriev_krivnov}
We will briefly discuss this very new example for localized-magnon states in Sec.~\ref{sec06}.
Next, in Sec.~\ref{sec07}, we deal with solid-state systems which should exhibit localized-magnon physics.
The natural mineral azurite Cu$_3$(CO$_3$)$_2$(OH)$_2$ is the most promising example for experimental solid-state observation of flat-band
physics.\cite{kikuchi,kikuchi_2,rule,azurite-parameters,effective_xy2,azurite2014}
We illustrate the theoretical predictions based on a diamond-chain Heisenberg model.
Another localized-magnon compound has been synthesized recently, Ba$_2$CoSi$_2$O$_6$Cl$_2$.\cite{tanaka}
We also briefly discuss this system in Sec.~\ref{sec07}.
Note, however, that these compounds do not fulfill the strict flat-band criteria, 
but show some characteristic features such as the magnetization jump.

In the second part of this review  we move to electron systems.
In Sec.~\ref{sec08} we consider two different classes of the flat-band Hubbard model 
and discuss their universal behavior at low electron densities and low temperatures.\cite{fb-el1,fb-el2,fb-el3}
Section~\ref{sec09} reviews recent work on the existence of the ground-state ferromagnetism in a particular two-dimensional flat-band Hubbard model 
which can be described by a new type of percolation called Pauli-correlated percolation.\cite{fb-el4}
In Sec.~\ref{sec10} we consider Hubbard systems 
for which a flat band does not lead to ground-state ferromagnetism, 
but the ground state becomes ferromagnetic, 
if the flat band acquires small dispersion and the on-site Coulomb repulsion is sufficiently large (dispersion-driven  ferromagnetism).\cite{dispersion-driven}
Furthermore, we discuss several related topics (Sec.~\ref{sec11}). 
First we give in Sec.~\ref{sec11} a brief account on a general construction scheme of a class of exact ground states 
of specific strongly correlated electron systems.\cite{gulacsi1,gulacsi2}
Then we discuss some recent results on flat-band systems with randomness,\cite{random1,random2,random3}
as well as the transport properties of flat-band two-terminal nanodevices.\cite{lopes}
In Sec.~\ref{sec12} we discuss possible experimental realization of theoretical scenarios 
for flat-band Hubbard models.\cite{tamura-quantum-dots-wires,nishino-3d-flatband,flat-band-experiment-polymers,lin-grapheneribbon,fb-nature,balents,finland1,finland2,noda,mukherjee}
Finally, in Sec.~\ref{sec13}, we close by a summary and sketch a perspective for future work.

\section{Flat Band and Localized Eigenstates}
\label{sec02}

\subsection{One- and many-particle localized eigenstates in frustrated Heisenberg antiferromagnets}
\label{sec021}

We begin with a brief account of the concepts used to discuss the localized-magnon states.
We consider the Heisenberg model in an external magnetic field
\begin{eqnarray}
\label{001}
H=\sum_{(ij)}J_{ij}{\bf{s}}_i\cdot{\bf{s}}_j-hS^z,
\;\;\;
S^z=\sum_{i=1}^Ns_i^z.
\end{eqnarray}
Here ${\bf{s}}_i=(s^x_i,s^y_i,s^z_i)$ is the spin-$\frac{1}{2}$ operator 
attached to the site $i$,
$J_{ij}>0$ denotes the antiferromagnetic exchange interaction between nearest-neighboring sites,
$h$ is the dimensionless external magnetic field,
the first sum in Eq.~(\ref{001}) runs over all neighboring sites of the $N$-site lattice.
$S^z$ commutes with the Hamiltonian $H$ and hence the $z$-component of the total spin is a good quantum number.
In the subspace with $S^z=\frac{N}{2}$ there is only one state, 
that is the fully polarized state $\vert {\rm{FM}}\rangle=\vert\uparrow\ldots\uparrow\rangle$
which plays the role of the vacuum state for magnons.
In the subspace with $S^z=\frac{N}{2}-1$ we have $N$ one-magnon states, and
their explicit form and their energy can be found exactly.
The one-magnon energies crucially depend on the lattice geometry.

Let us mention here that the concept of localized-magnon states is not
restricted to spin quantum number $s=\frac{1}{2}$ and isotropic Heisenberg systems,
see, e.g., Refs.~\refcite{lm1,kai}.
However, the related physical effects (see below) are most pronounced in the
extreme quantum case $s=\frac{1}{2}$ and vanish in the classical limit $s \to\infty$.

\begin{figure}[bt]
\centerline{\psfig{file=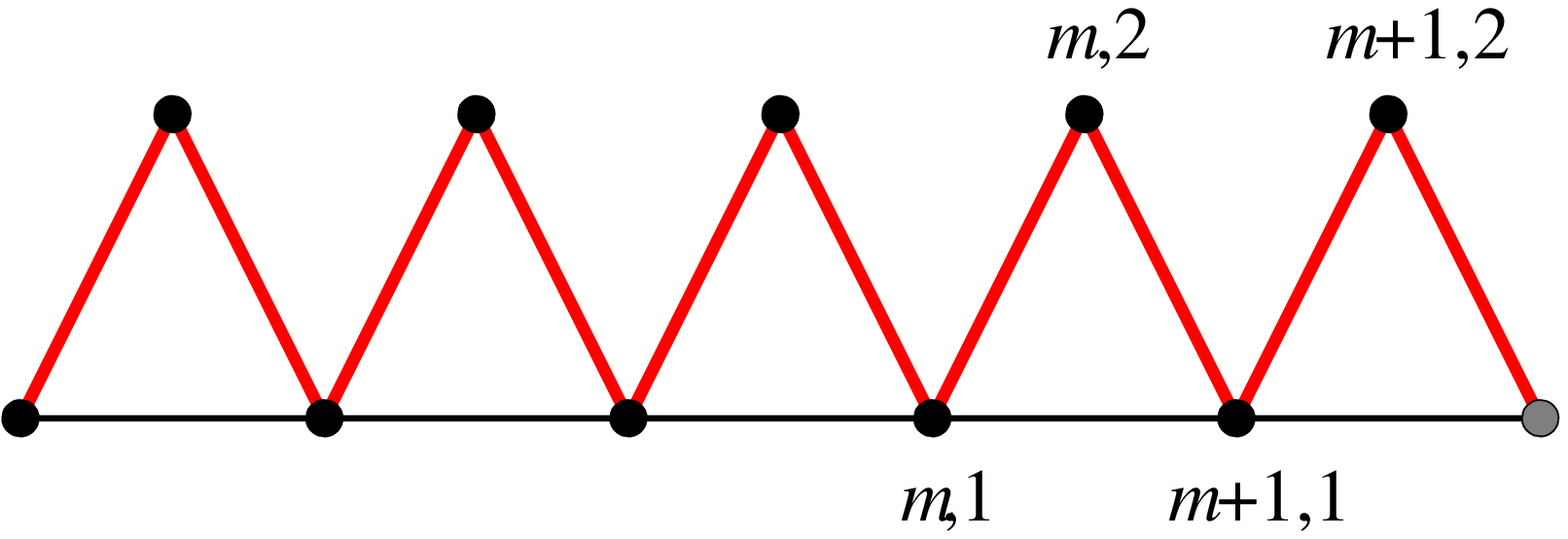,width=2.2in}\hspace{10mm}\psfig{file=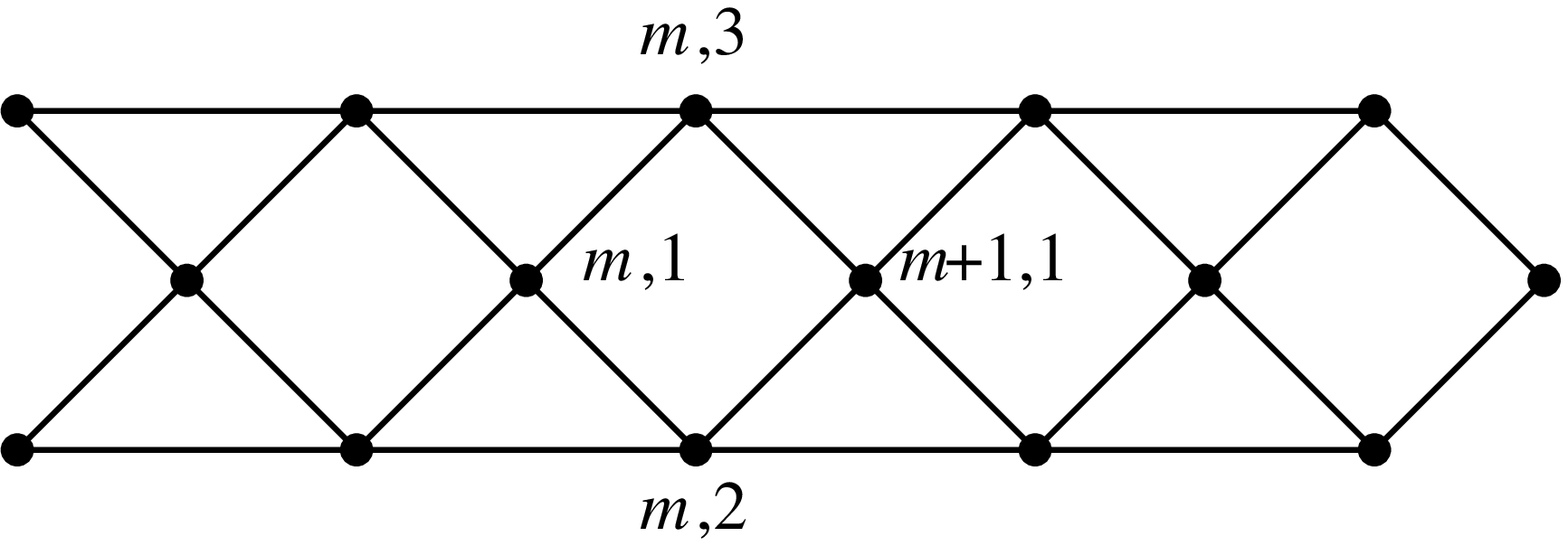,width=2.2in}}
\vspace{5mm}
\centerline{\psfig{file=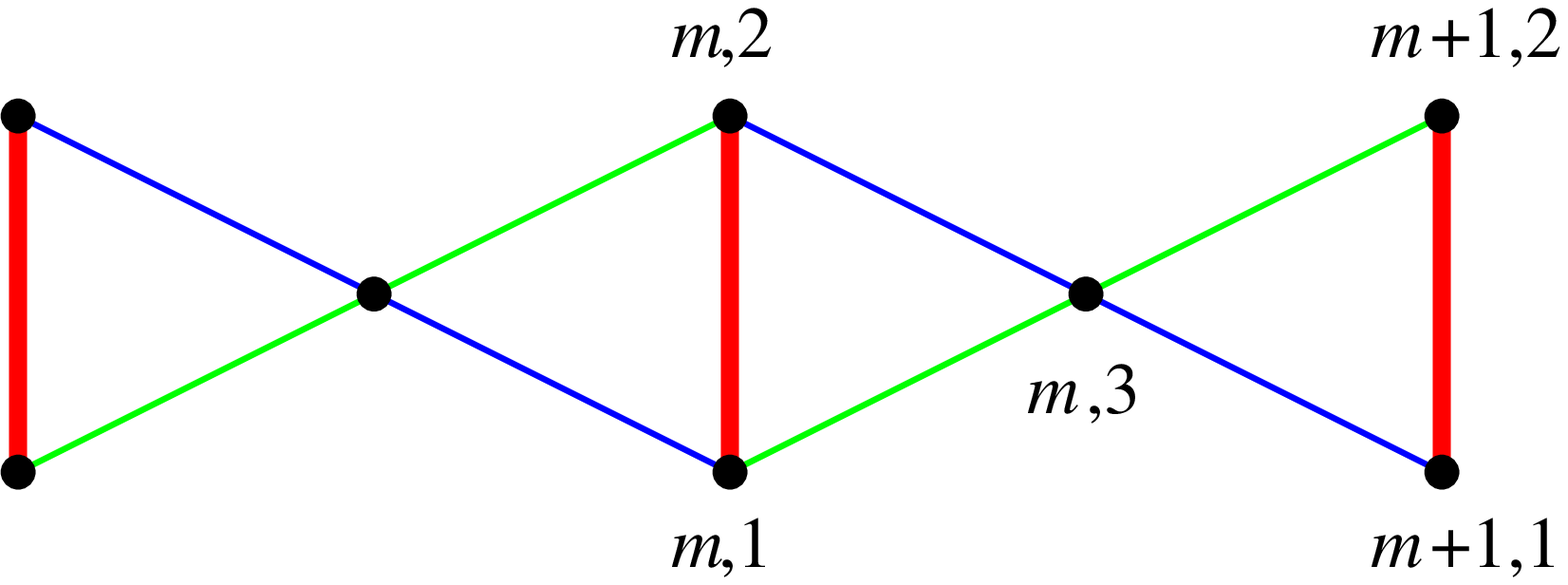,width=2.15in}\hspace{13.5mm}\psfig{file=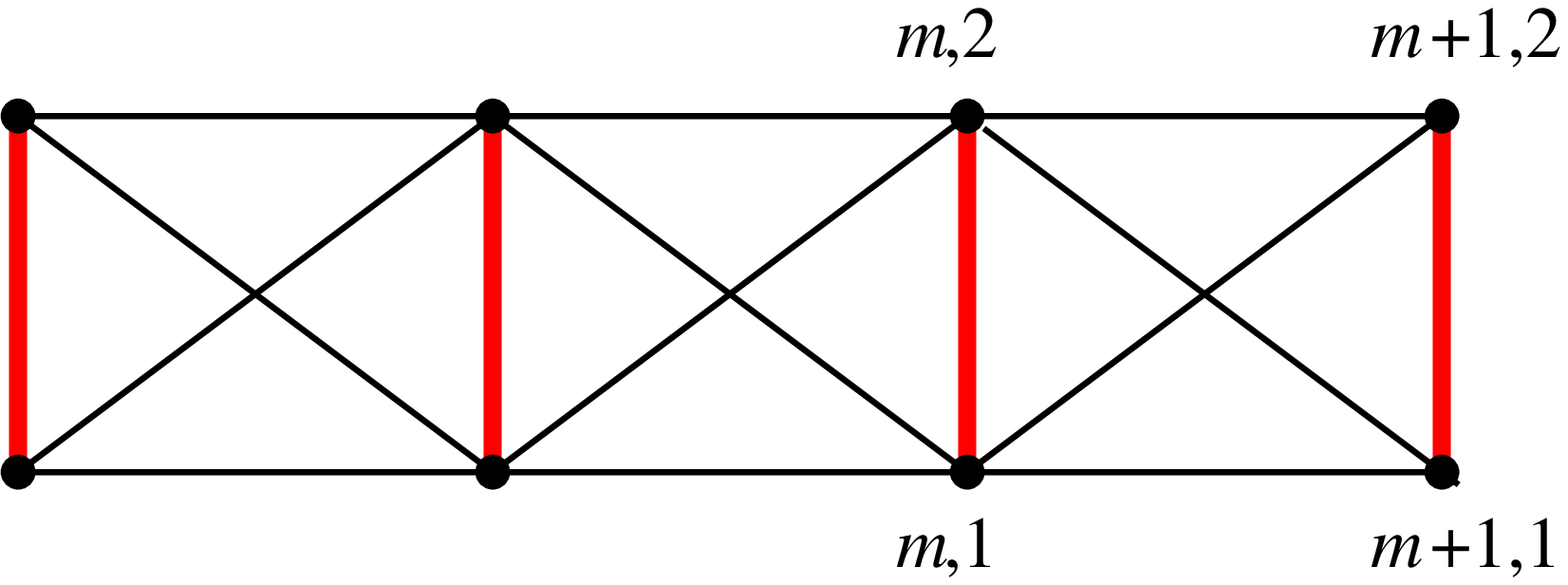,width=2.15in}}
\vspace{5mm}
\centerline{\psfig{file=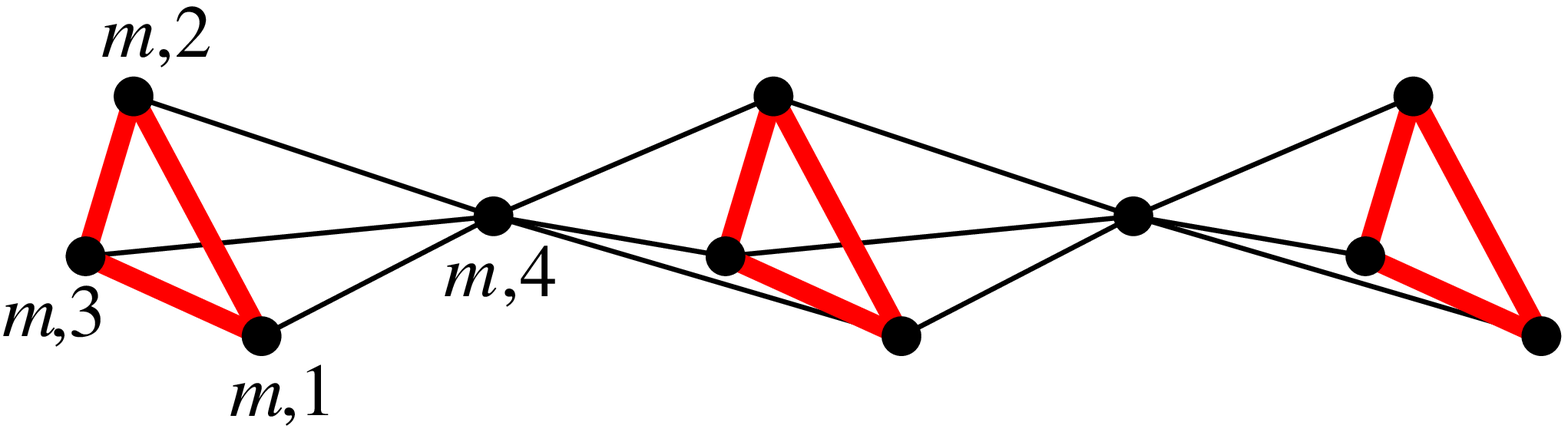,width=2.25in}\hspace{10mm}\psfig{file=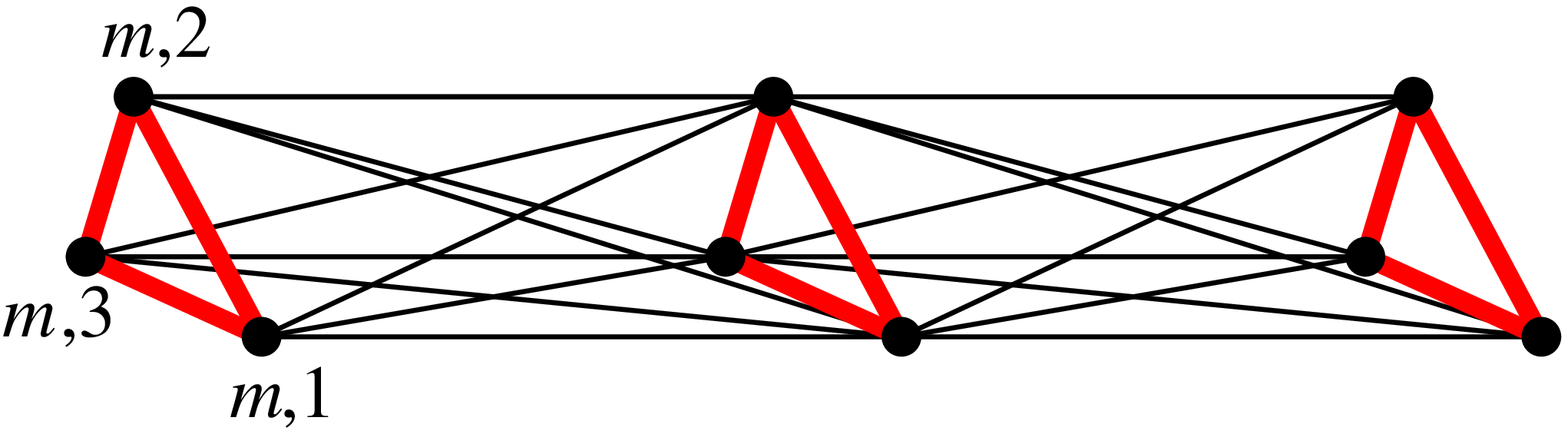,width=2.25in}}
\vspace*{8pt}
\caption{(Color online)
One-dimensional lattices which have a flat lowest-energy one-particle band:
The sawtooth lattice,
the kagome chain,
the frustrated diamond chain,
the frustrated two-leg ladder,
the double-tetrahedra chain, 
and
the frustrated (cylindrical) three-leg ladder.
The sites are enumerated by two indices, 
where the first one enumerates the cells and the second one enumerates the sites within
a cell.
Alternatively, a one-index numbering of sites is also possible; 
e.g., for the sawtooth chain by $m,1\to 2j$, $m,2\to 2j+1$ etc. and $j=0,\ldots,\frac{N}{2}-1$ 
or for the kagome chain by $m,1\to 3j$, $m,2\to 3j+1$, $m,3\to 3j+2$ etc. and $j=0,\ldots,\frac{N}{3}-1$.
While for the kagome chain all exchange (hopping) parameters $J$ ($t$) are
identical, there are two different bond strengths, 
$J_1$ ($t_1$) and $J_2$ ($t_2$), 
in the other lattices.
For the sawtooth chain a special flat-band condition is required, namely the exchange (hopping) parameter along the zigzag path $J_2$ ($t_2$)
has to be 2 ($\sqrt{2}$) times larger than the exchange (hopping) parameter $J_1$ ($t_1$) along the basal straight
line.
For  
the frustrated diamond chain,
the frustrated two-leg ladder,
the double-tetrahedra chain, 
and
the frustrated three-leg ladder 
the exchange (hopping)
parameter  $J_2$ ($t_2$) marked by the bold red lines 
must exceed a threshold value to have the flat band as the lowest one (note
that for  these lattices the bold red lines  at the same  time denote the trapping
cells).
For the diamond chain we also distinguish blue and green lines.
For the distorted diamond chain also considered here both bond strengths are  
different, where blue lines indicate exchange (hopping) parameters $J_1$ ($t_1$)
and green lines
exchange (hopping) parameters $J_3$ ($t_3$).
In case of $J_1 \ne J_3$ ($t_1 \ne t_3$) the lowest band becomes dispersive.
The lowest band is flat (flat-band geometry) if blue and green bonds mark identical bond
strengths, i.e.,
$J_1=J_3$ ($t_1=t_3$). 
}
\label{fig01}
\end{figure}

\begin{figure}[bt]
\centerline{\psfig{file=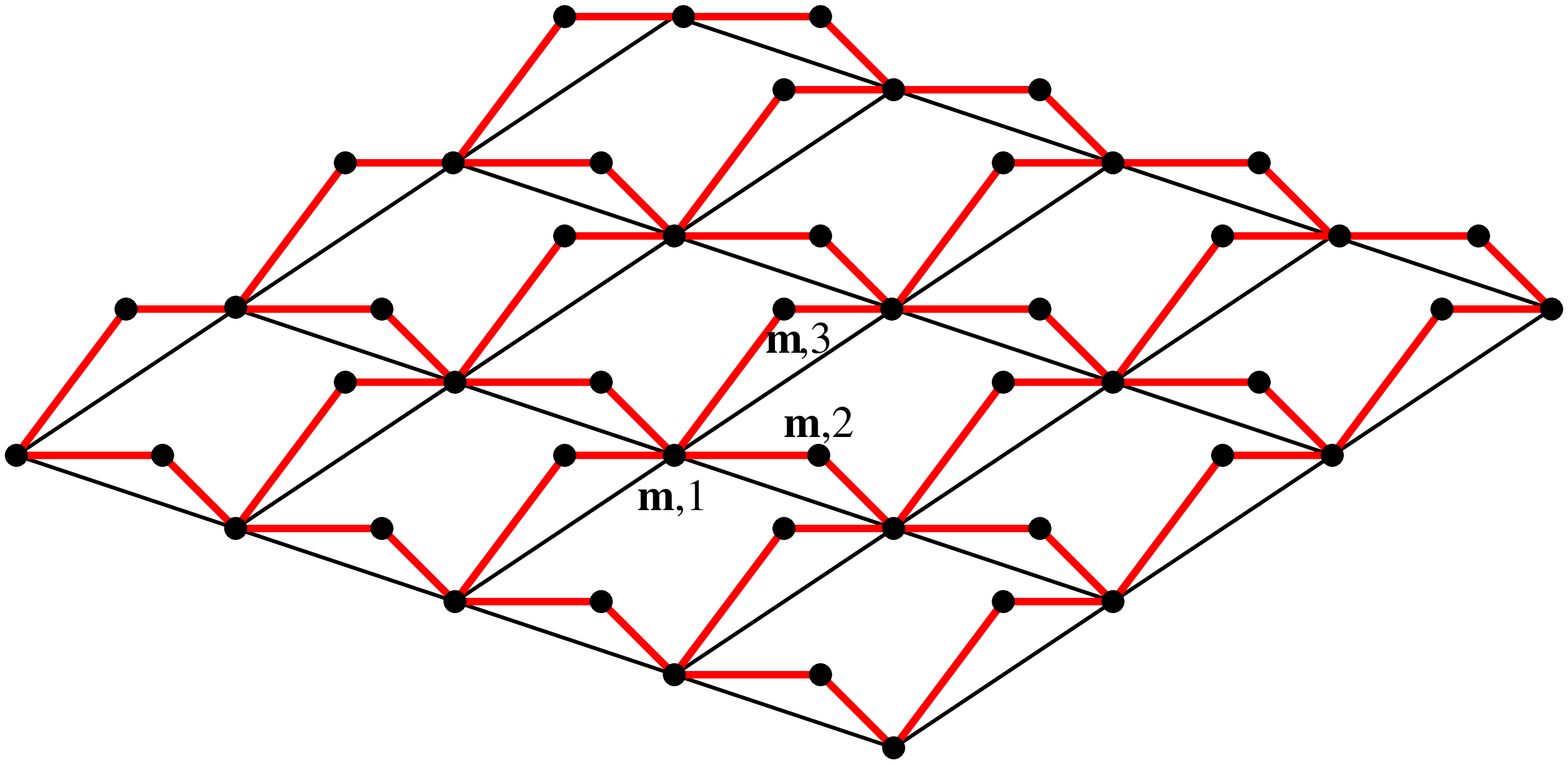,width=2.55in}\hspace{10mm}\psfig{file=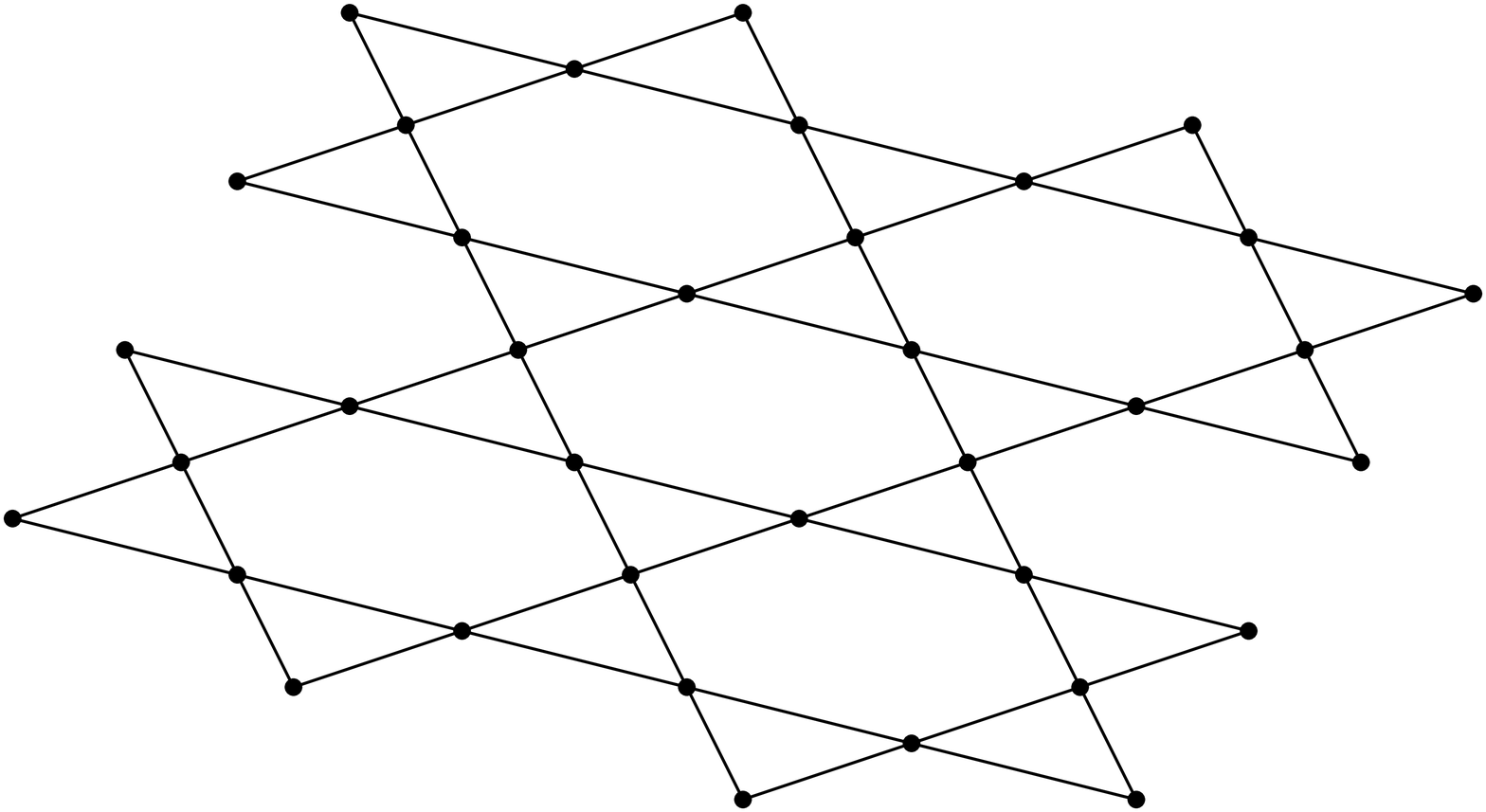,width=2.05in}}
\vspace{10mm}
\centerline{\psfig{file=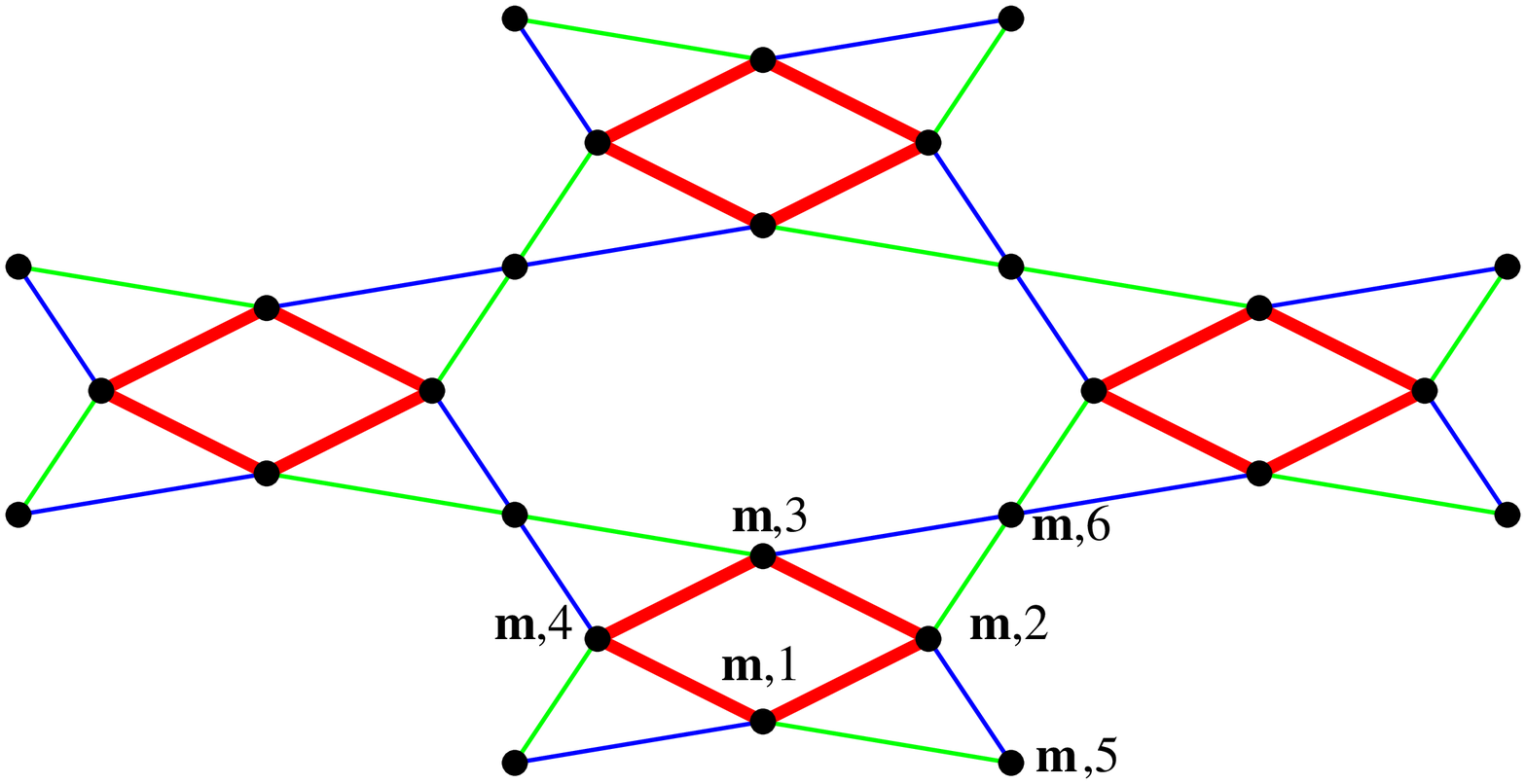,width=2.3in}\hspace{10mm}\psfig{file=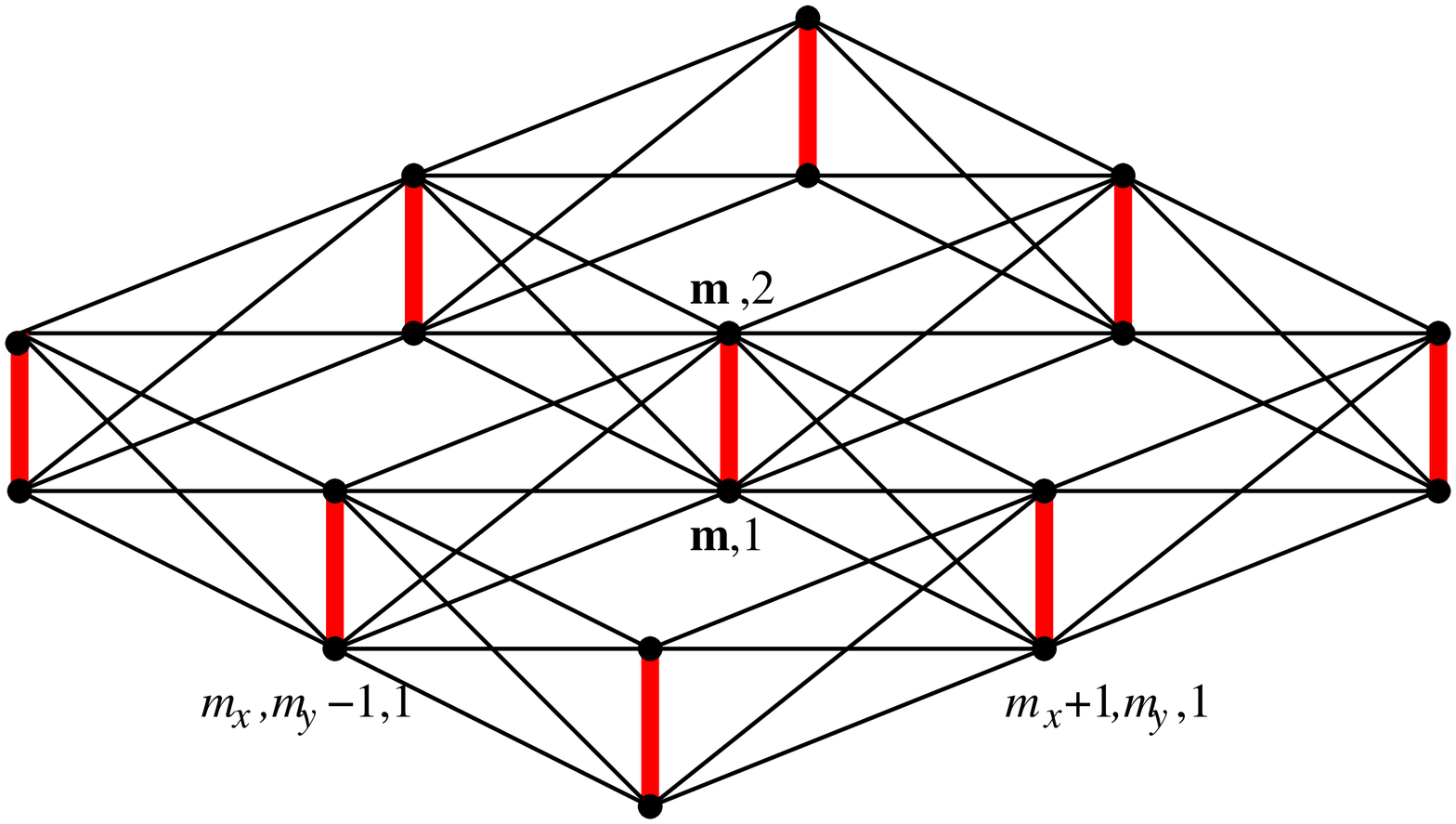,width=2.3in}}
\vspace*{8pt}
\caption{(Color online)
Two-dimensional lattices which have a flat lowest-energy one-particle band:
The Tasaki lattice,
the kagome lattice,
the square-kagome lattice,
and
the frustrated bilayer.
The sites are enumerated by two indices, 
where the first vector index enumerates the cells and the second one enumerates the sites within
a cell.
While for the kagome lattice all exchange (hopping) parameters $J$ ($t$) are
identical, there are two different bond strengths in the other lattices.
For the Tasaki lattice (two-dimensional counterpart of the sawtooth chain) 
a special flat-band condition is required, 
namely, the exchange (hopping) parameter along the zigzag path $J_2$ ($t_2$)
has to be $4$ ($2$) times larger than the exchange (hopping) parameter $J_1$ ($t_1$) along the basal 
lines.
For 
the square-kagome lattice
and
the frustrated bilayer 
the exchange (hopping)
parameter  $J_2$ ($t_2$) marked by the bold red lines 
must exceed a threshold value to have the flat band as the lowest one (note
that for  these lattices the bold red lines  at the same  time denote the trapping
cells).
For the square-kagome lattice we also distinguish blue and green lines.
For the flat-band geometry blue and green bonds mark identical bond strengths. 
For the distorted square-kagome lattice also considered here both bond strengths are  
different, where blue lines indicate exchange (hopping) parameters $J_1$ ($t_1$)
and green lines
exchange (hopping) parameters $J_3$ ($t_3$).
In case of $J_1 \ne J_3$ ($t_1 \ne t_3$) the lowest band becomes dispersive.}
\label{fig02}
\end{figure}

We consider lattices which support a completely dispersionless lowest-energy magnon band.
There is quite a lot of lattices having flat bands. 
In Figs.~\ref{fig01} and \ref{fig02} we show some of them:
The sawtooth lattice,\cite{sawtooth-gen}
the kagome chain,\cite{kagome-chains-gen}
the frustrated diamond chain,\cite{diamond}
the frustrated two-leg ladder,\cite{ladders-gen}
the double-tetrahedra chain,\cite{double_tetrahedra} 
and
the frustrated (cylindrical) three-leg ladder\cite{fr_three}
(one-dimensional lattices)
and
the Tasaki lattice,\cite{le2}
the kagome lattice,\cite{kagome-gen,kagome_new}
the square-kagome lattice,\cite{sqkag,ioannis} 
and
the frustrated bilayer\cite{bilayers-gen}
(two-dimensional lattices).
Other lattices which were also discussed in the localized-magnon context are:
The dimer-plaquette chain,\cite{dipla}
the star lattice,\cite{star}
the checkerboard lattice,\cite{checkerboard-gen}
the sorrel net,\cite{ioannis}
the pyrochlore lattice.\cite{pyrochlore-gen}
There are also lattices which where not considered in the localized-magnon context
so far, although, they belong to this class, e.g., the
triangulated kagome lattice.\cite{strecka_lattice} 
Traces of a flat band can be observed  also for magnetic molecules, 
see, e.g., Refs.~\refcite{fr_molecules,fr_molecules-review}.
We mention, that most of the flat-band lattices  were considered
also in the general context of frustrated quantum magnetism.  
For example, the spin-$\frac{1}{2}$ Heisenberg antiferromagnets 
on the kagome,\cite{kagome-gen,kagome_new} the star,\cite{star} and the square-kagome\cite{sqkag}  lattices
are known as examples of two-dimensional quantum spin systems with a magnetically disordered ground state.


To construct more flat-band lattices one may use Mielke's line-graph prescription\cite{le1} or Tasaki's decoration procedure.\cite{le2}
In this sense we can define two classes of flat-band lattices: One class is given by Mielke's line graphs
(e.g., the kagome chain, the kagome lattice or the checkerboard lattice)
whereas another one can be constructed by Tasaki's decoration procedure
(e.g., the sawtooth chain or the two-dimensional Tasaki lattice).
Note further, that some of the flat-band lattices belong to the class of models with local conservation laws
(e.g., the diamond chain, the dimer-plaquette chain, the frustrated ladder, and the frustrated bilayer).
In Refs.~\refcite{lm1} and \refcite{fb-el2} 
the construction starts from isolated cells which are further arranged in a infinite lattice.
An important 
feature of flat-band lattices considered here  is that the trapping cell is surrounded by
triangles, see Figs.~\ref{fig01} and \ref{fig02}. 
Just this triangular geometry of antiferromagnetic bonds
prohibits the magnon to escape from the trap due to destructive quantum interference. 
Obviously, antiferromagnetic couplings on a triangle are frustrated, 
i.e., frustration is a common feature of all lattices discussed in this paper. 
Interestingly, the same mechanism can lead to a complete localization of
electrons in 
a magnetic field (``Aharonov-Bohm cage'').\cite{ab_cage,dice_lattice}
One more scheme for generating flat-band lattices has been suggested very recently.\cite{random3}
It starts from a lattice with dispersive bands, then a set of Fano defects is added, and
finally by an appropriate transformations of the Fano lattice a
corresponding flat-band lattice can be obtained,
see Sec.~\ref{sec112}.\footnote{The presence of two types of states,
the localized (dispersionless) ones and the extended (dispersive) ones,
allows to use the universal language of the Fano-resonance phenomenon,
known from atomic physics,\cite{fano}
see Ref.~\refcite{fano-r}.}
It is also interesting to note
that flat bands are often associated with symmetries,
see Sec.~VII in Ref.~\refcite{zohar}.

The localized-magnon state can be written as follows
\begin{eqnarray}
\label{002}
 \vert {\rm {l.m.}}\rangle=\sum_{i \in {\cal C}} a_is_i^-\vert\uparrow\ldots\uparrow\rangle.
\end{eqnarray}
Here the sum runs over the sites belonging to one trapping cell $\cal C$
(e.g., the vertical bond for the diamond chain, the frustrated two-leg ladder or bilayer
or the hexagon/square for the kagome/square-kagome lattice).
We denote the energy of this state by $\varepsilon_1$ (the flat-band energy).
To cage the magnon in the trap ${\cal {C}}$ we have to fulfill two conditions,
namely, 
$\sum_{i \in {\cal{C}}} J_{li} a_i=0$ for $\forall l\notin {\cal{C}}$
and
$\sum_{l\notin {\cal{C}}}J_{li}={\rm{constant}}$ for $\forall i\in {\cal{C}}$,
see Ref.~\refcite{lm1}.

Localized magnons are building blocks for constructing exact many-magnon states.
Clearly, a state consisting of many localized magnons is an exact many-magnon state, 
if the localized magnons are sufficiently separated of each other, such that they do not interact. 
In case that neighboring trapping cells share a common site 
(e.g., the sawtooth chain or the kagome lattice) the occupation of neighboring cells is forbidden (hard-core rule), 
whereas for lattices with isolated trapping cells 
(e.g., the diamond chain but not the frustrated two-leg ladder or the frustrated bilayer) 
neighboring cells can be occupied 
(no hard-core rule or more precisely only a hard-monomer rule). 
According to the number of trapping cells ${\cal N} = aN$, 
where the proportionality factor $a$ depends on the lattice, 
there is a maximum number of localized magnons $n_{\rm max}\propto {\cal N}$
that corresponds to the complete occupation of the trapping cells 
(according to possible hard-core rules).

It can be shown that the multi-magnon eigenstates built by $n$ localized magnons 
under some conditions are the lowest-energy states in the subspaces with corresponding $S^z=\frac{N}{2}-n$.\cite{Schmidt}
Moreover, it can be shown almost in all cases rigorously\cite{linear_in} that the localized-magnon states are linearly independent
(but not necessarily orthogonal).

The multi-magnon state with maximum $n=n_{\rm max}$ is an exceptional state
called magnon crystal (complete ordered occupation of traps). It resembles a
Wigner crystal and it is typically protected by an excitation gap. As a
result it corresponds to a magnetization plateau state preceding the
magnetization jump associated with the localized-magnon states, cf. also
Sec.~\ref{sec062}. 

An important step to analyze the physics related to
localized-magnon states consists in the possibility to map the complete
manifold of the localized-magnon states on classical hard-core
lattice gases.\cite{prb2004,mike1,epjb2006}
For instance, for the sawtooth chain with ${\cal N}=\frac{N}{2}$ trapping cells
the mapping is onto the hard-dimer problem on a simple linear chain of
${\cal N}$ sites, and for the frustrated bilayer the corresponding hard-core
problem is that of  hard squares on a square lattice (see also
Sec.~\ref{sec03}).
Using then well-elaborated statistical mechanics of hard-core lattice gases
one easily calculates the low-temperature thermodynamics of the initial
quantum spin  system, see, e.g., Refs.~\refcite{prb2004,mike1,epjb2006}.

In the case of electron systems, the one-particle problem is quite similar.
More precisely, the one-electron problem corresponds to one-magnon problem with $XY$ interaction in the Hamiltonian (\ref{001}), 
i.e., with $s^x_is^x_j+s^y_is^y_j=\frac{1}{2}(s^+_is^-_j+s^-_is^+_j)$ 
instead of ${\bf{s}}_i\cdot{\bf{s}}_j$ in Eq.~(\ref{001}). 
However, the construction of many-particle states (i.e., the occupation rule
of the trapping cells) is different.
While for the Heisenberg spin system we have to avoid the interactions
between trapping cells 
to remain in the ground-state manifold,
in the case of the Hubbard system 
we have to avoid the on-site Hubbard repulsion between electrons with different spins 
in order not to run out of the set of ground states.
We discuss the one-electron and many-electron states of flat-band Hubbard systems in Secs.~\ref{sec08} and \ref{sec09}.

\subsection{Thermodynamics of flat-band antiferromagnets}
\label{sec022}

Localized-magnon states constitute a substantial part of all $2^N$ states of $N$-site spin-$\frac{1}{2}$
flat-band spin systems.
Moreover, in the presence of a high magnetic field around the saturation value $h_{\rm{sat}}$ they are the low-energy ones 
and in many cases separated from the other states by an energy gap.
Therefore they dominate the properties of the flat-band spin system in the low-temperature high-field regime.

The main observable effects at $T=0$ are:
A magnetization jump at the saturation field,\cite{lm1} 
a magnetic-field driven lattice instability,\cite{sp-Peierls} 
and a residual entropy.\cite{prb2004}
The ground-state magnetization jump between the values $M=\frac{N}{2}$ and
$M=\frac{N}{2}-n_{\max}$
(here $n_{\max}$ is the maximal possible number of localized magnons)
is in an obvious manner related to the structure of the energy spectrum of the spin system at hand.
Namely, the energy of the $n$-localized-magnon state in the presence of a magnetic field, 
$E_{{\rm{FM}}}-n\varepsilon_1 - h\left(\frac{N}{2}-n\right)$, 
becomes $n$-independent at the saturation field $h_{\rm{sat}}=\varepsilon_1$
and hence the magnetization $S^z=\frac{N}{2}-n$ acquires any value between $\frac{N}{2}$ and $\frac{N}{2}-n_{\max}$.
A state with  $n$ localized magnons has a huge degeneracy,
since the $n$ localized magnons can be placed over the lattice in many ways.
At the saturation field the ground-state degeneracy dramatically increases 
and corresponds to the grand-canonical partition function of an auxiliary classical gas of hard-core objects at zero value of the chemical potential.
This quantity grows exponentially with lattice size $N$ that leads to a residual entropy.

Moreover, an appropriate lattice distortion
fitting to the structure of the localized magnons and thus preserving the
localization condition    
leads to a  lowering of magnetic energy that is proportional to the displacement of
sites. By calculating the total energy which consists of the magnetic and elastic parts,
one can check whether the lattice distortion is favorable or not. 
It appears that, e.g., the kagome lattice exhibits a field-driven lattice deformation
(spin-Peierls transition) in the vicinity of the saturation field
which is accompanied by a hysteresis phenomenon.\cite{sp-Peierls}

At nonzero but low temperatures 
in the vicinity of the saturation field  an extra low-temperature maximum of the specific heat
occurs, indicating that the manifold of localized eigenstates
sets an additional low-energy scale.
Importantly, if the lattice dimension is two (or three)
an order-disorder phase transition occurs at low temperatures just below the saturation field.\cite{mike1,epjb2006}
The phase transition is related to the ordering of the hard-core objects on
an auxiliary lattice,
which mimics the ordering of the localized magnons.

Further details can be found in the original papers or review articles on localized-magnon systems.\cite{review1,review2}
As will be shown below  some of these features appear also in flat-band Hubbard systems
demonstrating that the construction and characterization of many-particle localized states 
is precisely at the heart of the fact that, to some extent, a unified consideration of spin and electron models is possible.
However, there are also important differences 
owing to the spin degree of freedom of electrons, the Pauli principle, and the on-site repulsion
which may lead to a coherent spin state at $T=0$, 
i.e., to the flat-band ferromagnetism of Mielke and Tasaki.\cite{tasaki_rev1,tasaki_rev2,tasaki_rev3}

\section{Emergent Ising Degrees of Freedom}
\label{sec03}

As it has been discussed already, the manifold of localized-magnon states can be completely characterized, 
i.e., their energies are obtained trivially from the flat-band energy $\varepsilon_1$
and their degeneracies can be obtained as a canonical partition function of a certain hard-core lattice-gas model.
In many cases the energy levels above the localized-magnon manifold are
unknown. 
However, in some cases one can construct the lowest-energy excited states,
too. By including these  lowest-energy excited states into a lattice-gas approach
one can substantially extend the localized-magnon description. 
Instead of a lattice gas with infinite repulsion (hard-core rule) we then arrive at a lattice gas with finite
repulsion.\cite{ising_degrees}

To be specific,
we consider the spin-$\frac{1}{2}$ Heisenberg antiferromagnet (\ref{001})
on the frustrated two-leg ladder (see Fig.~\ref{fig01}) and the frustrated bilayer (see Fig.~\ref{fig02}).
For these models one can introduce the total spin on a vertical $J_2$-bond  
${\bf{t}}_{m}={\bf{s}}_{m,1}+{\bf{s}}_{m,2}$, 
where $m=1,\ldots,{\cal{N}}=\frac{N}{2}$ runs either over the sites of a simple chain
(in case of the ladder) or the sites of square lattice (bilayer case).
Then the Hamiltonian of the models becomes
\begin{eqnarray}
\label{003}
H=\sum_m\left[\frac{J_2}{2}\left({\bf{t}}^2_m-\frac{3}{2}\right)-ht_m^z\right]+J_1\sum_{(ml)}{\bf{t}}_m\cdot {\bf{t}}_l,
\end{eqnarray}
where the second sum runs over the nearest-neighbor bonds of the linear
chain or the square lattice, respectively.
As it is evident from Eq.~(\ref{003}),
${\bf{t}}_{m}^2=t_m(t_m+1)$, $m=1,\ldots,{\cal{N}}$ are good quantum numbers
and the Hamiltonian of the models depends on the set of quantum numbers $\{t_m\}$, $t_m=0,1$.
As a result, a subset of $2^{{\cal{N}}}$ low-lying eigenstates can be calculated exactly\cite{ladders-gen}
and their contribution to thermodynamics can be estimated using a classical lattice-gas model.\cite{ising_degrees}

Let us consider the relevant eigenstates in more detail.
First we consider states which consist of $n$ singlets on the vertical bonds $m_1,\ldots,m_n$
(i.e., ${\bf t}^2_{m_i}=0$) 
and $N-n$ fully polarized triplets 
(i.e., ${\bf t}^2_m=2$, $t^z_m=1$) 
on the remaining vertical bonds.
If we forbid the occupation of neighboring vertical bonds by singlets (hard-core rule)
we are faced with independent localized-magnon states and the singlet on a vertical bond is the localized magnon.
The energy of the independent localized-magnon states is $E^{\rm{lm}}_n=E_{\rm{FM}}-n\varepsilon_1$,
$\varepsilon_1=h_1=h_{\rm{sat}}$
with $E_{\rm{FM}}={\cal{N}}J_1+{\cal{N}}\frac{J_2}{4}$, $\varepsilon_1=J_2+2J_1$ (ladder) 
and $E_{\rm{FM}}=2{\cal{N}}J_1+{\cal{N}}\frac{J_2}{4}$, $\varepsilon_1=J_2+4J_1$ (bilayer).
The degeneracy of the independent localized-magnon states $g_{\cal{N}}(n)$ is given by the canonical partition function 
${\cal{Z}}_{\rm{hc}}(n,{\cal{N}})$ of $n$ hard-core objects (hard dimers or hard squares) 
on the lattice (simple chain or square lattice) of ${\cal{N}}$ sites. 
The independent-localized magnons are linearly independent\cite{linear_in} 
and constitute the ground-state manifold for $S^z=N-1,\ldots,\frac{N}{2}$ 
if $J_2\ge 2J_1$ (ladder) or $J_2\ge 4J_1$ (bilayer).

Allowing the localized magnons to be nearest neighbors, we arrive at another set of eigenstates, the interacting localized-magnon states. 
For magnetization values $S^z={\cal{N}}-n$, $n=2,\ldots,\frac{{\cal{N}}}{2}$,
the energies of the interacting localized-magnon excited states are:
$E^\nu_n=E^{\rm{lm}}_n+\nu J_1$, 
where $\nu$ is the number of pairs of neighboring localized magnons.
The interacting localized-magnon states are the low-lying excited states for these values of $S^z$ if $J_2$ is large
enough, i.e.,  $J_2>J^c_2$
(strong-coupling regime). 
From exact-diagonalization data for finite systems we have found for $J^c_2$ above which the strong-coupling regime holds the
values
$J^c_2\approx 3.00 J_1$ (ladder) 
and
$J^c_2\approx 4.65J_1$ (bilayer).
Obviously, for the models at hand the localized-magnon ground states are separated by a gap from excitations in the strong-coupling regime, 
i.e., for $J_2>J^c_2$. 
Note, however, that the existence of an excitation gap is not necessarily a common feature of localized-magnon systems.

For lower values of the magnetization, 
$S^z=\frac{{\cal{N}}}{2}-r$, $r= 1,\ldots,\frac{{\cal{N}}}{2}$, 
where no independent localized-magnon states exist, 
the class of interacting localized-magnon states contains the ground-state manifold as well as low-lying excited states in the strong-coupling regime.
The ground-state manifold is built by $n=\frac{{\cal{N}}}{2} +r$ ($r=
1,\ldots,\frac{{\cal{N}}}{2}$) localized magnons,
where, 
e.g., 
$\frac{{\cal{N}}}{2}$ magnons occupy one sublattice of the underlying lattice (simple chain or the square lattice) completely, 
and the remaining $r$ localized magnons sit on the other sublattice. 
The energy of this state is 
$E_{\frac{{\cal{N}}}{2}+r}=-{\cal{N}}\frac{J_2}{4}-rJ_2$, $J_2=h_2$. 
The low-lying excited states are constructed from the ground state by rearranging the localized magnons 
to increase the number of neighboring magnons. 
Then each new pair of neighboring localized magnons increases the energy by $J_1$.

Note that the interacting localized-magnon states can be visualized as partially overlapping hard-core objects 
in contrast to the independent localized-magnon states which can be visualized as non-overlapping hard-core objects. 
We are interested in the partition function $Z(T,h,N)$ of the frustrated quantum spin system 
which, after taking into account that in the strong-coupling regime 
the independent and interacting localized-magnon states dominate at low temperatures and high fields,
can be written within the frames of a lattice-gas model of classical particles with finite nearest-neighbor repulsion $V=J_1$
\begin{eqnarray}
\label{004}
Z(T,h,N)\approx\sum_{n_1=0,1}\ldots \sum_{n_{\cal{N}}=0,1}
e^{-\frac{E_{\rm{FM}}-h{\cal{N}}+(h-h_1)\sum_m n_m+J_1\sum_{(ml)} n_mn_l}{T}}
\nonumber\\
=e^{-\frac{E_{\rm{FM}}-h{\cal{N}}}{T}}\Xi_{\rm{lg}}(T,\mu,{\cal{N}}),
\end{eqnarray}
where 
$\Xi_{\rm{lg}}(T,\mu,{\cal{N}})=\sum_{n_1=0,1}\ldots \sum_{n_{\cal{N}}=0,1}e^{-\frac{{\cal{H}}(\{n_m\})}{T}}$ 
is the grand-canonical partition function of the classical lattice-gas model with the Hamiltonian
${\cal{H}}(\{n_m\})=-\mu\sum_m n_m+J_1\sum_{(ml)} n_mn_l$, $\mu=h_1-h$,
and $h_1=\varepsilon_1$ is the saturation field $h_{\rm{sat}}$.
Introducing the Ising variables $\sigma_m=2n_m-1$ (i.e., $n_m=\frac{1+\sigma_m}{2}$) 
one arrives at a classical antiferromagnetic Ising model in a uniform magnetic field.
Extensive studies of the relation between the frustrated quantum spin models and the classical lattice-gas models for finite systems (up to $N= 32$) show 
that the effective description 
[i.e., the (non-frustrated) Ising model]
can reproduce the thermodynamic properties of the initial frustrated quantum spin model in the strong-coupling regime.\cite{ising_degrees} 

\begin{figure}[bt]
\centerline{\psfig{file=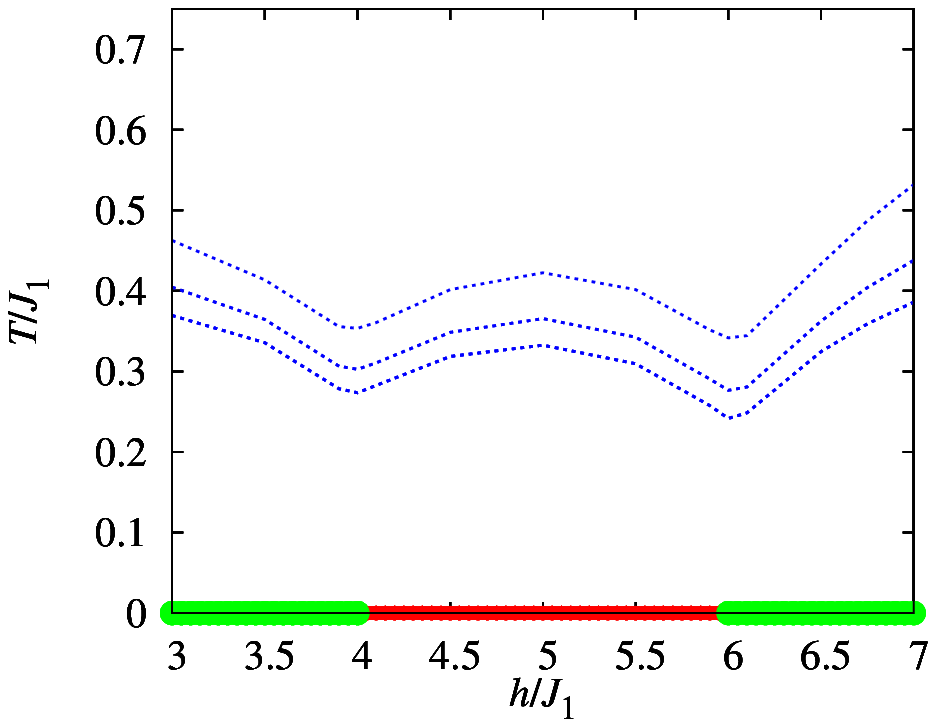,width=2.25in}\hspace{10mm}\psfig{file=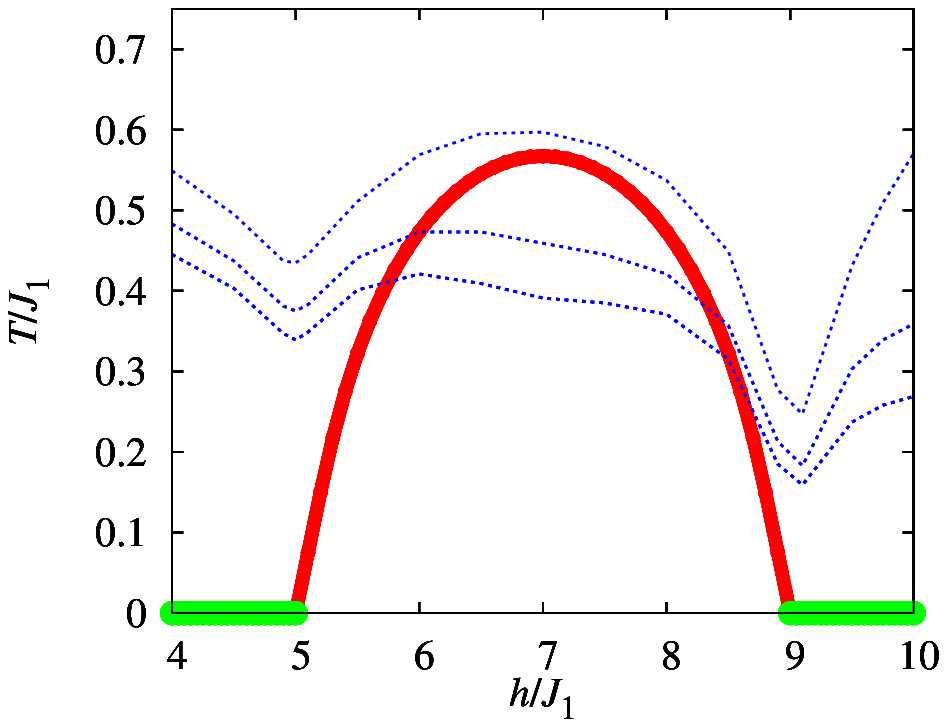,width=2.25in}}
\vspace*{8pt}
\caption{(Color online)
Phase diagrams as they follow from the lattice-gas description 
of the $s=\frac{1}{2}$ Heisenberg antiferromagnet in a magnetic field 
on the frustrated two-leg ladder  with $J_1=1$, $J_2=4$ 
(left)
and 
on the frustrated bilayer  with $J_1=1$, $J_2=5$
(right). 
A staggered occupation of vertical bonds by localized magnons 
(a kind of ``antiferromagnetic'' long-range order) 
occurs along the red line between $h_2=J_2=4$ and $h_1=J_2+2J_1=6$ at $T=0$ (ladder) 
or below the red critical line $T_c(h)$ with starting point  $h_2=J_2=5$, $T=0$ and endpoint
$h_1=J_2+4J_1=9$, $T=0$ (bilayer),
whereas a uniform occupation of vertical bonds 
(a kind of ``ferromagnetic'' long-range order) 
occurs along the green lines $h < h_2$ at $T=0$ and $h_1< h$ at $T=0$. 
The remaining part of the phase diagrams corresponds to a disordered phase. 
Moreover, we show the lines below which the exact-diagonalization data and the lattice-gas predictions 
for the specific heat of the finite system of
$N=16$ sites coincide with the accuracy up to 5\%, 2\%, and 1\% 
(blue dashed lines from top to bottom).}
\label{fig03}
\end{figure}
\begin{figure}[bt]
\centerline{\psfig{file=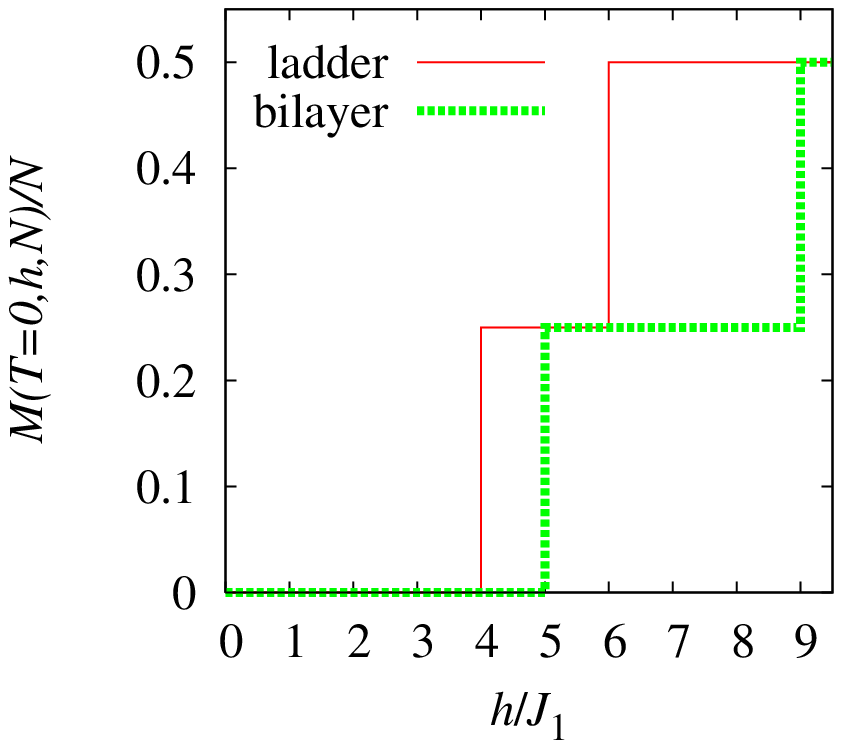,width=2.25in}\hspace{10mm}\psfig{file=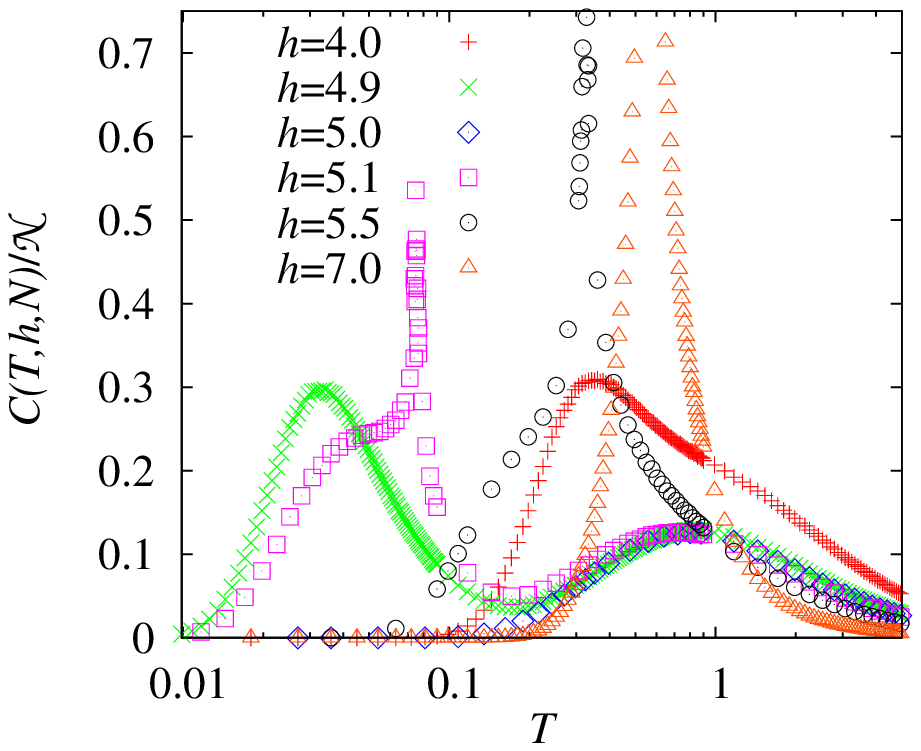,width=2.25in}}
\vspace*{8pt}
\caption{(Color online)
Zero-temperature magnetization curves for the ladder ($J_1=1$, $J_2=4$) and bilayer ($J_1=1$, $J_2=5$) models (left) 
and  specific heat as a function of temperature for the bilayer model 
for various magnetic fields below the saturation field (right). 
The plateau at
$\frac{M}{N}=\frac{1}{4}$ is between $h_2=J_2$ and the saturation field $h_1$.
The specific-heat data were obtained by Monte Carlo simulations of the classical lattice-gas model with finite repulsion $V=J_1$
for systems of up to $800\times 800$ sites exploiting usual Metropolis algorithm
with up to 3\,000\,000 Monte Carlo steps.} 
\label{fig04}
\end{figure}

Since the emergent classical models may be easily studied either analytically (ladder) or using classical Monte Carlo simulations
(bilayer),
a lot of results for these models are available 
(see, e.g., Refs.~\refcite{baxter,sq_lat_afm_in_a_field}).
Hence we can discuss the properties of the frustrated quantum antiferromagnets
at hand on the basis of this knowledge.
In Figs.~\ref{fig03} and \ref{fig04} we report the phase diagram of both models in the half plane
``field--temperature'', 
the low-temperature magnetization curves and the temperature dependence of the specific heat below the saturation field for the 
bilayer. 
An interesting low-temperature behavior in the magnetic-field region $h_2<h<h_1$ is related to an ordering of localized magnons.
Indeed, both auxiliary lattices are bipartite ones
and in the two-dimensional case (i.e., for the square lattice) 
the occupation of the two sublattices exhibits a finite-temperature order-disorder phase transition
corresponding to that one which is present in the square-lattice Ising antiferromagnet in a uniform magnetic
field.\cite{sq_lat_afm_in_a_field}
Crossing the critical line $T_c(h)$
the thermodynamic quantities for the frustrated bilayer exhibit singularities: 
The specific heat shows a logarithmic singularity, 
the staggered magnetization
(order parameter) within the (``antiferromagnetically'') ordered phase
decays to zero with the critical exponent $\beta=\frac{1}{8}$, and 
the staggered susceptibility diverges with the exponent $\gamma=\frac{7}{4}$.
These field and temperature dependences of various thermodynamic quantities
clearly indicate the ordering of the localized magnons on vertical bonds of the frustrated bilayer
which may occur for $T < T_{{\rm{Onsager}}}=\frac{1}{2 \ln(\sqrt{2}+1)}J_1\approx 0.567 296J_1$. 
The investigations  presented in Ref.~\refcite{ising_degrees} can be obviously extended to other
two-dimensional lattices, 
for example, frustrated bilayers consisting of two triangular, honeycomb, kagome
etc. layers.
Furthermore,
these findings may be of relevance 
to understand the physical properties of the recently synthesized spin dimer magnet Ba$_2$CoSi$_2$O$_6$Cl$_2$.\cite{tanaka}
Although, this compound has a pronounced $XY$ anisotropy of the antiferromagnetic Heisenberg interactions
and certainly also a deviation from ideal flat-band geometry 
remnant features of the discussed bilayer system could be present,
cf. also Sec.~\ref{sec072}.

\section{Chiral Localized Magnons}
\label{sec04}

Most of the investigations of localized-magnon systems consider the case of a bipartite trapping cell, 
e.g., a single bond, a square or a hexagon.
Such cells have a non-degenerate one-magnon ground state 
which remains the ground state of the whole lattice after suitable connection of the cells into the infinite lattice.
However, one can also consider a non-bipartite trapping cell,
e.g., a triangular trap. 
The triangular trap has a twofold degenerate ground state that leads to new effects.\cite{chiral1,chiral2,double_tetrahedra,fr_three}
This degeneracy is associated to the chirality.
The chirality operator for the triangle may be introduced as follows:
\begin{eqnarray}
\label{005}
\chi=\frac{4}{\sqrt{3}}({\bf{s}}_1\cdot [{\bf{s}}_2\times {\bf{s}}_3])
\end{eqnarray}
(see the corresponding triangular cells in Fig.~\ref{fig01}).
As a result,
the lowest-energy magnon band having the energy
$\varepsilon_{1,2}(\kappa)=-\frac{3}{2}J_2-J_1=-\varepsilon_{1,2}$
(double-tetrahedra chain, see Fig.~\ref{fig01})
or
$\varepsilon_{1,2}(\kappa)=-\frac{3}{2}J_2-3J_1=-\varepsilon_{1,2}$
(frustrated triangular tube, see Fig.~\ref{fig01})
is twofold degenerate, 
i.e., the one-magnon flat-band states are $2{\cal{N}}$-fold degenerate.
The flat-band states are located within triangles and can be written explicitly as follows:
\begin{eqnarray}
\label{006}
\vert +\rangle_m=\frac{1}{\sqrt{3}}\left(s_{m,1}^-+\omega s_{m,2}^-+\omega^2 s_{m,3}^-\right)\vert\uparrow\ldots\uparrow\rangle,
\nonumber\\
\vert -\rangle_m=\frac{1}{\sqrt{3}}\left(s_{m,1}^-+\omega^2 s_{m,2}^-+\omega s_{m,3}^-\right)\vert\uparrow\ldots\uparrow\rangle,
\end{eqnarray}
where $\omega=e^{\frac{2\pi}{3}i}$.
Using Eqs.~(\ref{005}) and (\ref{006}) one can check that $\chi_m\vert\pm\rangle_m=\pm \vert\pm\rangle_m$.
Moreover,
the flat-band states become the lowest-energy ones if $J_2>2J_1$.

We pass  to the low-temperature high-field thermodynamics which is conditioned by the set of independent localized-magnon states.
The formula for the (dominating) contribution of the localized-magnon states to the partition
function reads
\begin{eqnarray}
\label{007}
Z(T,h,N) \approx \sum_{n=0}^{n_{\max}} g_{\cal{N}}(n) e^{-\frac{E_{\rm{FM}}-\frac{N}{2}h-n(h_1-h)}{T}},
\end{eqnarray}
where $h_1=h_{\rm{sat}}=\varepsilon_{1,2}$ is the saturation field  and
$g_{\cal{N}}(n)$
is the number of localized states of $n$ magnons on a lattice of ${\cal{N}}$
cells.
For $g_{\cal{N}}(n)$ 
now we have to take into account also the chiral degrees of freedom.
That yields
$g_{\cal{N}}(n)=2^n{\cal{C}}_{\cal{N}}^n$ 
for the double-tetrahedra chain
and
$g_{\cal{N}}(n)=2^n{\cal{Z}}_{\rm{hd}}(n,{\cal{N}})$ for the frustrated triangular tube,
where ${\cal{C}}_{\cal{N}}^n={{\cal{N}}\choose{n}}$ is the binomial coefficient 
and ${\cal{Z}}_{\rm{hd}}(n,{\cal{N}})$ denotes the canonical partition function of $n$ hard dimers on a ${\cal{N}}$-site linear chain.
Here the factor $2^n$ stems from the chirality.
The sum in Eq.~(\ref{007}) can be easily calculated\cite{chiral1}
and, as a result, one gets the thermodynamics of the models at hand in the low-temperature high-field regime.
The considered spin systems exhibit the typical features of localized-magnon systems: 
The magnetization jump, the residual entropy, and the  extra low-temperature peak in the specific heat.
Furthermore, similar to the case of the frustrated two-leg ladder (see the previous section),
one can elaborate for the frustrated three-leg ladder
a soft lattice-gas model instead of the hard-dimer model 
to take into account the low-energy excitations.\cite{chiral1} 
Based on finite-size calculations
the critical value of $J_2$ above which the strong-coupling regime holds
was estimated as $J_2^c\approx 2.68J_1$.
Note that in contrast to the two-leg ladder, 
the lattice-gas model for the three-leg ladder does not possess the symmetry around $h_1$ and $h_2$,
because of the chirality.

\begin{figure}[bt]
\centerline{\psfig{file=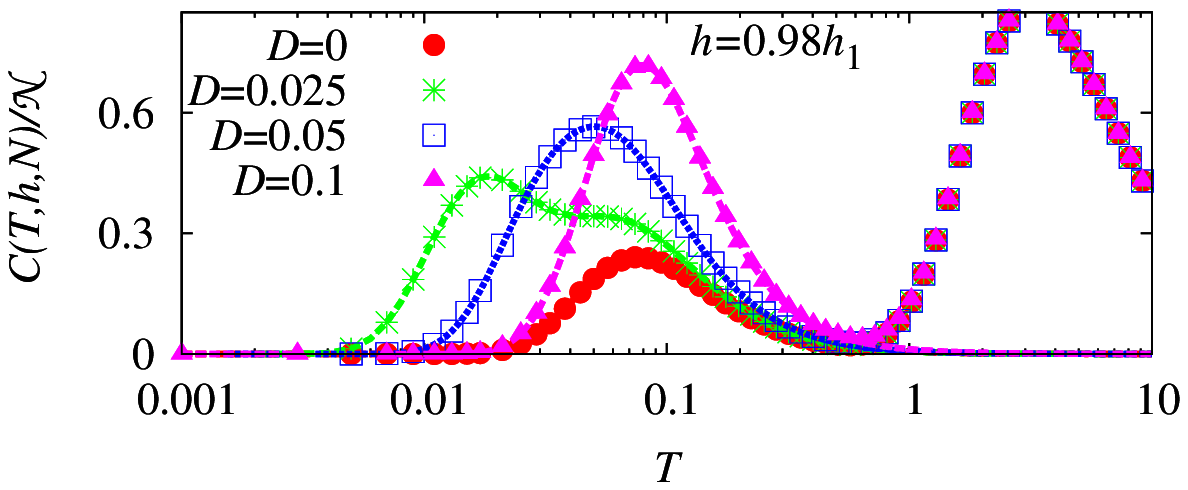,width=2.4in}\hspace{5mm}\psfig{file=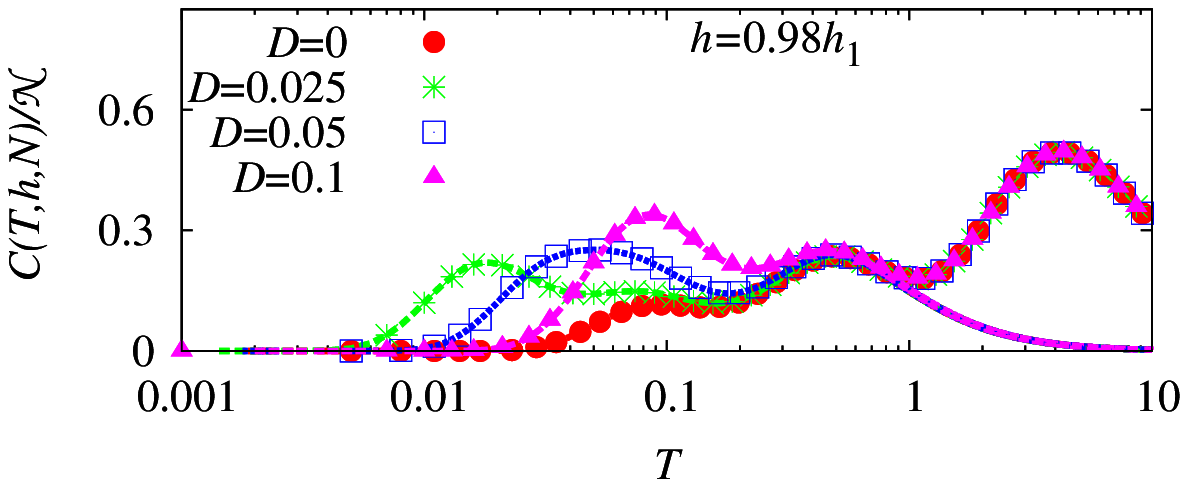,width=2.4in}}
\vspace*{8pt}
\caption{(Color online)
Low-temperature dependence of the specific heat for the double-tetrahedra chain (left) and frustrated triangular tube (right)
slightly below the saturation field $h=0.98h_1$ (${\cal{N}}=4$, $J_1=1$, $J_2=5$,
$D=0,\,0.025,\,0.05,\,0.1$).
Exact-diagonalization data (symbols) are compared to analytical predictions (lines).
}
\label{fig05}
\end{figure}

The degeneracy owing to the chirality can be lifted,
e.g., by a Dzyaloshinskii-Moriya interaction.
If we add a small perturbation 
in form of a Dzyaloshinskii-Moriya interaction
${\bf{D}}=D{\bf{e}}_z=(0,0,D)$ between neighboring spins of each triangular trap,
the partition function is again given by Eq.~(\ref{007}),
however,
with 
$g_{\cal{N}}(n)=(2\cosh\frac{\sqrt{3}D}{2T})^n{\cal{C}}_{\cal{N}}^n$ 
for the double-tetrahedra chain
and
$g_{\cal{N}}(n)=(2\cosh\frac{\sqrt{3}D}{2T})^n{\cal{Z}}_{\rm{hd}}(n,{\cal{N}})$ 
for the frustrated triangular tube.
The introduced perturbation leads to a splitting of the ground-state levels
and, therefore, one more low-energy scale
arises. 
As a result, the low-temperature features close to the saturation field $h_{\rm{sat}}=h_1$ are more subtle. 
For example, 
the temperature profiles of the specific heat show more tiny features which can be seen in Fig.~\ref{fig05}.
For a special set of parameters the temperature dependence of the specific heat
may exhibit three (four) maxima for the double-tetrahedra chain (frustrated triangular tube).
The low-temperature maxima are excellently described within the presented effective low-energy theory.

The considerations of this section refer to the antiferromagnetic Heisenberg model.
Later on in Sec.~\ref{sec081} we will discuss how  they can be straightforwardly applied to the Hubbard
model.\cite{chiral1,chiral2,double_tetrahedra}

\section{Beyond the Ideal Flat-Band Geometry}
\label{sec05}

The localized-magnon picture emerges only in the case of strict localization of magnons
or, in other words, when the lowest magnon band is strictly flat.
Of course, this requires certain relations between exchange couplings 
which hardly can be satisfied in a real-life case.
Among the reasons for that one may mention nonideal geometry or some extra
exchange couplings  which lead to a violation of the localization condition.
However, one may expect to find a magnetic compound which is quite close to the localized-magnon prototype.
Therefore it is important to consider spin models with a small deviation from
the strictly dispersionless case, i.e.,
the former flat band acquires a slight dispersion, 
and to figure out how the localized-magnon features are modified by (weak)
dispersion.

In what follows we discuss the effect of small deviations from the conditions for localization
considering three models which belong to the so-called monomer universality class,\cite{epjb2006}
namely,
the diamond chain (see Fig.~\ref{fig01}), 
the dimer-plaquette chain, 
and the square-kagome lattice (see Fig.~\ref{fig02}).\cite{deviations1,deviations2}
For large exchange bonds on the trapping cell, $J_2$,
an appropriate approach is the strong-coupling approximation\cite{ladders-gen,str-coup-appr}
which, in fact, was elaborated for a distorted frustrated diamond spin chain in
Refs.~\refcite{effective_xy2} and \refcite{aa}.
Let us discuss this approach in some detail considering as an example a distorted frustrated diamond chain, see Fig.~\ref{fig01}.
Inspired by the azurite geometry, we assume that $J_1 \ne J_3$ and set $J_1+J_3=2J<J_2$
(for a more general nonideal geometry see Ref.~\refcite{deviations2}).
The starting point of the strong-coupling approximation is the simple problem
of two spins interacting via a $J_2$ bond. 
In a high magnetic field 
the state $\vert u\rangle=\vert\uparrow_1\uparrow_2\rangle$ 
has the lowest energy $\frac{J_2}{4}-h$ 
whereas just below the ``bare'' saturation field $h_0=J_2$  
the state $\vert d\rangle=\frac{1}{\sqrt{2}}(\vert\uparrow_1\downarrow_2\rangle-\vert\downarrow_1\uparrow_2\rangle)$ (localized magnon)
becomes the lowest-energy one with the energy $-\frac{3J_2}{4}$. 
The collection of  spin dimers on $J_2$-bonds and the isolated spins at the site $m,3$, see Fig.~\ref{fig01}, at $h=h_0$ 
constitutes  the ``main'' Hamiltonian $H_{\rm{main}}$
whereas the remaining $J_1$ and $J_3$ interactions represent the perturbation $V=H-H_{\rm{main}}$.
We introduce the model space which is spanned by the ground states of the main Hamiltonian $\vert\varphi_0\rangle$
and the projector onto this space $P=\vert\varphi_0\rangle\langle\varphi_0\vert$
\begin{eqnarray}
\label{008}
P=\otimes_m P_m,
\;\;\;
P_m={\cal{P}}_m\otimes \left(\vert\uparrow_{3}\rangle\langle \uparrow_{3}\vert\right)_m,
\;\;\;
{\cal{P}}_m=\left(\vert u\rangle\langle u\vert+\vert d\rangle\langle d\vert\right)_m.
\end{eqnarray}
We are interested in an effective Hamiltonian $H_{\rm{eff}}$ which acts in the model space only
but gives the ground-state energy of the Hamiltonian $H$.
The effective Hamiltonian can be found within  perturbation theory\cite{klein,fulde,essler}
\begin{eqnarray}
\label{009}
H_{\rm{eff}}=PHP+PV\sum_{\alpha\ne 0}\frac{\vert\varphi_\alpha\rangle\langle\varphi_\alpha\vert}{\varepsilon_0-\varepsilon_\alpha}VP+\ldots ,
\end{eqnarray}
where $\vert\varphi_\alpha\rangle$, $\alpha\ne 0$, denotes the excited states of $H_{\rm{main}}$.
Moreover,
it is convenient to introduce (pseudo)spin-$\frac{1}{2}$ operators
\begin{eqnarray}
\label{010}
T^z=\frac{1}{2}\left(\vert u\rangle\langle u\vert - \vert d\rangle\langle d\vert\right),
\;\;\;
T^+=\vert u\rangle\langle d\vert,
\;\;\;
T^-=\vert d\rangle\langle u\vert
\end{eqnarray}
for each cell $m$.
As a result, one arrives at the spin-$\frac{1}{2}$ isotropic $XY$ model in a
$z$-aligned field with the Hamiltonian
\begin{eqnarray}
\label{011}
H_{\rm{eff}}={\cal{N}}{\sf{C}}-{\sf{h}}\sum_mT_m^z+{\sf{J}}\sum_{(mn)}\left(T_m^xT_n^x+T_m^yT_n^y\right),
\end{eqnarray}
where the last sum runs over all nearest-neighbor bonds on a simple chain.
The parameters of the effective Hamiltonian are given by 
${\sf{h}}=h-h_1-\frac{(J_3-J_1)^2}{4J_2}$,
${\sf{J}}=\frac{(J_3-J_1)^2}{4J_2}$,
and $h_1=J_2+J$, $J=\frac{J_3+J_1}{2}$, see Refs.~\refcite{aa,effective_xy2,deviations2}.
For the dimer-plaquette chain and the square-kagome lattice the effective model is again given by the Hamiltonian in Eq.~(\ref{011}),
however, with specific values of parameters, see Ref.~\refcite{deviations2}.

\begin{figure}[bt]
\centerline{\psfig{file=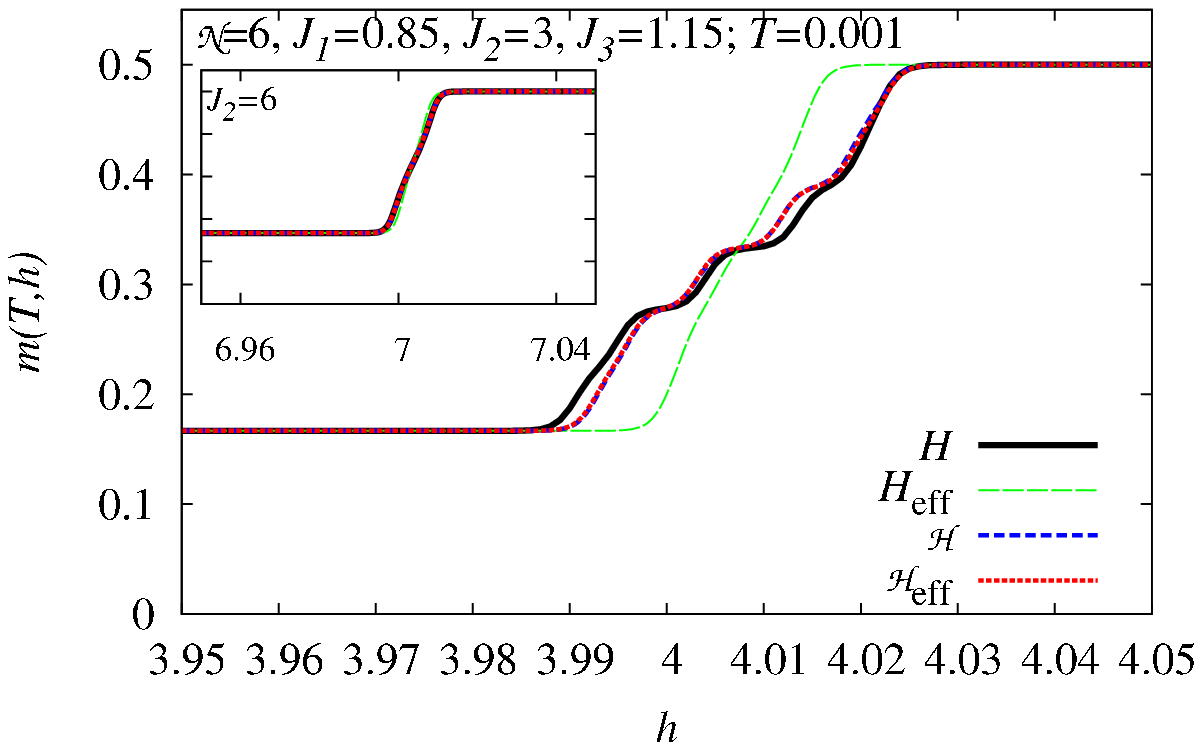,width=2.4in}\hspace{3mm}\psfig{file=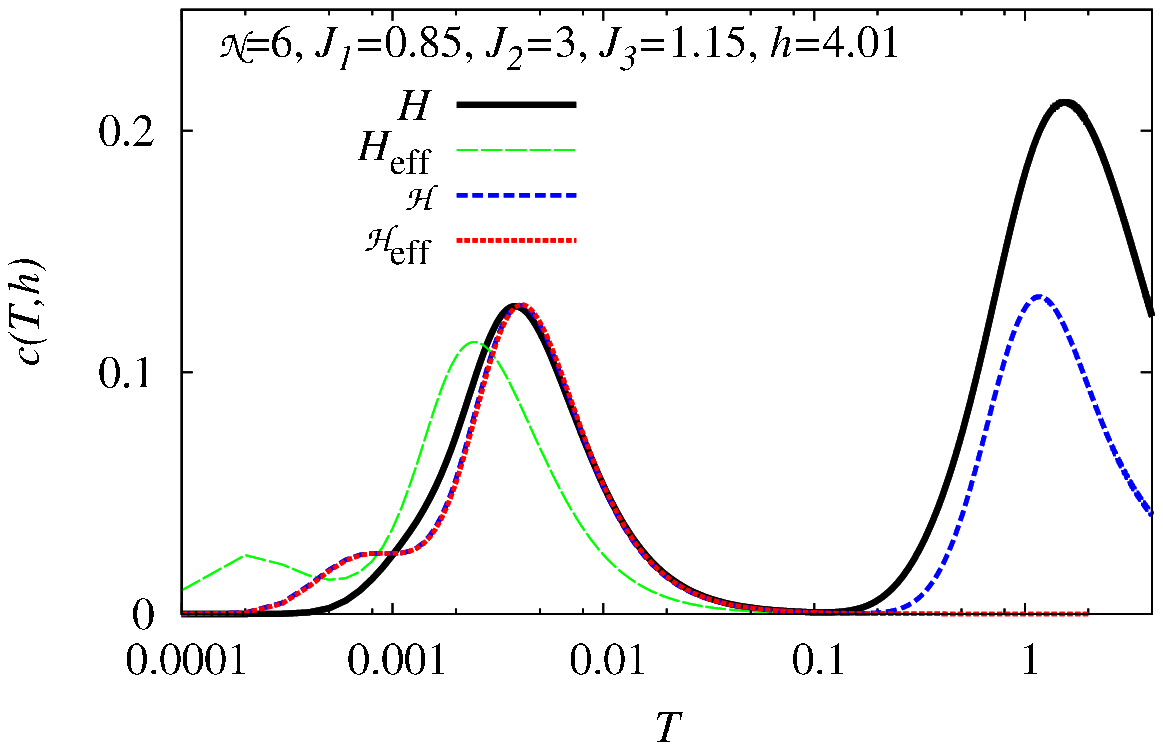,width=2.4in}}
\vspace*{8pt}
\caption{(Color online)
Magnetization curves (left) and low-temperature dependence of the specific heat (right) for a distorted frustrated diamond chain
(${\cal{N}}=6$, $J_1=0.85$, $J_2=3$, $J_3=1.15$).
Exact-diagonalization data (black bold solid curves) versus effective-model predictions,
namely,
strong-coupling approach (\ref{011}) (green long-dashed curves)
and 
localized-magnon approach (\ref{012}) (blue short-dashed curves) and (\ref{013}) (red dotted curves).}
\label{fig06}
\end{figure}

In Fig.~\ref{fig06} we illustrate the quality of the strong-coupling predictions (long-dashed green lines).
For this purpose we compare strong-coupling-approach predictions with exact-diagonalization data for the initial model.
The comparison of data for the magnetization and the specific heat reported in Fig.~\ref{fig06} shows
that the strong-coupling theory gives correct qualitative predictions 
but it does not provide quantitative details.
That is not astonishing after all,
since we start from the limit of small $J_1$ and $J_3$ in comparison to $J_2$ and
consider only the second-order term in Eq.~(\ref{009}).

The strong-coupling calculation does not use the special features of localized-magnon
picture, where we assume $J_1=J_3=J<\frac{J_2}{2}$.
Starting from the localized-magnon description we need only $\vert J_3-J_1\vert$ to be
small.
Hence, one may expect that such an approach would contain more relevant physics than the strong-coupling one already in the ``main''
(i.e., unperturbed) 
problem. To follow this idea we may apply Eq. (\ref{009}) for the following splitting of the Hamiltonian $H$:
$H_{\rm{main}}$ corresponds to the ideal geometry
(i.e., with $J=\frac{J_1+J_3}{2}$ instead of $J_1$ and $J_3$ at $h=h_1$ with $h_1$ denoting the saturation field for the ideal-geometry case)
and $V=H-H_{\rm{main}}$.
However, we are immediately faced with a difficulty in calculating the second term in the r.h.s. of Eq.
(\ref{009}),
since only the ground states of $H_{\rm{main}}$ (i.e., the set of localized-magnon states)
are known,
whereas the excited states $\vert\varphi_\alpha\rangle$, $\alpha\ne 0$, remain unknown.
To overcome this obstacle we start from the Hamiltonian
\begin{eqnarray}
\label{012}
{\cal{H}}={\cal{P}}H{\cal{P}},
\;\;\;
{\cal{P}}=\otimes_m{\cal{P}}_m.
\end{eqnarray}
Here the projector ${\cal{P}}_m$ [see Eq.~(\ref{008})] restricts the states of the cell $m$ to only two states $\vert u\rangle$ and $\vert d\rangle$.
After this approximation we can split the Hamiltonian ${\cal{H}}$
into the main part ${\cal{H}}_{\rm{main}}$ which corresponds to the ideal geometry
and the perturbation ${\cal{V}}={\cal{H}}-{\cal{H}}_{\rm{main}}$.
Now all eigenstates of ${\cal{H}}_{\rm{main}}$ are known and we can easily calculate the second term in the r.h.s. of Eq. (\ref{009}).
As a result,
for the considered frustrated quantum antiferromagnets we arrive at the spin-$\frac{1}{2}$ $XXZ$ Heisenberg model
\begin{eqnarray}
\label{013}
{\cal{H}}_{\rm{eff}}={\cal{N}}{\sf{C}}-{\sf{h}}\sum_mT_m^z
+\sum_{(mn)}\left[{\sf{J}}\left(T_m^xT_n^x+T_m^yT_n^y\right)+{\sf{J}}^zT_m^zT_n^z\right].
\end{eqnarray}
The parameters of ${\cal{H}}_{\rm{eff}}$ for the distorted diamond chain are
given by
\begin{eqnarray}
\label{014}
{\sf{C}}=-h-\frac{J_2}{4}+\frac{J}{2}-\frac{(J_3-J_1)^2}{16J_2}\left(\frac{1}{1-\frac{J}{J_2}}+1\right)
\nonumber\\
=-h-\frac{J_2}{4}+\frac{J}{2}-\frac{(J_3-J_1)^2}{16J_2}\left(2 +\frac{J}{J_2}+\ldots\right),
\nonumber\\
{\sf{h}}=h-h_1-\frac{(J_3-J_1)^2}{4J_2},
\;\;\;
h_1=J_2+J,
\;\;\;
J=\frac{J_3+J_1}{2},
\nonumber\\
{\sf{J}}=\frac{(J_3-J_1)^2}{4J_2}\frac{1}{1-\frac{J}{J_2}}
=\frac{(J_3-J_1)^2}{4J_2}\left(1+\frac{J}{J_2}+\ldots\right),
\nonumber\\
{\sf{J}}^z=\frac{(J_3-J_1)^2}{4J_2}\left(\frac{1}{1-\frac{J}{J_2}}-1\right)
=\frac{(J_3-J_1)^2}{4J_2}\left(\frac{J}{J_2}+\ldots\right).
\end{eqnarray}
To compare with the 
strong-coupling approach  we give in Eq. (\ref{014}) except the full expressions also the leading
terms of an  expansion in $\frac{J}{J_2}$. Obviously,
the lowest-order terms agree with the strong-coupling approximation
(\ref{011}). Note that the next-order terms in Eq. (\ref{014})  correspond to
the next-order terms within the strong-coupling scheme obtained in Ref.~\refcite{aa}.
The effective Hamiltonian parameters for the dimer-plaquette chain and the square-kagome lattice 
(even for a more general nonideal geometry)
can be found in Ref.~\refcite{deviations2}.
From Fig.~\ref{fig06} one can clearly see that the perturbation-theory approach based on the
localized-magnon picture 
yields much better results than the strong-coupling approach, and it can provide even
a quantitative description of the initial model.
We mention, that the
ideal-geometry limit of the effective model gives an alternative (but identical) description of the thermodynamics 
previously discussed, e.g., in the reviews~\refcite{review2,review3}.

We conclude that the nonideal geometry leads to new effective (unfrustrated) quantum spin models 
which are well known in statistical mechanics.
Hence, we may use the broad knowledge on their properties to understand the low-temperature high-field behavior of the frustrated quantum antiferromagnets at hand.
The most interesting model is the two-dimensional square-kagome lattice,
since the effective models exhibit the famous Berezinskii-Kosterlitz-Thouless (BKT) transition.\cite{bkt}

Let us 
briefly outline the relevant results concerning the BKT transition.
Since the early 1970ies it is known
that the classical two-dimensional isotropic $XY$ model undergoes a transition 
from bound vortex-antivortex pairs at low temperatures 
to unpaired vortices and antivortices at some critical temperature $T_{\rm{BKT}}$.\cite{bkt} 
Below $T_{\rm{BKT}}$ (superfluid phase) the system is characterized by quasi-long-range order,
i.e., correlations decay algebraically at large distances without emerging of a nonvanishing
order parameter.
Above $T_{\rm{BKT}}$ (normal phase) the system is disordered with an exponential increase of the correlation length $\xi$ 
as $T\to T_{\rm{BKT}}+0$,
$\xi\propto e^{\frac{b}{\sqrt{\tau}}}$ with $\tau=\frac{T-T_{\rm{BKT}}}{T_{\rm{BKT}}}$.
The critical temperature for the classical square-lattice isotropic $XY$ model (without
magnetic field) is  $T_{\rm{BKT}} \approx 0.893\vert{\sf{J}}\vert$.\cite{bkt_tc}
In the quantum spin-$\frac{1}{2}$ case the BKT critical behavior also occurs, 
however, the critical temperature is lower,
$T_{\rm{BKT}}\approx 0.34\vert {\sf{J}}\vert$.\cite{bkt_q}
The quantum model is gapless with an excitation spectrum which is linear in the momentum.
The specific heat $c(T)$ shows $T^2$ behavior as $T\to 0$, increases very rapidly around $T_{\rm{BKT}}$,
and exhibits a finite peak at temperature $T^*$ somewhat above $T_{\rm{BKT}}$.
This kind of the low-temperature thermodynamics survives for not too large $z$-aligned magnetic field $\vert{\sf{h}}\vert <2\vert{\sf{J}}\vert$.\cite{bkt_q}
Also for spin-$\frac{1}{2}$ square-lattice $XXZ$ model with dominating isotropic $XY$ interaction in a $z$-aligned magnetic field 
the BKT transition appears.\cite{bkt_xxz}

\begin{figure}[bt]
\centerline{\psfig{file=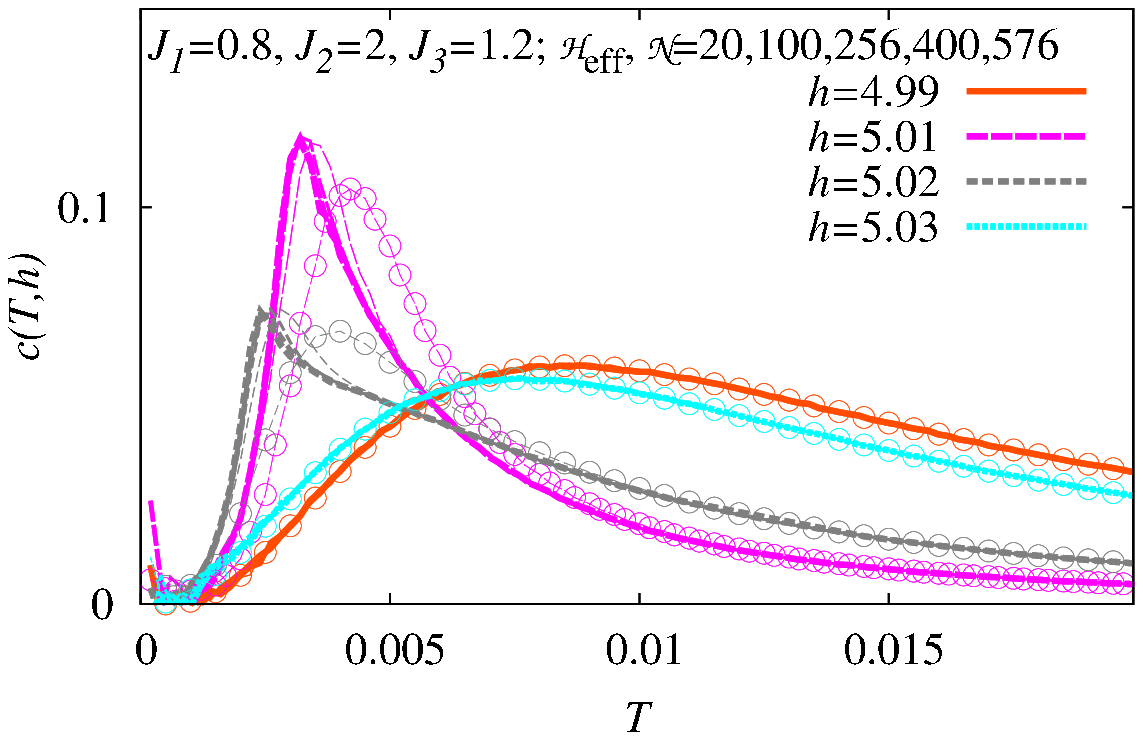,width=2.25in}\hspace{10mm}\psfig{file=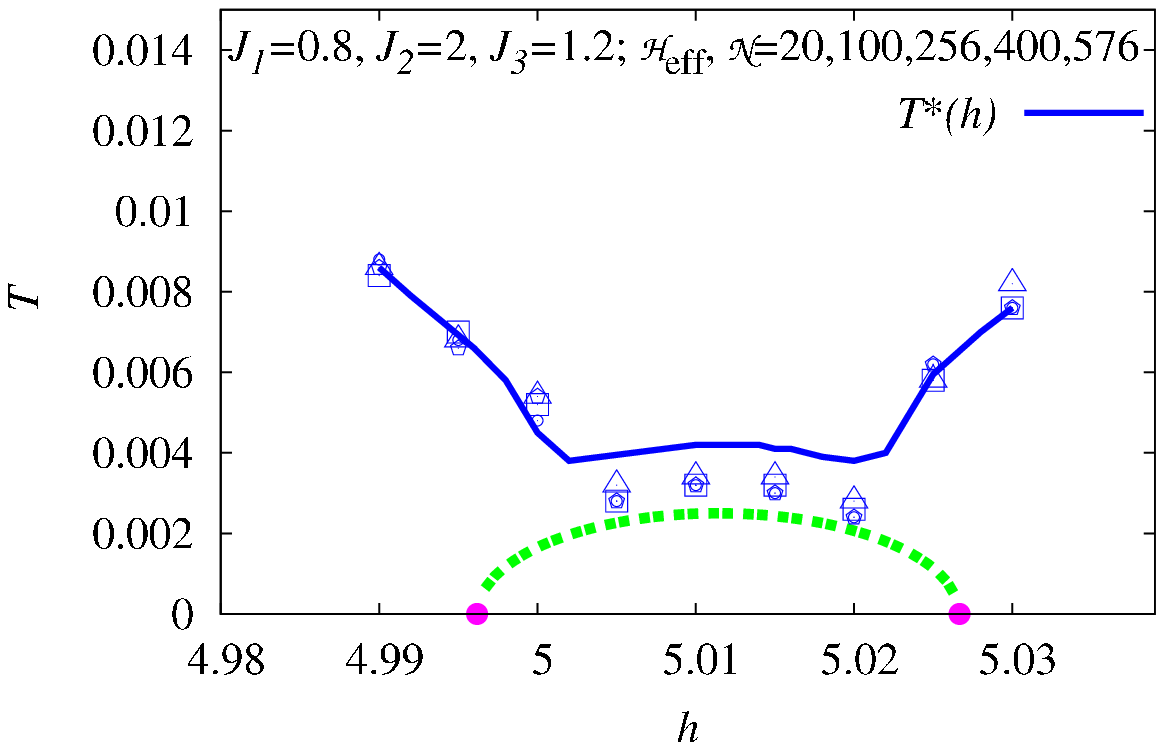,width=2.25in}}
\vspace*{8pt}
\caption{(Color online)
Distorted square-kagome lattice with $J_1=0.8$, $J_2=2$,
$J_3=1.2$.
Left panel: Low-temperature specific heat per site $c(T,h)$ at high
fields 
obtained by quantum Monte Carlo simulations for the effective Hamiltonian
(\ref{013}). The line thickness increases with
increase of ${\cal N}$. For comparison, we also show exact-diagonalization data for
${\cal N}=20$ by very thin curves with circles.
Right panel: Sketch of the phase diagram in the ``field--temperature''
half plane.
The sketch of the phase boundary (thick dashed
green line) is obtained from the position of the maximum $T^*$ in $c(T,h)$
that is shown by the thick solid blue curve (${\cal N}=20$) and the blue symbols 
(triangles: ${\cal N}=100$, 
squares: ${\cal N}=256$, 
pentagons: ${\cal N}=400$, 
circles: ${\cal N}=576$).
The endpoints of  phase boundary
at $h\approx 4.996$ and $h\approx 5.027$ are indicated by filled violet circles.}
\label{fig07}
\end{figure}

Now we transfer this knowledge 
to the distorted square-kagome Heisenberg antiferromagnet.
We use the position of the maximum $T^*$ in the specific heat as an indicator of the BKT transition
to construct a sketch of the phase diagram of the model, see Fig.~\ref{fig07}.
From the temperature dependence of the specific heat (see left panel of
Fig.~\ref{fig07})  calculated  by exact diagonalizations and quantum Monte Carlo simulations
for the effective model (\ref{013})
one observes an agreement with the expected $T^2$ behavior for $T\to 0$ in a small region below the saturation field
(and an exponential decay outside this region).
For gapped systems there is a broad maximum in the specific heat.
In the gapless field region the maximum occurs at lower temperatures and it is more pronounced becoming
peak-like.
Furthermore, in the gapless field region the temperature profiles show noticeable finite-size effects.
A tentative BKT-transition line $T_{{\rm{BKT}}}(h)$ 
(green short-dashed curve in the right panel of Fig.~\ref{fig07})
drawn on the basis of the specific heat data 
provides a generic phase diagram of the distorted square-kagome Heisenberg antiferromagnet at low temperatures and high fields.
The largest transition temperature appears for zero effective field ${\sf{h}}$,
which corresponds to 
$h=h_1+\frac{(J_3-J_1)^2}{8J_2}\frac{1}{1-\frac{J}{4J_2}}$ ($h_1=2J_2+J$, $J=\frac{J_3+J_1}{2}$) of the initial
frustrated model.
A deviation from this field yields a decrease of $T_{\rm{BKT}}$, and finally
$T_{\rm{BKT}}$ becomes zero.
Thus, in the highly frustrated quantum Heisenberg antiferromagnet on the square-kagome lattice with deviations from ideal flat-band geometry 
the BKT  transition may appear in a certain region of finite magnetic-field values,
i.e., the model exhibits a magnetic-field-driven BKT transition.

Interestingly, 
recent experimental and theoretical studies of the complex spin-dimer material C$_{36}$H$_{48}$Cu$_2$F$_6$N$_8$O$_{12}$S$_2$
(having a spin interaction network corresponding to a two-dimensional distorted honeycomb
lattice)
have given evidence of a field-induced BKT physics in this system.\cite{tutsch}

\section{Closely Related Studies on Spin Systems}
\label{sec06}

\subsection{Bose condensation in flat bands} 
\label{sec061}

One interesting study related to localized-magnon physics 
that has been recently presented by S.~D.~Huber and E.~Altman in Ref.~\refcite{huber-altman} 
deals with Bose condensation in flat bands.
The authors of Ref.~\refcite{huber-altman} consider a Hamiltonian of the form
\begin{eqnarray}
\label{015}
H=\sum_{(ij)}\vert t_{ij}\vert\left(b_i^\dagger b_j+{\rm{H.c.}}\right)
+\frac{U}{2}\sum_i b_i^\dagger b_i^\dagger b_i b_i,
\end{eqnarray}
where $b_i^\dagger$ ($b_i$) is a bosonic creation (annihilation) operator on lattice site $i$. 
Usually weakly interacting bosons form a condensate in the lowest eigenstate of the kinetic-energy operator.
However,
if the hopping parameters $\vert t_{ij}\vert$ provide a flat lowest band,
the nature of the ground state will be determined by the interaction $U$
acting within the hugely degenerate flat-band-states manifold.
S.~D.~Huber and E.~Altman consider the one-dimensional sawtooth lattice and the two-dimensional kagome lattice.\cite{huber-altman}
Both lattices have a lowest flat band in the one-particle spectrum.
Such bosonic models with flat bands are of direct relevance for ultracold atoms in optical lattices\cite{huber_exp1}
or diluted frustrated spin-$1$ magnets with fully polarized spins.\cite{huber_exp2}
Exact many-body ground states can be constructed by occupying some subsets
of spatially non-overlapping trapping cells, i.e., at sufficiently low lattice fillings
(densities of bosons) $\nu$.
Again, the multiple possibilities to put the bosons into the non-overlapping localized states lead to a massive degeneracy of the ground state. 
There is a critical density,
$\nu_{\max}=\frac{n_{\max}}{N}=\frac{1}{4}$ ($N$ is the number of lattice sites) for the sawtooth
chain 
and $\nu_{\max}=\frac{1}{9}$ for the kagome lattice 
at which a close-packed crystal state is formed.
So far, there is a close correspondence to the spin systems discussed in the preceding sections.
S.~D.~Huber and E.~Altman now consider lattice fillings slightly above
close-packed filling $\nu_{\max}$.
The wave function of any additional boson must overlap spatially with
those of other bosons and then an exact construction
of eigenstates is not possible.
The question addressed in Ref.~\refcite{huber-altman} is:
What is the structure of the ground states at lattice fillings slightly above $\nu_{\max}$?
In particular, do
the added bosons condense to form a superfluid on top of the
close-packed crystal state?
We will  briefly discuss this question  now.
The strategy to attack the problem follows a previous study of the quantum Hall effect:\cite{girvin}
The authors project the Hamiltonian (\ref{015}) onto the flat band 
and obtain this way  an effective low-energy Hamiltonian which depends only on the (weak)
on-site interaction $U$.
The resulting low-energy model is defined on a new lattice,
it is free of frustration and can be analyzed further using standard methods.

We start with the sawtooth-chain model.
On the one-particle level
the Hamiltonian (\ref{015}) corresponds to the spin-$\frac{1}{2}$ isotropic $XY$ Hamiltonian,
since the commutation relations are not important.
One easily finds that the localized states exist if $t_2=\sqrt{2}t_1$. They have the form
$\vert V_j\rangle=-\frac{1}{2}(b_{A,j-1}^\dagger -\sqrt{2}b_{B,j}^\dagger+b_{A,j}^\dagger)\vert {\rm{vac}}\rangle$
and their energy is $-2t_1=-2t$.
To achieve a better comparison with the original paper of
S.~D.~Huber and E.~Altman 
we use here the indices of
Ref.~\refcite{huber-altman}.
These indices $A,j$ and $B,j$ correspond to $m,2$ and $m,1$, respectively, in Fig.~\ref{fig01} and the lattice has $2{\cal{N}}$ sites.
On the many-particle level the particle statistics becomes important. 
One can construct many-boson ground states of the interacting Hamiltonian by occupying only non-overlapping $\vert V_j\rangle$ states.
In particular,
the crystal state [charge-density wave (CDW) state,
all non-overlapping trapping cells are occupied] is given by
$\vert {\rm{CDW}}_{\frac{1}{4}}\rangle=\prod_{j=1}^{\frac{{\cal{N}}}{2}}V_{2j}^\dagger \vert {\rm{vac}}\rangle$.
The states $\vert V_j\rangle$ are linearly independent and complete in the flat-band subspace, 
however, they are not orthogonal.
It is convenient to introduce an alternative set of localized states within the flat-band subspace which form an orthonormal basis 
\begin{eqnarray}
W_i^\dagger \vert {\rm{vac}}\rangle
=
\sum_j \left[w_A^*(r_j-r_i)b^\dagger_{A,j} + w_B^*(r_j-r_i)b^\dagger_{B,j}\right] \vert {\rm{vac}}\rangle,
\nonumber\\
w_A(r_j)=\frac{1}{2\pi}\int_{-\pi}^\pi d\kappa\cos\frac{\theta_\kappa}{2} e^{i\kappa r_j-i\frac{\kappa}{2}},
\;\;\;
w_B(r_j)=-\frac{1}{2\pi}\int_{-\pi}^\pi d\kappa\sin\frac{\theta_\kappa}{2} e^{i\kappa
r_j}
\end{eqnarray}
[for explicit expression for $\theta_\kappa$ see Eq.~(4) in Ref.~\refcite{huber-altman}].
Although these Wannier states (or $W$-states) locally have the structure of the $V$-states, 
the $W$-states decay exponentially with a localization length of $\xi\approx\log (2.15) a$,\cite{huber-altman}
where $a$ is the lattice constant.

Now additional particles are introduced into the system and
a small interaction $U<\Delta$ is assumed 
(for the opposite limit $U>\Delta$, 
see Ref.~\refcite{phillips}), 
where $\Delta=2t$ is the energy gap from the flat band to the next band.
To implement the projection onto the flat band,
we express the bosonic operators in terms of the Wannier basis,
$b_{A(B),j}^\dagger=\sum_m w_{A(B)}^*(r_j-r_m)W^\dagger_m+{\rm{higher\,bands}}$,
and neglect the contribution from the higher bands.
The resulting Hamiltonian contains a renormalized on-site interaction
$\tilde{U}\approx 0.40U$,
further-range interactions as well as assisted-hopping terms,
where all effective parameters are proportional to $U$.
Since the renormalized on-site interaction       
$\tilde{U}$ is the largest parameter,           
S.~D.~Huber and E.~Altman replace the on-site repulsion by a hard-core constraint and arrive at a spin-$\frac{1}{2}$ chain model
\begin{eqnarray}
\label{017}
H_{\rm{eff}}
=\sum_i
\left[
I^z\sigma_i^z\sigma_{i+1}^z 
+\frac{I^{\rm{dw}}}{2}\left(\sigma_i^z+\frac{1}{2}\right)\left(\sigma^+_{i\pm 1}\sigma^-_{i\pm 2} +{\rm{H.c.}}\right)
\right.
\nonumber\\
\left.
+I^{\rm{sp}}\left(\sigma_i^z+\frac{1}{2}\right)\left(\sigma^+_{i-1}\sigma^-_{i+ 1} +{\rm{H.c.}}\right)
\right]
\end{eqnarray}
with the set of parameters
\begin{eqnarray}
\label{018}
\tilde{U}\approx 0.40U (\to\infty),
\;\;\;
I^z\approx 0.112 U,
\;\;\;
I^{\rm{dw}} \approx - 0.025 U,
\;\;\;
I^{\rm{sp}} \approx - 0.011 U.
\end{eqnarray}

We turn to the kagome-lattice model.
Diagonalizing the tight-binding part of Hamiltonian we get three bands,
$\varepsilon_0({\bf{k}})=-2t$,
$\varepsilon_{\pm}({\bf{k}})=t\{1\pm\sqrt{3+2[\cos k_1+\cos (k_1-k_2) +\cos k_2]}\}$.
Localized states are located within single hexagons.
Importantly,
there is a band touching at the $\Gamma$ point, i.e., the flat band is not
gapped. 
As a result, the Wannier states are not exponentially localized but have power-law tails
(the slowest decay is $\sim \frac{1}{\vert{\bf{r}}\vert}$).
Projecting onto the flat band 
we are faced with a further complication, 
because of the vanishing gap in the one-boson spectrum between the flat band and the second
band:
In the case of the kagome lattice we do not have the small parameter $\frac{U}{\Delta}$.
However, the authors of Ref.~\refcite{huber-altman} argue that for fillings $\nu\gtrsim \frac{1}{9}$
the interactions open an effective band gap which finally controls the projection
onto the flat band. Again a hard-core constraint is
supposed, and, finally an effective Hamiltonian in terms of spin-$\frac{1}{2}$ operators on the triangular lattice is
derived:
\begin{eqnarray}
\label{019}
H_{\rm{eff}}
\approx
\sum_i\sum_{\alpha=1}^6
\left\{
\frac{I^z}{2}\sigma^z_{{\bf{r}}_i}\sigma^z_{{\bf{r}}_i+{\bf{a}}_\alpha}
+\left(\sigma^z_{{\bf{r}}_i}+\frac{1}{2}\right)
\left[
I_1^{xy}\sigma^+_{{\bf{r}}_i+{\bf{a}}_\alpha} \sigma^-_{{\bf{r}}_i+{\bf{a}}_{\alpha+2}}
\right.
\right.
\nonumber\\
\left.
\left.
+I_2^{xy} \sigma^+_{{\bf{r}}_i+{\bf{a}}_\alpha}
\left(\sigma^-_{{\bf{r}}_i+{\bf{a}}_{\alpha}+{\bf{a}}_{\alpha+1}}+\sigma^-_{{\bf{r}}_i+{\bf{a}}_{\alpha}+{\bf{a}}_{\alpha-1}}\right)
+I_3^{xy}\sigma^+_{{\bf{r}}_i+{\bf{a}}_\alpha} \sigma^-_{{\bf{r}}_i+{\bf{a}}_{\alpha+3}}
\right.
\right.
\nonumber\\
\left.
\left.
+I_4^{xy}\sigma^+_{{\bf{r}}_i+{\bf{a}}_\alpha} \sigma^-_{{\bf{r}}_i+2{\bf{a}}_{\alpha}}
+{\rm{H.c.}}
\right]
\right\}
+\ldots
\end{eqnarray}
with the set of parameters
\begin{eqnarray}
\label{020}
\tilde{U}=0.14 U \;(\to\infty),
\;
I^z\approx 0.028 U,
\;
\frac{I_1^{xy}}{I^z}\approx -\frac{1}{3},
\;
\frac{I_2^{xy}}{I^z}\approx \frac{1}{6},
\;
\frac{I_3^{xy}}{I^z}\approx -\frac{1}{8},
\;
\frac{I_4^{xy}}{I^z}\approx \frac{1}{16}.
\;\;\;
\end{eqnarray}
Here the vectors ${\bf{a}}_\alpha$, $\alpha=1,\ldots, 6$, connect the nearest neighbors on the triangular lattice,
see Ref.~\refcite{huber-altman}.

The effective spin-$\frac{1}{2}$ Hamiltonians (\ref{017}) (one-dimensional case) and (\ref{019}) (two-dimensional case)  
are much easier to study 
compared to the initial models.
In the one-dimensional case a bosonization technique can be applied 
whereas in the two-dimensional case a mean-field treatment can be elaborated.
The results of the investigation of the fate of the CDW state at lattice filling above $\nu_{\max}$ are different for the sawtooth chain and the kagome lattice.
For the sawtooth chain the long-range order of the CDW state will be immediately destroyed by proliferation of free domain walls upon increasing the density of bosons,
i.e., the exact ground state $\vert {\rm{CDW}}_{\frac{1}{4}}\rangle$ melts due to delocalization of domain walls.
The universal low-energy properties at filling $\nu=\frac{1}{4}+\delta\nu$ are described by
a commensurate-incommensurate transition.
For the kagome lattice the doping of bosons into interstitial sites of the CDW is energetically
favorable versus adding bosons as domain walls. 
Since the interstitial bosons do not destroy the CDW,
the long-range order of the CDW state is expected to survive and to coexist with a condensate formed by mobile interstitial bosons,
i.e., one deals with a supersolid phase.
In Ref.~\refcite{huber-altman} one can find further details concerning the sawtooth chain and the kagome lattice above $\nu_{\max}$ 
obtained on the basis of effective Hamiltonians (\ref{017}) and (\ref{019}),
as well as discussions about possible realizations of the kagome system 
using ultracold atoms and frustrated quantum magnets in high magnetic fields.
A similar study for the frustrated two-leg ladder is reported in Ref.~\refcite{tovmasyan}
(see also Ref.~\refcite{hosho}).

\subsection{Valence-bond plateau states of the antiferromagnetic kagome spin-lattice} 
\label{sec062}

\begin{figure}[bt]
\centerline{\psfig{file=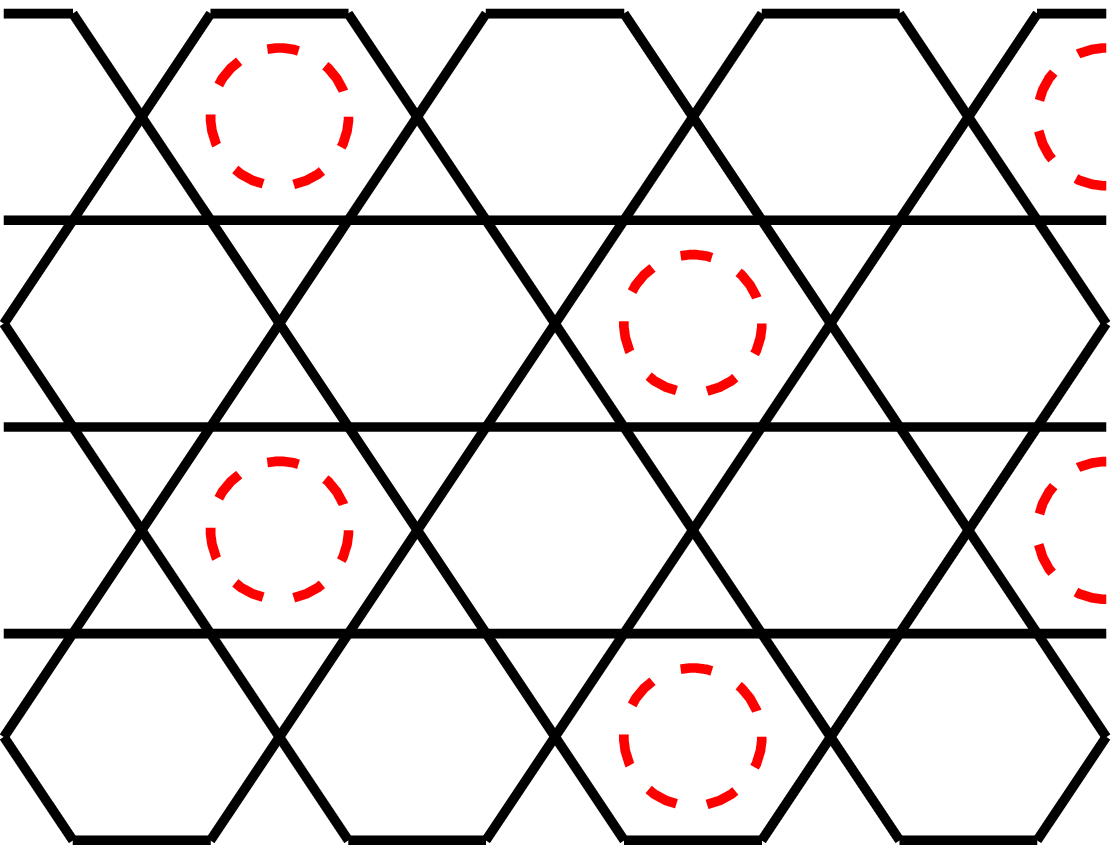,width=2.15in}\hspace{10mm}\psfig{file=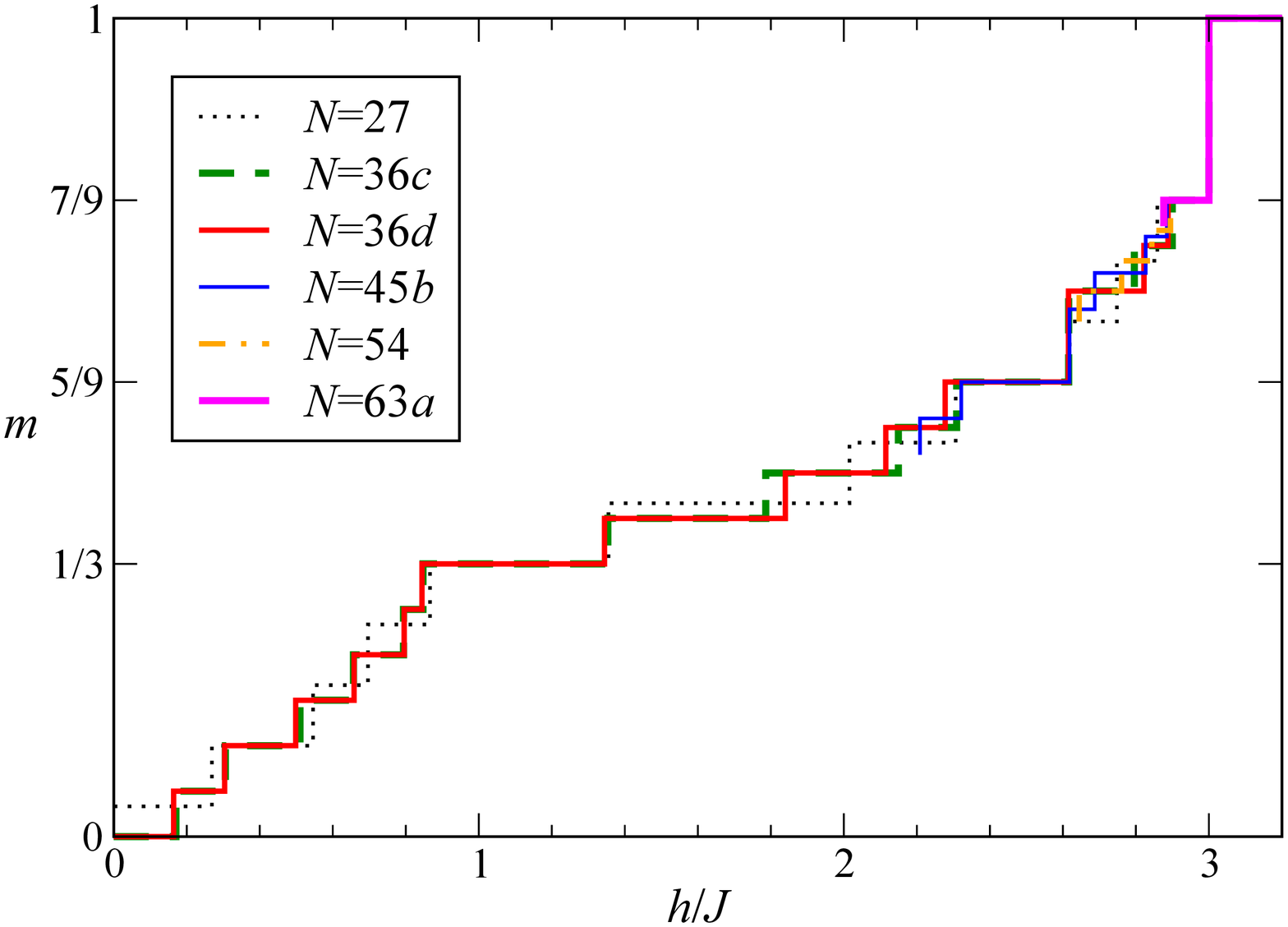,width=2.35in}}
\vspace*{8pt}
\caption{(Color online)
Magnon-crystal state (left) and magnetization curve (right)
for the quantum antiferromagnet on various finite kagome lattices.}
\label{fig08}
\end{figure}

The spin-$\frac{1}{2}$ Heisenberg antiferromagnet on the kagome lattice is one
of the most interesting and challenging problems in highly frustrated magnetism.
Although numerous studies of the kagome-lattice antiferromagnet exist, 
the physics of the model in zero field is still not fully
understood, see, e.g.,
Refs.~\refcite{kagome_flat_1,kagome_flat_2,kagome-gen,kagome_new}.
The application of a magnetic field probes the low-energy states of subspaces with various values of $S^z$ 
and leads to new intriguing features, see Fig.~\ref{fig08}.
Thus, the model exhibits a well-pronounced plateau in the (ground-state) magnetization curve at $\frac{1}{3}$ of the saturation
magnetization.\cite{lm1,lnp645,Hida2001,Richter2004,earlier}
Note, however, that quite recently the existence of this $\frac{1}{3}$-plateau was
questioned.\cite{Sakai}
It is noteworthy that recent experiments  on two kagome
compounds\cite{Okamoto2011}
have found indications for a plateau close to (but not exactly at)
magnetization
$m=\frac{1}{3}$.

In Ref.~\refcite{earlier} it was argued 
that the states of the $\frac{1}{3}$-plateau in the Heisenberg model and the
corresponding one in the Ising
limit belong to the same phase. In particular, a quantum valence-bond-crystal
state was proposed as a candidate for the $\frac{1}{3}$-plateau ground state.   
On the other hand, 
it is known that there is a $\frac{7}{9}$-plateau preceding the jump to saturation due to localized-magnon
states.\cite{lm1,sp-Peierls}
The  $\frac{7}{9}$-plateau ground state can be rigorously constructed: It is a
threefold degenerate close-packed magnon-crystal state\cite{lm1,sp-Peierls,capponi}
that is also of quantum valence-bond nature. Its structure is sketched in
Fig.~\ref{fig08} (left panel):
In the background of polarized (``up'') spins,  on every third hexagon marked by a dashed circle one flipped (``down'') spin
is distributed over the sites of the hexagon.
The width of the $\frac{7}{9}$-plateau  is about
 $0.1J$.\cite{sp-Peierls,nishimoto,capponi} 
Furthermore, the numerical data for the magnetization curve shown in
 Fig.~\ref{fig08} (right panel) give evidence for another plateau at
 $m=\frac{5}{9}$.
Having the valence-bond picture of the $m=\frac{3}{9}=\frac{1}{3}$ and  $m=\frac{7}{9}$ plateau
 states in mind,
in Ref.~\refcite{capponi} a unified description of the magnetization-plateau states
 at $m=\frac{3}{9}$, $\frac{5}{9}$ and $\frac{7}{9}$ was proposed, that
also provides a clarification of  earlier controversial proposals concerning
kagome plateaus.\cite{earlier,Sakai}

For the unified description of plateau states
the exact wave function for the $m=\frac{7}{9}$ plateau  is generalized 
to corresponding valence-bond-crystal wave functions for the candidate plateaus at $m=\frac{1}{3}$ and $\frac{5}{9}$.
As mentioned above the independent localized-magnon state for $m=\frac{7}{9}$ is the so-called magnon-crystal state
(see Fig.~\ref{fig08})
\begin{eqnarray}
\vert\Psi_{\rm{VBC}}^{\frac{7}{9}}\rangle
=\prod_j\vert L,\downarrow\rangle_j\prod_l\vert\uparrow\rangle_l \; .
\end{eqnarray}
Here the first product runs over an ordered pattern of all non-overlapping
trapping cells (hexagons) denoted by the dashed circles 
and the second product runs over the remaining sites.
The localized-magnon state of a hexagon is
\begin{eqnarray}
\vert L,\downarrow\rangle=\vert\vert\downarrow\uparrow\uparrow\uparrow\uparrow\uparrow\rangle\rangle_{\pi}.
\end{eqnarray}
We have introduced here the momentum eigenstates for a hexagon
$\vert\vert\sigma_0\ldots\sigma_5\rangle\rangle_k
=\frac{1}{\sqrt{{\cal{N}}}}\sum_{r=0}^5 e^{ikr}\vert\sigma_r\ldots\sigma_{5+r}\rangle$,
where $\sigma_n=\uparrow,\downarrow$,
$n+r$ has to be read modulo 6,
and ${\cal{N}}$ is a normalization factor ensuring
$_k\langle\langle \sigma_0\ldots\sigma_5 \vert\vert\sigma_0\ldots\sigma_5\rangle\rangle_k=1$
(${\cal{N}}=6$ unless the state repeats under less than 6 translations).
The magnon-crystal state is the threefold degenerate ground state in the subspace with
$S^z=\frac{7}{9}\frac{N}{2}$ 
and its energy per site at $h=0$ is $e_{\rm{VBC}}^{\frac{7}{9}}=\frac{J}{6}$.

Note that the quantum valence-bond crystal
state proposed in Ref.~\refcite{earlier}
for the $\frac{1}{3}$-plateau is described by a similar wave function:
The global pattern is again as sketched in Fig.~\ref{fig08}, 
but the dashed circles were supposed to represent a combination of the
two N\'{e}el states of a hexagon.
This picture works well approaching the Ising limit of the model, but for
the Heisenberg model it is favorable\cite{capponi}  to consider the 
true ground state of the Heisenberg model on the hexagon, $\vert
L,\downarrow\downarrow\downarrow\rangle$, instead of the  combination of the
two N\'{e}el states.
The corresponding threefold-degenerate valence-bond-crystal state reads
\begin{eqnarray}
\vert\Psi_{\rm{VBC}}^{\frac{3}{9}}\rangle
=\prod_j\vert L,\downarrow\downarrow\downarrow\rangle_j\prod_l\vert\uparrow\rangle_l
\end{eqnarray}
with
\begin{eqnarray}
&& \vert L,\downarrow\downarrow\downarrow\rangle
=\frac{1}{\sqrt{195+51\sqrt{13}}} 
\Big[
3\vert\vert\downarrow\downarrow\downarrow\uparrow\uparrow\uparrow\rangle\rangle_\pi
\nonumber\\
&& +\frac{3(3+\sqrt{13})}{2}
\left( \vert\vert\downarrow\downarrow\uparrow\uparrow\downarrow\uparrow\rangle\rangle_\pi
-\vert\vert\downarrow\downarrow\uparrow\downarrow\uparrow\uparrow\rangle\rangle_\pi\right)
+\sqrt{3}(4+\sqrt{13})\vert\vert\uparrow\downarrow\uparrow\downarrow\uparrow\downarrow\rangle\rangle_\pi
\Big].
\end{eqnarray}
Although this state is not an exact eigenstate of the Hamiltonian
it is a good representation of the true ground state in the subspace with
$S^z=\frac{3}{9}\frac{N}{2}$.
Its energy per site at $h=0$ is $e_{\rm{VBC}}^{\frac{1}{3}}=-\frac{J}{9}-\frac{\sqrt{13}J}{18}\approx -0.311\,419\,515\, 3 J$.

\begin{figure}[bt]
\centerline{\psfig{file=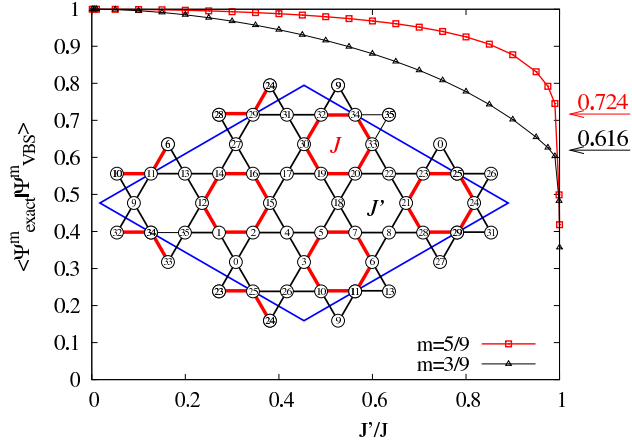,width=3.25in}}
\vspace*{8pt}
\caption{(Color online)
Exact-diagonalization data for the overlaps
$\langle\Psi_{\rm{exact}}^{m}\vert\Psi_{\rm{VBC}}^{m}\rangle$ of the valence-bond-crystal
states with the exact ground states for $m=\frac{5}{9}$ and $m=\frac{3}{9}$ as function of ratio of bond
strengths $\frac{J^\prime}{J}$, 
where $J$ is bond strength of the trapping cells 
hosting localized magnons in the magnon-crystal state
and $J^\prime$ is bond strength of all other bonds. 
The arrows at the right side indicate the overlap of a symmetric linear combination of the
three degenerate valence-bond-crystal states with the true ground state
for the uniform kagome lattice, i.e., at $J^\prime=J$.}
\label{overlap}
\end{figure}

Inspired by the valence-bond-crystal states for the plateaus at $m=\frac{7}{9}$ and $\frac{3}{9}$,
it is natural to consider a corresponding valence-bond state at $m=\frac{5}{9}$:
\begin{eqnarray}
\vert\Psi_{\rm{VBC}}^{\frac{5}{9}}\rangle
=\prod_j\vert L,\downarrow\downarrow\rangle_j\prod_l\vert\uparrow\rangle_l
\end{eqnarray}
with
\begin{eqnarray}
\vert L,\downarrow\downarrow\rangle
=\frac{1}{2\sqrt{5}} 
\left[
(\sqrt{5}-1)\vert\vert\uparrow\uparrow\uparrow\uparrow\downarrow\downarrow\rangle\rangle_0
-(\sqrt{5}+1)\vert\vert\uparrow\uparrow\uparrow\downarrow\uparrow\downarrow\rangle\rangle_0
+2\sqrt{2}\vert\vert\uparrow\uparrow\downarrow\uparrow\uparrow\downarrow\rangle\rangle_0
\right]. \;\;\;\;
\end{eqnarray}
Its energy per site at $h=0$ is $e_{\rm{VBC}}^{\frac{5}{9}}=-\frac{\sqrt{5}J}{18}\approx -0.124\,225\,998\, 7 J$.
Again, this is not the true ground state although it is not far from it.

In Ref.~\refcite{capponi} large-scale exact-diagonalization studies are compared with theoretical predictions based on the variational wave functions for the plateau states.
This comparison demonstrates that the introduced model states provide a good description  of the physics within the plateau regimes.
In particular, the comparison of the (connected) spin and dimer correlation function,
$\langle s^z_is^z_j\rangle_c=\langle s^z_is^z_j\rangle-\langle s^z_i\rangle\langle s^z_j\rangle$
and
$\langle ({\bf{s}}_i\cdot {\bf{s}}_j)({\bf{s}}_k\cdot {\bf{s}}_l)\rangle_c
=\langle ({\bf{s}}_i\cdot {\bf{s}}_j)({\bf{s}}_k\cdot {\bf{s}}_l)\rangle
-\langle ({\bf{s}}_i\cdot {\bf{s}}_j)\rangle\langle({\bf{s}}_k\cdot {\bf{s}}_l)\rangle$,
calculated using exact diagonalizations and the introduced valence-bond-crystal picture
as well as  the analysis of the low-lying excitations above the plateau ground states, yield 
clear evidence that the valence-bond-crystal picture holds also for the $\frac{3}{9}$- and
the $\frac{5}{9}$-plateau states.
To support this conclusion we present in Fig.~\ref{overlap}
exact-diagonalization data for the overlaps
$\langle\Psi_{\rm{exact}}^{m}\vert\Psi_{\rm{VBC}}^{m}\rangle$ of the
valence-bond-crystal
states with the exact ground states for $m=\frac{5}{9}$ and $m=\frac{3}{9}$ as function of
ratio of bond
strengths $\frac{J^\prime}{J}$ for a finite lattice of $N=36$ sites. 
Here $J$ is the bond strength of the trapping cells 
corresponding to the valence-bond-crystal state 
and $J^\prime$ is the bond strength of all other bonds.
Indeed the overlap is large in almost the whole region  $0 \le \frac{J^\prime}{J} \lesssim 1$. 
Only in the close vicinity of the uniform kagome limit ($J^\prime=J$) it
drops down to
$\langle\Psi_{\rm{exact}}^{\frac{5}{9}}\vert\Psi_{\rm{VBC}}^{\frac{5}{9}}\rangle\approx 0.418$
and
$\langle\Psi_{\rm{exact}}^{\frac{3}{9}}\vert\Psi_{\rm{VBC}}^{\frac{3}{9}}\rangle\approx 0.357$.
Taking at $J^\prime=J$ instead of only one valence-bond-crystal state 
a symmetric linear combination of the three degenerate valence-bond-crystal states 
gives even a large overlap 
$\langle\Psi_{\rm{exact}}^{\frac{5}{9}}\vert\Psi_{\rm{symVBC}}^{\frac{5}{9}}\rangle\approx 0.724$
and
$\langle\Psi_{\rm{exact}}^{\frac{3}{9}}\vert\Psi_{\rm{symVBC}}^{\frac{3}{9}}\rangle\approx 0.616$.
Note that for $m=\frac{7}{9}$ the overlap is always unity, since
$\vert\Psi_{\rm{VBC}}^{\frac{7}{9}}\rangle$ is an exact eigenstate.
We may compare the calculated overlaps with corresponding data for the
N\'{e}el-long-range ordered $s=\frac{1}{2}$ square-lattice Heisenberg antiferromagnet.
For a finite lattice with $N=32$  sites
the overlap of a symmetric linear combination of the two N\'{e}el states is   
 $\langle\Psi_{\rm{exact}}\vert\Psi_{\rm{symNeel}}\rangle\approx 0.203$,
 i.e., it is even lower than those for the valence-bond-crystal model states.
Thus the obtained large overlaps may serve as another indication of
 valence-bond crystalline order in the plateaus of the kagome
 antiferromagnet.

Finally we note that, complementary to the approach of
Ref.~\refcite{capponi},
S. Nishimoto et al.\cite{nishimoto} used another route to the same problem.
They developed a grand-canonical density-matrix-renormalization-group numerical technique
to obtain the magnetization curve and to identify the nature of the ground states.
The density-matrix-renormalization-group results\cite{nishimoto} agree well with
the valence-bond ansatzes and the exact-diagonalization data of
Ref.~\refcite{capponi}.
Moreover, Ref.~\refcite{nishimoto} predicts an exotic (tiny) quantum plateau at $m=\frac{1}{9}$.

\subsection{Localized-magnon ground states in zero magnetic field: 
            The sawtooth chain with ferro- and antiferromagnetic interactions}
\label{sec063}

The sawtooth-chain (also dubbed as delta-chain or one-dimensional Tasaki-lattice) 
Heisenberg antiferromagnet is a well studied example exhibiting localized-magnon
physics. If the exchange coupling along the zigzag path $J_2$ 
is two times larger than the exchange coupling along the basal straight line $J_1$, see Fig.~\ref{fig01},
the lower one-magnon band is dispersionless and localized-magnon physics emerges at low temperatures around the saturation field
$h_{\rm{sat}}=2J_2=4J_1$.
Interestingly,  very recently it has been found 
that 
the spin-$\frac{1}{2}$ Heisenberg model on this lattice 
with {\bf ferromagnetic} zigzag $J_2<0$ and antiferromagnetic basal $J_1=\frac{1}{2}\vert J_2\vert>0$ interactions
has a close relation to localized-magnon physics.\cite{dmitriev_krivnov}
(We notice here that in Ref.~\refcite{dmitriev_krivnov} 
the ferromagnetic zigzag interaction is denoted by $J_1<0$ and the antiferromagnetic basal interaction is denoted by $J_2=\frac{1}{2}\vert J_1\vert>0$.)
Note that a $s=\frac{1}{2}$ ferro-antiferromagnetic sawtooth-chain Heisenberg system may be used for
the description of the compound 
[Cu(bpy)(H$_2$O)][Cu(bpy)(mal)-(H$_2$O)](ClO$_4$)$_2$.\cite{inagaki}

Since the results of Ref.~\refcite{dmitriev_krivnov} may open a new window 
to find exact localized many-body ground states in quantum spin systems, 
we will illustrate here the main findings for the ferromagnetic-antiferromagnetic sawtooth chain in some detail.
We consider the Hamiltonian of the spin-$\frac{1}{2}$ sawtooth-chain Heisenberg model
\begin{eqnarray}
\label{027}
H=J_2\sum_{n=1}^{\frac{N}{2}}\left({\bf{s}}_{2n-1}\cdot {\bf{s}}_{2n} + {\bf{s}}_{2n}\cdot {\bf{s}}_{2n+1}-\frac{1}{2}\right)
+J_1\sum_{n=1}^{\frac{N}{2}}\left({\bf{s}}_{2n-1}\cdot {\bf{s}}_{2n+1} -\frac{1}{4}\right).
\end{eqnarray}
(Following Ref.~\refcite{dmitriev_krivnov} we use here the one-index numbering of sites, see Fig.~\ref{fig01},
i.e., $m,1\to 2n-1$, $m,2\to 2n$ and $n=1,\ldots,\frac{N}{2}$.)
For dominating ferromagnetic exchange $J_2$  the ground state is the fully polarized ferromagnetic state. 
At a critical point $\alpha=\frac{J_1}{\vert J_2\vert}=\frac{1}{2}$ 
a transition to a partly polarized ferrimagnetic  ground state takes place.
Just at the critical point the ground-state manifold of the model 
is given by a macroscopically degenerate set of localized-magnon-like eigenstates, 
where the ground-state degeneracy is even  larger than that for the antiferromagnetic
sawtooth chain, where the latter one was calculated in Ref.~\refcite{prb2004}.
A striking difference between the ferromagnetic-antiferromagnetic and the purely antiferromagnetic sawtooth chain 
is the appearance of the localized ground states at zero field for the ferromagnetic-antiferromagnetic case.  

To construct the ground states at the critical point $\alpha=\frac{1}{2}$ it is convenient 
to put $J_2=-1$ and to represent the Hamiltonian (\ref{027}) as a sum of local Hamiltonians:
\begin{eqnarray}
\label{028}
H=\sum_{i=1}^{\frac{N}{2}} H_i,
\;\;\;
H_i
=
-\left({\bf{s}}_{i_1} + {\bf{s}}_{i_3}\right)\cdot {\bf{s}}_{i_2}
+\frac{1}{2}{\bf{s}}_{i_1} \cdot {\bf{s}}_{i_3}
+\frac{3}{8}.
\end{eqnarray}
Here $i$ enumerates up-triangles which constitute the sawtooth-chain lattice
whereas  $i_1$ and $i_3$ are the (left and right) basal sites and $i_2$ is the apical site in the $i$th triangle.
The eigenvalues of $H_i$ are: 0 ($S=\frac{3}{2}$), 0 ($S=\frac{1}{2}$), and $\frac{3}{2}$ ($S=\frac{1}{2}$),
and, therefore, the ground-state energy of $H$ (\ref{028}) is bounded from below by $\sum_i 0=0$.
Since the ferromagnetic state with maximal total spin $S_{\max}=\frac{N}{2}$ is an eigenstate of $H$ [see Eq.~(\ref{027})] with zero energy,
the ground-state energy of $H$ equals this lower bound, i.e., it is zero.

Let us now illustrate how the complete (and linearly independent) ground-state manifold can be
constructed.\cite{dmitriev_krivnov}
It is convenient to consider the sawtooth chain with both
open and periodic boundary conditions.
First we consider the one-magnon subspace with $S^z=\frac{N}{2}-1$.
There are $\frac{N}{2}$ linear independent localized states of the form
$\varphi_i\vert \uparrow\ldots\uparrow\rangle$ 
with 
\begin{eqnarray}
\label{029}
\varphi_i=s_{2i}^-+2s_{2i+1}^-+s_{2(i+1)}^-,
\;\;\;
i=1,\ldots ,{\cal{N}},
\;\;\;
{\cal{N}}=\frac{N}{2}
\end{eqnarray}
(periodic boundary conditions).
Note here that similar states exist for the antiferromagnetic sawtooth
chain, however, with the important difference, that the corresponding
operator,
$\varphi^{\rm afm}_i=s_{2i}^- - 2s_{2i+1}^-+s_{2(i+1)}^-$, contains a minus sign.
For open boundary conditions (and odd $N$) the first and the last states in
this sequence are generated by the operators
$\varphi_1=2s_{1}^-+s_{2}^-$ and $\varphi_{{\sf{N}}}=s_{N-1}^-+2s_{N}^-$,
respectively,
where ${\sf{N}}=\frac{N+1}{2}$.
It is easy to prove that
$H_l\varphi_l \vert \uparrow\ldots\uparrow\rangle=0$ and
$H_{l+1}\varphi_l \vert \uparrow\ldots\uparrow\rangle=0$ hold.  
For all other $i$ we can use the commutation 
to find $H_{i}\varphi_l \vert \uparrow\ldots\uparrow\rangle=\varphi_l
H_{i}\vert
\uparrow\ldots\uparrow\rangle=0$.
Hence $\varphi_l \vert \uparrow\ldots\uparrow\rangle$ is an eigenstate of
$H$ with zero energy.
The states created by operators $\varphi_i$ are eigenstates
of $H$ and $S^z$, but not of ${\bf{S}}^2$.
Rather a state $\varphi_i \vert \uparrow\ldots\uparrow\rangle$ is a mixture
of eigenstates of ${\bf{S}}^2$ with $S=\frac{N}{2}$ and
$S=\frac{N}{2}-1$.
On the other hand, states created by $\varphi^{\rm afm}_i$ are eigenstates
of ${\bf{S}}^2$.
Note, that the symmetric 
linear combination $\sum_i\varphi_i\vert \uparrow\ldots\uparrow\rangle=2S^-\vert \uparrow\ldots\uparrow\rangle$ 
belongs to the subspace with $S=\frac{N}{2}$. There are $\frac{N}{2}-1$ other
linear combinations of $\varphi_i\vert \uparrow\ldots\uparrow\rangle$ 
(but not the states themselves)
which belong to the subspace with $S=\frac{N}{2}-1$.
It is convenient to introduce another set of linearly independent operator functions
(instead of $\varphi_i$
defined above), 
namely
\begin{eqnarray}
\label{030}
\Phi(m)=\sum_{i=1}^{m}\varphi_i,
\;\;\;
m=1,\ldots ,{\sf{N}}.
\end{eqnarray}
The function $\Phi({\sf{N}})\vert \uparrow\ldots\uparrow\rangle$ belongs to the subspace with $S=\frac{N}{2}$,
while the remaining $\frac{N}{2}-1$ functions yield contributions to the subspace with $S=\frac{N}{2}-1$.

Next we consider the two-magnon subspace with $S^z=\frac{N}{2}-2$ and imply for simplicity open boundary conditions.
Bearing in mind the two-magnon localized states of the antiferromagnetic
sawtooth chain,
one immediately finds that $\varphi_i\varphi_j \vert \uparrow\ldots\uparrow\rangle$ with $j\ge i+1$ are exact ground states of the Hamiltonian $H$ with 
zero energy,
since they are exact eigenstates of each local Hamiltonian $H_l$ with zero energy.
The number of such hard-dimer states is given by the binomial coefficient
${\cal{C}}_{{\sf{N}}-1}^{2}=\frac{1}{2}({\sf{N}}-1)({\sf{N}}-2)$ 
(this is the canonical partition function of two dimers on the open chain of ${\sf{N}}$ sites).
Another choice for the two-magnon ground states with zero energy is given by
$\Phi(m_1)\Phi(m_2) \vert \uparrow\ldots\uparrow\rangle$ with $1\le m_1<m_2\le {\sf{N}}-1$
(there are also  ${\cal{C}}_{{\sf{N}}-1}^{2}$ states of this kind).
There are additional zero-energy ground states in the subspace with $S^z=\frac{N}{2}-2$ 
which belong to the subspace with $S=\frac{N}{2}-1$ and $S=\frac{N}{2}$:
The operator function $\Phi({\sf{N}})$ and $m_1\le {\sf{N}}-1$ yield the state 
$\Phi(m_1)\Phi({\sf{N}})\vert \uparrow\ldots\uparrow\rangle = 2S^-\Phi(m_1)\vert \uparrow\ldots\uparrow\rangle$ 
which belongs to the subspace with $S=\frac{N}{2}-1$ and $S=\frac{N}{2}$.
The number of such states is ${\cal{C}}_{{\sf{N}}-1}^{1}$.
One more state in the two-magnon subspace
is given by
$\Phi({\sf{N}})\Phi({\sf{N}})\vert \uparrow\ldots\uparrow\rangle = (2S^-)^2\vert \uparrow\ldots\uparrow\rangle$ 
which belongs to the subspace with $S=\frac{N}{2}$.
In summary, the total number of the exact ground states in the two-magnon subspace is
${\cal{C}}_{{\sf{N}}-1}^{0}+{\cal{C}}_{{\sf{N}}-1}^{1}+{\cal{C}}_{{\sf{N}}-1}^{2}>{\cal{C}}_{{\sf{N}}-1}^{2}$.

We pass to the $k$-magnon subspace.
We choose the $k$-magnon states in the form
\begin{eqnarray}
\Phi(m_1)\Phi(m_2)\ldots \Phi(m_{k-1})\Phi(m_k) \vert \uparrow\ldots\uparrow\rangle,
\nonumber\\
1\le m_1<m_2<\ldots < m_{k-2} <m_{k-1}< m_k\le {\sf{N}}-1
\label{31}
\end{eqnarray}
(this set belongs to the subspaces $S\ge\frac{N}{2}-k$,
i.e., these states carry components of ${\bf{S}}^2$ eigenstates with $S=\frac{N}{2}-k$ 
and also components of ${\bf{S}}^2$ eigenstates with higher $S$);
\begin{eqnarray}
\Phi(m_1)\Phi(m_2)\ldots \Phi(m_{k-1})\Phi({\sf{N}}) \vert \uparrow\ldots\uparrow\rangle,
\nonumber\\
1\le m_1<m_2<\ldots <m_{k-2} < m_{k-1}\le {\sf{N}}-1
\label{32}
\end{eqnarray}
(this set belongs to the subspaces $S\ge\frac{N}{2}-k+1$);
\begin{eqnarray}
\Phi(m_1)\Phi(m_2)\ldots \Phi({\sf{N}})\Phi({\sf{N}}) \vert \uparrow\ldots\uparrow\rangle,
\nonumber\\
1\le m_1<m_2<\ldots <m_{k-2} \le {\sf{N}}-1
\label{33}
\end{eqnarray}
(this set belongs to the subspaces $S\ge\frac{N}{2}-k+2$);
etc. and finally
\begin{eqnarray}
\Phi({\sf{N}})\Phi({\sf{N}})\ldots \Phi({\sf{N}})\Phi({\sf{N}}) \vert \uparrow\ldots\uparrow\rangle
\label{34}
\end{eqnarray}
(this set belongs to the subspace $S=\frac{N}{2}$).
All the states are eigenstates of the Hamiltonian with the zero-energy eigenvalue. 
The total number of these states is
\begin{equation}
g^{\rm OBC}_{N}(k)={\cal{C}}_{{\sf{N}}-1}^{0}+\ldots +{\cal{C}}_{{\sf{N}}-1}^{k-2}+{\cal{C}}_{{\sf{N}}-1}^{k-1}+{\cal{C}}_{{\sf{N}}-1}^{k},
\label{35}
\end{equation}
see Table~\ref{tab1}.
It is important to note that  the degeneracy $g^{\rm OBC}_{N}(k)$ calculated for states of the form (\ref{31}) -- (\ref{34}) 
is larger than that of the hard-dimer states constructed by the operators $\varphi_i$ defined above.
The total number of degenerated ground states of the chain with open boundary conditions is 
\begin{eqnarray}
W^{\rm OBC}
&=&
2\sum_{k=0}^{{\sf{N}}-1}g^{\rm OBC}_{N}(k)
=
2\sum_{k=0}^{{\sf{N}}-1}\left({\sf{N}}-k\right){\cal{C}}_{{\sf{N}}-1}^{k}
\\
&=&
2\left[{\sf{N}}2^{{\sf{N}}-1}-\left({\sf{N}}-1\right)2^{{\sf{N}}-2}\right]
=
({\sf{N}}+1)2^{{\sf{N}}-1}
\underset{N\to\infty}{\longrightarrow}
e^{{\sf{N}}\ln 2 }
\underset{N\to\infty}{\longrightarrow}
e^{\frac{N}{2}\ln 2}. \nonumber
\end{eqnarray}
The resulting residual entropy (per site)
 at zero magnetic field
for open boundary conditions is
$\lim_{N\to\infty} s^{\rm OBC}_0= \lim_{N\to\infty}\frac{\ln W^{\rm OBC}}{N} = {\frac{1}{2}\ln 2}$.

\begin{table}
\tbl{Exact-diagonalization data for degeneracies: Open chain of $N=17$ sites (i.e., ${\sf{N}}=9$).}
{\begin{tabular}{@{}cccccccccc@{}} \Hline 
\\[-1.8ex] 
$S^z=S$  & $\frac{17}{2}$ & $\frac{15}{2}$ & $\frac{13}{2}$ &  $\frac{11}{2}$ &  $\frac{9}{2}$ &  $\frac{7}{2}$ &  $\frac{5}{2}$ & $\frac{3}{2}$ & $\frac{1}{2}$ \\[0.8ex] 
\hline \\[-1.8ex]
$k$      &              0 &              1 &              2 &              3  &              4 &              5 &              6 &             7 &             8 \\[0.8ex] 
\hline \\[-1.8ex]
$g^{\rm OBC}_{17}$ &              1 &              8 &             28 &             56  &             70 &             56 &             28 &             8 &             1 \\[0.8ex] 
\Hline \\[-1.8ex] 
\end{tabular}}
\label{tab1}
\end{table}

Let us also briefly consider the case of periodic boundary conditions.
Now instead of ${\sf{N}}=\frac{N+1}{2}$ we have ${\cal{N}}=\frac{N}{2}$.
Using operators (\ref{029}) we can construct the ground states in the subspace $S^z=\frac{N}{2}-k$ in the form
$\varphi_{i_1}\ldots\varphi_{i_k} \vert \uparrow\ldots\uparrow\rangle$.
The number of these hard-dimer localized-magnon states equals to $\frac{{\cal{N}}}{{\cal{N}}-k}{\cal{C}}_{{\cal{N}}-k}^{k}$
(i.e., equals to the canonical partition function of $k$ hard dimers on a simple periodic chain of ${\cal{N}}$ sites).
As for open boundary conditions, again the hard-dimer states do not exhaust all possible ground-state eigenstates in the subspace with $S^z=\frac{N}{2}-k$, $k\ge 2$.
For example, for $k=2$ we can check by inspection that 
$\varphi_i(\varphi_{i-1}+\varphi_i+\varphi_{i+1}) \vert \uparrow\ldots\uparrow\rangle$
is a zero-energy eigenstate of the local Hamiltonians $H_i$, $H_{i+1}$, $H_{i-1}$, and of all other ones.
The authors of Ref.~\refcite{dmitriev_krivnov} conjecture 
(and check by exact diagonalizations for finite chains up to $N=28$) 
that the number of the ground states in the subspace with $S^z=\frac{N}{2}-k$ for $k=0,1,2\ldots,\frac{N}{4},\ldots,\frac{N}{2}$ is given by the formula
\begin{eqnarray}
\label{035}
g^{\rm PBC}_{N}(k)
=
\left\{
\begin{array}{ll}
{\cal{C}}_{\cal{N}}^k,                                          & 0\le k\le \frac{{\cal{N}}}{2},\\
{\cal{C}}_{\cal{N}}^{\frac{{\cal{N}}}{2}}+\delta_{k,{\cal{N}}}, & \frac{{\cal{N}}}{2}\le k\le {\cal{N}}
\end{array}
\right.
\end{eqnarray}
(the Kronecker delta corresponds to the resonating-valence-bond singlet eigenstate\cite{hamada}),
see Table~\ref{tab2}.
From Eq. (\ref{035}) we get the residual entropy (per site) at zero magnetic field $s_0=\frac{\ln W}{N}$ for periodic boundary conditions
\begin{eqnarray}
W^{\rm PBC}
=2\sum_{k=0}^{{\cal{N}}-1}g^{\rm PBC}_{N}(k) + g^{\rm PBC}_{N}({\cal{N}})
\underset{N\to\infty}{\longrightarrow} e^{\frac{N}{2}\ln 2},
\nonumber\\ 
\lim_{N\to\infty} s^{\rm PBC}_0={\frac{1}{2}\ln 2} =\lim_{N\to\infty} s^{\rm OBC}_0.
\end{eqnarray}
Although, on first glance  Eqs.~(\ref{35}) and (\ref{035}) look quite different 
they yield finally for $N\to \infty$ the same result for the residual entropy.

\begin{table}
\tbl{Exact-diagonalization data for degeneracies: Periodic chain of $N=20$ sites (i.e., ${\cal{N}}=10$).}
{\begin{tabular}{@{}cccccccccccc@{}} \Hline 
\\[-1.8ex] 
$S^z=S$  & 10 & 9 &  8 &  7 &  6 &  5 & 4 & 3 & 2 & 1 &  0 \\[0.8ex] 
\hline \\[-1.8ex]
$k$      &  0 & 1 &  2 &  3 &  4 &  5 & 6 & 7 & 8 & 9 & 10 \\[0.8ex] 
\hline \\[-1.8ex]
$g^{\rm PBC}_{20}$ &  1 & 9 & 35 & 75 & 90 & 42 & 0 & 0 & 0 & 0 &  1 \\[0.8ex] 
\Hline \\[-1.8ex] 
\end{tabular}}
\label{tab2}
\end{table}

In Ref.~\refcite{dmitriev_krivnov}
analytical arguments are supported by extensive exact-diagonalization studies.
In particular, the gaps in $k$-magnon subspaces were calculated.
It was found that the gaps for the $k$-magnon states with $k\ge 2$ decrease rapidly with increasing $k$.
As a result (and by contrast to the antiferromagnetic sawtooth chain) the 
contribution of the excited states to the partition function cannot be neglected even for very low temperatures.

\begin{figure}[bt]
\centerline{\psfig{file=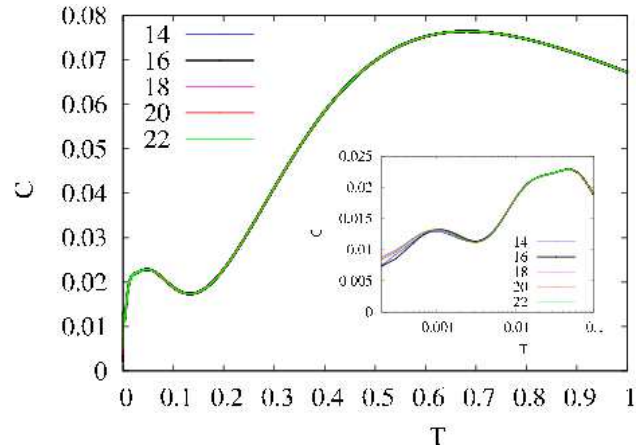,width=3.25in}}
\vspace*{8pt}
\caption{(Color online)
Exact-diagonalization data for the specific heat per site of the ferro-antiferromagnetic sawtooth Heisenberg chain at the critical point 
($J_1=\frac{1}{2}$, $J_2=-1$). 
The inset shows the low-temperature part in logarithmic scale. 
In the main panel the curves for $N=14,\ldots,22$ practically coincide indicating small finite-size effects down to $T\approx 0.001$.} 
\label{C_T_Krivnov}
\end{figure}

The thermodynamics of the ferro-antiferromagnetic sawtooth Heisenberg chain at the critical point
exhibits some interesting low-temperature features. 
There are additional low-temperature maxima in the specific heat, see Fig.~\ref{C_T_Krivnov}.  
The susceptibility diverges for $T \to 0$ according to $\chi =c_{\chi}T^{-\alpha}$ 
with $c_{\chi} \approx 0.317$ and $\alpha \approx 1.09$.
This divergence is naturally related to the ground states with finite magnetic moment. 
However, the exponent $\alpha$ is significantly lower than that for the one-dimensional Heisenberg ferromagnet $\alpha_{\rm{FM}}=2$ 
that is valid  on the left side of the critical point, 
i.e., for $J_1 < \frac{1}{2}\vert J_2\vert$ (see, e.g., Ref.~\refcite{FM_crit}). 
The low-temperature behavior of the entropy per spin is given by  
$\frac{S(T)}{N}=\frac{1}{2}\ln 2+c_{s}T^{\lambda }$
with $c_{s}\approx 0.245$ and $\lambda \approx 0.12$.
The existence of a residual entropy leads also to an enhanced magnetocaloric effect.

It is useful to summarize the differences in localized-magnon physics 
in the ferro-antiferromagnetic sawtooth chain\cite{dmitriev_krivnov} and the antiferromagnetic sawtooth chain.\cite{prb2004}
Most importantly, the localized-magnon states are ground states in zero magnetic field, 
but not at the (typically large) saturation field.
All the effects related to the localized-magnon states are therefore better accessible in experiment, 
in particular,
the enhanced magnetocaloric effect is observed when switching off the applied magnetic field.
The excitation gaps above the degenerate ground-state manifold are very small. 
These low-lying excitations together with the ground-state manifold   
lead to characteristic power-law dependences for several thermodynamic quantities at low temperatures.

\section{Solid-State Localized-Magnon Systems}
\label{sec07}

The properties of localized-magnon systems are useful for explaining the low-temperature physics of frustrated quantum
antiferromagnetic compounds at high magnetic fields.
Most promising hitherto known candidates to show localized-magnon physics are 
the spin-$\frac{1}{2}$ (distorted) diamond-chain compound azurite Cu$_3$(CO$_3$)$_2$(OH)$_2$\cite{kikuchi}
and
the spin-$\frac{1}{2}$ $XY$-like two-dimensional dimer magnet Ba$_2$CoSi$_2$O$_6$Cl$_2$.\cite{tanaka}
The recent experiments on these systems raise hopes for direct comparison with theoretical models elaborated within the localized-magnon context.

\subsection{Azurite}
\label{sec071}

The natural mineral azurite Cu$_3$(CO$_3$)$_2$(OH)$_2$ is a frustrated magnet displaying interesting magnetic
behavior.\cite{kikuchi,kikuchi_2,rule,azurite-parameters,effective_xy2,azurite2014}
In particular, 
H. Kikuchi et al.\cite{kikuchi} discovered a magnetization plateau at $\frac{1}{3}$ of the saturation value 
followed by a steep (jumpwise) increase to the saturation value in the low-temperature magnetization curve.
For the magnetic properties, the most important feature of azurite are diamond-like chains which are formed by spin-$\frac{1}{2}$ 
copper ions.
There have been several  attempts to derive a microscopic model for magnetic properties of azurite.
In the Letter by H. Jeschke et al.\cite{azurite-parameters} the authors performed first-principle density-functional-theory calculations 
as well as model computations using the density-matrix-renormalization-group
and exact-diagonalization methods
and obtained a microscopic model for the frustrated quantum magnet azurite.
As a result, they proposed an effective generalized spin-$\frac{1}{2}$ diamond-chain model
which provides a consistent description of a wide range of experiments.
The Hamiltonian for the generalized diamond chain shown in Fig.~\ref{fig01} is as follows
\begin{eqnarray}
\label{spin_azu}
H&=&\sum_{m=1}^{{\cal{N}}}
\left[
J_2 {\bf{s}}_{m,1}\cdot {\bf{s}}_{m,2}
+J_1 \left({\bf{s}}_{m,2}+{\bf{s}}_{m+1,1}\right)\cdot{\bf{s}}_{m,3}
+J_3 \left({\bf{s}}_{m,1}+{\bf{s}}_{m+1,2}\right)\cdot{\bf{s}}_{m,3}
\right.
\nonumber\\
&& \left.
+J_m {\bf{s}}_{m,3}\cdot {\bf{s}}_{m+1,3}
-g\mu_{\rm{B}} {\sf{H}} \left(s^z_{m,1}+s^z_{m,2}+s^z_{m,3}\right) 
\right]; 
\;\:\: 
({\bf{s}}_{m,i})^2=\frac{3}{4}.
\end{eqnarray}
Here the sum runs over all ${\cal{N}}$ cells and the total number of spins is $N=3{\cal{N}}$.
${\sf{H}}$ is the external magnetic field and $\mu_{\rm{B}}$ is the Bohr magneton.
The exchange constants and the $g$-factor are as follows:\cite{azurite-parameters}
\begin{eqnarray}
J_1=15.51\,{\rm{K}},
\;\;\;
J_2=33\,{\rm{K}},
\;\;\;
J_3=6.93\,{\rm{K}},
\;\;\;
J_m=4.62\,{\rm{K}},
\;\;\;
g=2.06.
\end{eqnarray}
This model does not take into account a magnetic anisotropy which is clearly present in azurite\cite{kikuchi_2}
and interchain couplings which are responsible for the low-temperature
magnetic ordering
of azurite.
Nevertheless, the proposed model is a convenient starting point for state-of-the-art numerical many-body calculations and derivations of effective low-energy theories.

The set of parameters for the spin model of azurite does not fit exactly 
to the localized-magnon conditions, since $J_1\ne J_3$ and $J_m\ne 0$.
However, we are not far from the localized-magnon (i.e., flat-band) point in the parameter space
and one may expect to see some remnant features of the localized magnons at low temperatures and high fields.
Starting from  Eq.~(\ref{spin_azu}) it is possible to derive a simpler spin model
in the low-temperature high-field regime which mimics the properties of
azurite.\cite{deviations2}
By means of perturbation theory, see Sec.~\ref{sec05}, we arrive at the (pseudo)spin-$\frac{1}{2}$ $XXZ$ Heisenberg chain with the Hamiltonian
\begin{eqnarray}
\label{low_energy}
{\cal{H}}_{\rm{eff}}=\sum_{m=1}^{\cal{N}}
\left[
{\sf{C}}-{\sf{h}}T_m^z+{\sf{J}}\left(T_m^xT_{m+1}^x+T_m^yT_{m+1}^y\right)+{\sf{J}}^zT_m^zT_{m+1}^z
\right],
\end{eqnarray}
cf. Eq.~(\ref{013}).
Using the strong-coupling approach\cite{aa} supplemented by
an analysis of the exact-diagonalization data of the initial full model (\ref{spin_azu}) 
for excitations on the $\frac{1}{3}$-plateau and the other experimentally accessible 
features of the magnetization curve,\cite{effective_xy2}
one may find
\begin{eqnarray}
{\sf{h}}=(1.384{\sf{H}}-44.860)\,{\rm{K}},
\;\;\;
{\sf{J}}=1.249\,{\rm{K}},
\;\;\;
{\sf{J}}^z=0.565\,{\rm{K}}.
\end{eqnarray}
Here ${\sf{H}}$ is the experimentally applied magnetic field measured in teslas.
On the other hand, 
within the localized-magnon perturbation theory (and neglecting $J_m=0$) one may find\cite{deviations2}
\begin{eqnarray}
{\sf{h}}=(1.384{\sf{H}}-44.778)\,{\rm{K}},
\;\;\;
{\sf{J}}=0.845\,{\rm{K}},
\;\;\;
{\sf{J}}^z=0.287\,{\rm{K}},
\end{eqnarray}
cf. Eqs.~(\ref{013}) and (\ref{014}).
The effective low-energy Hamiltonian (\ref{low_energy}) may be useful for making theoretical predictions for low-temperature high-field properties of azurite.

Among the observable properties of azurite which are related to localized-magnon physics there are 
the plateau at $\frac{1}{3}$ of the saturation magnetization and the further steep increase
to saturation  reached at about 32.5 T\cite{kikuchi,kikuchi_2}
as well as the
inelastic neutron scattering spectra on the $\frac{1}{3}$-plateau (${\sf{H}}\approx 14$ T), 
see Ref.~\refcite{rule},
illustrating a band of dimer excitations whose dispersion is strongly suppressed by
the competition of $J_1$ and $J_3$.
As it was predicted in Ref.~\refcite{effective_xy2} (but not measured yet)
azurite should also exhibit a large magnetocaloric effect in the high-field region.

The successful description of available experimental data for azurite requires further theoretical studies.
In particular,
an accurate treatment of 
a magnetic anisotropy
(magnetization curve depends on the direction of the external magnetic field)
and
interchain couplings
(azurite orders below about 2 K)
is clearly necessary.
These issues are not accounted within the discussed above models and remain to be explored.

\subsection{Spin dimer magnet Ba$_2$CoSi$_2$O$_6$Cl$_2$}
\label{sec072}

Very recently,\cite{tanaka} 
Ba$_2$CoSi$_2$O$_6$Cl$_2$ single crystals have been synthesized,
their crystal structure has been determined,
and the magnetic properties have been investigated.
The magnetism of this compound is determined by the Co$^{2+}$ ions.
Structural studies show that Ba$_2$CoSi$_2$O$_6$Cl$_2$ is a two-dimensional coupled spin dimer system 
with an exchange network corresponding to the frustrated bilayer shown in Fig.~\ref{fig02}.
Furthermore, the Co$^{2+}$ ion is in an octahedral environment, 
and for temperatures much lower than 250 K its effective magnetic moment is represented by fictitious spin-$\frac{1}{2}$ operators.
When the octahedral environment exhibits tetragonal symmetry 
the effective interaction between fictitious spins is described by the spin-$\frac{1}{2}$ $XXZ$ Heisenberg model
with a strong easy-plane anisotropy, i.e.,
the effective exchange interaction is $XY$-like.
In Ref.~\refcite{tanaka} it was argued that the interaction of $z$-components is only about of $15\%$ of the isotropic $XY$ interaction.
Importantly,
the interdimer exchange interactions are strongly frustrated
and satisfy the relations
$J_{11}=J_{22}$,
$J_{11}+J_{22}=J_{12}+J_{21}$,
see Fig.~\ref{fig02}.
The authors of Ref.~\refcite{tanaka} estimate that the intradimer exchange interaction $J_2$ 
is about 4.6 times larger than the interdimer interactions, i.e.,
 $J_2\ge 4J$ which is required to realize the localized-magnon scenario.
Certainly, one has to expect (small) deviations from the ideal frustrated-bilayer geometry, 
however, the main peculiarity is a strong $XY$-like magnetic anisotropy.

Let us briefly discuss some of the experimental results reported in Ref.~\refcite{tanaka}.
The performed electron-spin-resonance measurements on Ba$_2$CoSi$_2$O$_6$Cl$_2$ at low temperatures allow to obtain the $g$-factors.
The magnetization process was measured  at $T=1.3$ K up to a magnetic field of ${\sf {H}}=70$ T. 
The saturation of the Co$^{2+}$ spin occurs at ${\sf {H}}_{\rm{sat}}\simeq 57$ and $41$ T for ${\sf{H}}\parallel c^*$
(i.e., parallel to the $c^*$ axis)
and ${\sf{H}}\parallel ab$
(i.e., parallel to the $ab$ plane), 
respectively;
$g_{c^*}\simeq 2.0$ and $g_{ab}\simeq 3.86$.
For both field directions, there is a wide zero-magnetization plateau indicating a
gapped
non-magnetic ground state. The authors argue that two Co$^{2+}$ spins located on the bases of neighboring CoO$_4$Cl 
pyramids are coupled to form an antiferromagnetic dimer singlet.
The plateau at zero magnetization is followed 
by a jumpwise change of the magnetization to a next plateau at half of the saturation magnetization.
Then again a jumpwise change to the saturation magnetization is found. 
The transition fields where the jumps take place are 
${\sf{H}}_c^{\parallel}=46.8$ T and ${\sf{H}}_{\rm{sat}}^{\parallel}=56.7$ T for ${\sf{H}}\parallel c^*$
and
${\sf{H}}_c^{\perp}=32.0$ T and ${\sf{H}}_{\rm{sat}}^{\perp}=41.0$ T for ${\sf{H}}\parallel ab$.
The singlet ground state is stabilized in a wide field range below ${\sf{H}}_c$.
Thus the measured magnetization curves for Ba$_2$CoSi$_2$O$_6$Cl$_2$ reported in Ref.~\refcite{tanaka} 
strongly resemble the theoretically predicted magnetization curve for the frustrated bilayer,\cite{ising_degrees} 
see the left panel in Fig.~\ref{fig04}.

Summarizing their experimental findings, the authors of Ref.~\refcite{tanaka} come to the conclusion
that Ba$_2$CoSi$_2$O$_6$Cl$_2$ is a two-dimensional coupled spin-dimer system with $XY$-like exchange interactions.
A stepwise magnetization process with plateaus at zero magnetization and at $\frac{1}{2}$
of the saturation, irrespective of the magnetic field directions,
together with the absence of a structural phase transition,
show that the frustration for the interdimer exchange interactions is almost perfect.
The $\frac{1}{2}$-plateau state is almost exactly given by the alternate product of singlet and triplet dimers,
which is called in Ref.~\refcite{tanaka} a ``Wigner crystal of triplons''\footnote{The $s^z=1$ component of the spin triplet is called the triplon.}
or a ``Wigner crystal of magnons''.
Obviously this Wigner crystal is just the magnon-crystal
state, i.e., the state where the trapping cells are completely occupied by
localized magnons in accordance with the hard-core rule, cf. Sec.~\ref{sec021} and Sec.~\ref{sec062}.

The above illustrated experimental results lead to the exciting suggestion 
that Ba$_2$CoSi$_2$O$_6$Cl$_2$ is a solid-state realization of the quantum Heisenberg antiferromagnet on a frustrated bilayer lattice
discussed in Sec.~\ref{sec03} (see also Ref.~\refcite{ising_degrees}).
Certainly, some differences are also evident: 
The model discussed in Sec.~\ref{sec03} has isotropic exchange interactions and ideal flat-band geometry.
An extension of the localized-magnon picture for nonideal geometry has been discussed in Sec.~\ref{sec05}
and apparently can be done for the frustrated bilayer too (at least within the strong-coupling
approach).
Within the framework of the strong-coupling approach it is straightforward  to take into account
also the anisotropy of the exchange interactions.\cite{anisotropy_jmmm}

The most intriguing feature of the frustrated bilayer is a phase transition which occurs 
if the magnetic field is in the range $h_2<h<h_1$ (see Sec.~\ref{sec03}).
In the terminology of the Ref.~\refcite{tanaka} that is the range ${\sf{H}}_c<{\sf{H}}<{\sf{H}}_{\rm{sat}}$. 
Therefore it would be very interesting to explore the magnetic field versus temperature phase diagram of Ba$_2$CoSi$_2$O$_6$Cl$_2$
performing, for example, precise measurements of the specific heat
and seeking for traces of the phase transition.
However, it might be that the required critical field ${\sf{H}}^{\parallel}_c=46.8$ T
is too high to
perform the corresponding thermodynamic measurements,
whereas for the other field direction, when ${\sf{H}}^{\perp}_c=32.0$ T, theoretical predictions should be elaborated yet.
[For the isotropic (i.e., $XXX$) Heisenberg interaction the field orientation is irrelevant.]
One may also expect that the critical fields of Ba$_2$CoSi$_2$O$_6$Cl$_2$ may vary under pressure,
this way putting them into a better accessible region.

\section{Flat-Band Hubbard Model at Low Temperatures and Low Electron Densities}
\label{sec08}

We consider now electronic flat-band Hubbard models.
Initially the Hubbard model was introduced to understand the origin of metallic
ferromagnetism.\cite{origin_FM}
Indeed, to avoid the repulsive on-site interaction $U$ a parallel
alignment of electron spins is favorable. 
On the other hand, 
typically the kinetic energy prevents the emergence of ferromagnetism.
The suppression of the kinetic energy may open the route to ferromagnetism in the Hubbard
model.
Flat bands may lead to a fascinating example of ferromagnetism in electronic systems due to complete quenching of the kinetic energy. For the class of such models A.~Mielke\cite{le1} and H.~Tasaki\cite{le2} proved
rigorously the existence of a fully polarized ferromagnetic ground state at certain electron numbers.

To be specific, we consider the standard repulsive one-orbital Hubbard model with the Hamiltonian
\begin{eqnarray}
\label{044}
H=\sum_{\sigma=\uparrow,\downarrow}H_{0,\sigma} + U\sum_i n_{i,\uparrow}n_{i,\downarrow},
\nonumber\\
H_{0,\sigma}=\sum_{(ij)}t_{ij}\left(c_{i,\sigma}^\dagger c_{j,\sigma}+c_{j,\sigma}^\dagger c_{i,\sigma}\right)
+\mu\sum_i n_{i,\sigma},
\nonumber\\
n_{i,\sigma}=c_{i,\sigma}^\dagger c_{i,\sigma},
\;\;\;
\{c_{i,\sigma_i},c_{j,\sigma_j}^\dagger\}=\delta_{\sigma_i,\sigma_j}\delta_{i,j},
\;\;\;
\{c_{i,\sigma_i},c_{j,\sigma_j}\}=0
\end{eqnarray}
with $t_{ij}>0$.
We have chosen in Eq.~(\ref{044}) nonstandard sign conventions for the hopping integrals $t_{ij}$ and the chemical potential $\mu$
to have closer analogy to the Heisenberg antiferromagnets in a magnetic field considered
in the previous sections.
Thus, hopping integrals $t_{ij}$ correspond to the
antiferromagnetic exchange
bonds  $J_{ij}$ and chemical potential $\mu$ corresponds to the applied
magnetic field $h$. 
There are several reviews on  Mielke-Tasaki Hubbard models\cite{tasaki_rev1,tasaki_rev2,tasaki_rev3}
and, therefore, we touch below only some aspects which were less emphasized previously.
Our basic strategy is to look for the correspondence between localized-magnon and localized-electron systems
stressing the similarities and contrasting the differences.

We begin with lattices,
which, although having lowest-energy flat bands, do not show ferromagnetism in the ground state.
Then we pass to some  one-dimensional chains which
exhibit  Mielke-Tasaki ground-state ferromagnetism. 
In all cases we completely characterized the many-electron ground states up
to a macroscopic number of electrons $n=n_{\rm max} \propto N$ corresponding to half filling of the
flat  band.
They constitute a set of independent ferromagnetic clusters.
Furthermore, we calculate their degeneracy.
As a result,
we obtain the thermodynamics of the standard Hubbard model on these lattices
at low temperatures and low electron densities.

\subsection{Frustrated diamond chain, frustrated ladder etc.}
\label{sec081}

In this section we summarize some results of Refs.~\refcite{fb-el2,chiral1,chiral2}.
There is a class of lattices which have a lowest-energy flat band, 
but do not belong to the class of Mielke's or Tasaki's flat-band ferromagnets.
Examples are
the frustrated diamond chain,  the frustrated two-leg ladder,
the double-tetrahedra chain, the frustrated three-leg ladder (see
Fig.~\ref{fig01}), the dimer-plaquette chain etc. in one dimension
or the square-kagome lattice (see Fig.~\ref{fig02}), the modified checkerboard lattice etc. in higher dimensions.
The localized states which represent the flat-band states are located within traps,
which, in contrast to Mielke's or Tasaki's lattices, do not have common sites, i.e., the traps are isolated from each other. 

We consider as an example the frustrated diamond Hubbard chain, see
Fig.~\ref{fig01}. 
The one-electron spectrum consists of three bands:
\begin{eqnarray}
\varepsilon_{1}=-t_2+\mu,
\;\;\;
\varepsilon_{2,3}(\kappa)=\frac{t_2}{2}\mp\sqrt{\frac{t_2^2}{4}+4t_1^2(1+\cos\kappa)}+\mu.
\end{eqnarray}
If $t_2>2t_1$ the flat band with energy
$\varepsilon_{1}$ is the lowest one.
The corresponding eigenstates can be localized on the vertical bonds.
Explicitly they can be written as
\begin{eqnarray}
l_{m,\sigma}^\dagger\vert{\rm{vac}}\rangle
=\frac{1}{\sqrt{2}}\left(c_{m,1,\sigma}^\dagger-c_{m,2,\sigma}^\dagger\right)\vert{\rm{vac}}\rangle
\end{eqnarray}
with $m$ running over all ${\cal{N}}=\frac{N}{3}$ trapping cells,
see Fig.~\ref{fig01}.
The application of $n$ distinct operators $l_{m,\sigma}^\dagger$ to the vacuum state $\vert{\rm{vac}}\rangle$ 
yields $n$-electron eigenstates with the energy $n\varepsilon_1$ 
which are the ground states in the $n$-electron subspace.
For $U>0$ their degeneracy is easily calculated as
$g_{{\cal{N}}}(n)=2^n{\cal{C}}_{\cal{N}}^n$, 
where ${\cal{C}}_{\cal{N}}^n={{\cal{N}}\choose{n}}=\frac{{\cal{N}}!}{n!(n-{\cal{N}})!}$
is the binomial coefficient.
Obviously, the traps (vertical bonds) are disconnected 
and therefore each traps can be occupied by an electron with arbitrary spin
$\uparrow$ or $\downarrow$. Surely, within the ground-state manifold there is
the fully polarized ferromagnetic state, but non-magnetic
states are predominant and averaging over all $n$-electron ground states
yields\cite{fb-el2}
$\frac{\langle \bm{S}^2\rangle_n}{N^2}
= \frac{3n}{4N^2}
\underset{N\to\infty}{\longrightarrow}
0$.
These arguments can be repeated for the other lattices mentioned above.
We only have to change accordingly 
the flat-band energy 
and 
the critical value of $t_2$ 
above which the described localized-electron picture holds.
In the case of the systems having triangular traps 
we have to take into account the additional chiral degrees of freedom to determine the degeneracy, see Sec.~\ref{sec04}.
As a result, 
each cell may be either empty, 
or occupied by an electron with spin $\sigma=\uparrow,\downarrow$ and one of two values of the chirality $\chi=\pm$,
or occupied by two electrons with different chiralities which belong to a spin-triplet state,
that is,
$g_1(0)=1$, $g_1(1)=4$, $g_1(2)=3$.
This circumstance will manifest itself in the explicit formula for the grand-canonical partition function, 
see Eq.~(\ref{ggg_2}) below.

\begin{figure}[bt]
\centerline{\psfig{file=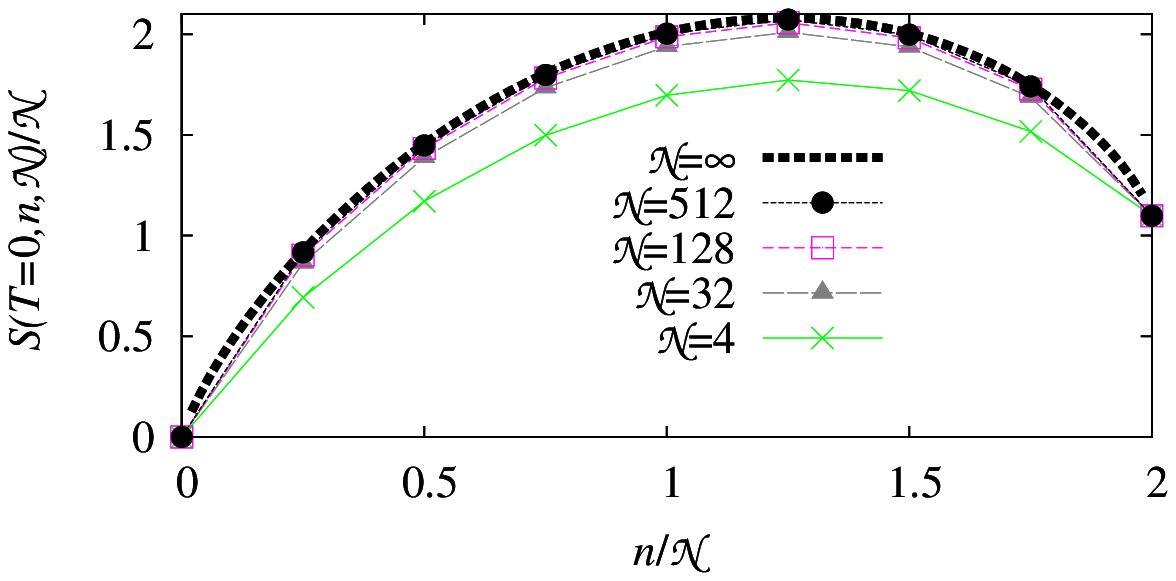,width=2.4in}\hspace{5mm}\psfig{file=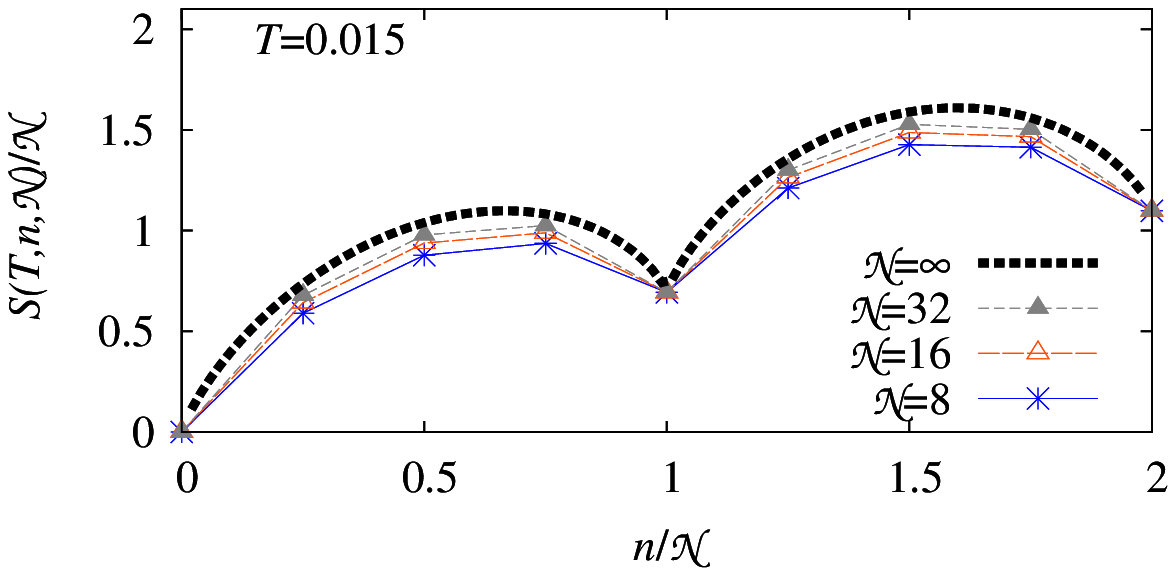,width=2.4in}}
\vspace*{8pt}
\caption{(Color online) 
Entropy $\frac{S(T,n,{\cal{N}})}{{\cal{N}}}$ versus electron density $\frac{n}{{\cal{N}}}$ 
for the Hubbard model on the double-tetrahedra chain as well as on the frustrated three-leg ladder with $t_2>2t_1$
($T=0$, left)
and in the presence of small imaginary component of the hopping integral between neighboring sites of the triangular traps $ig$, 
$g=0.1$
(slightly perturbed system)
($T\to 0$, right). Note that the curves for both systems are identical.}
\label{fig09}
\end{figure}

The thermodynamic properties of all considered systems are similar.
At low temperatures and for the chemical potential around $\mu_0=\mu-\varepsilon_1$
they are dominated by the huge number 
of 
localized-electron states. Their 
contribution to the grand-canonical partition function is given by 
\begin{eqnarray}
\label{ggg_1}
 \Xi(T,\mu,N)=\sum_{n=0}^{\cal{N}} g_{\cal{N}}(n) e^{-\frac{n\varepsilon_1}{T}}
 =\left(1+2z\right)^{\cal{N}},
\;\;\;z=e^{-\frac{\varepsilon_1}{T}} .
\end{eqnarray}
For the double-tetrahedra chain or the triangular tube due to the chiral
degrees of freedom we get 
\begin{eqnarray}
\label{ggg_2}
\Xi(T,\mu,N)=\hspace{-1mm} \sum_{n_1=0,1,2}\hspace{-1mm}\ldots\hspace{-1mm}\sum_{n_{\cal{N}}=0,1,2} 
\hspace{-2mm} g_1(n_1)\ldots g_{1}(n_{\cal{N}} )z^{n_1+\ldots +n_{\cal{N}}}
 =\left(1+4z+3z^2\right)^{\cal{N}}.\;
\end{eqnarray}
Knowing the grand-thermodynamical potential 
$\Omega(T,\mu,N)=-T\ln\Xi(T,\mu,N)$
one can easily obtain all thermodynamic quantities.
For example,
the entropy per cell is given by $s(T,\mu)=-\frac{\partial \Omega(T,\mu,N)}{{\cal{N}}\partial T}$,
the specific heat per cell is given by $c(T,\mu)=T\frac{\partial s(T,\mu)}{\partial T}$,
the average number of electrons is given by $\overline{n}(T,\mu)=\frac{\partial \Omega(T,\mu,N)}{\partial \mu}$.
We can also obtain thermodynamic quantities in the canonical ensemble such
as $s(T,n)$ or $c(T,n)$.
The thermodynamic quantities for the electron system show many similarities to the corresponding
ones of the spin system.
Thus, 
the entropy $s(T,\mu_0)$ remains finite as $T\to 0$ (residual entropy),
the grand-canonical specific heat $c(T,\mu)$ exhibits an extra low-temperature peak if $\mu$ is below or above $\mu_0$,
and the ground-state dependence of the average electron density versus the chemical potential exhibits a jump at $\mu=\mu_0$.
More usual canonical quantities also provide fingerprints of localized-electron states.
Thus, 
the entropy as a function of the  electron density at low temperatures
is especially interesting in the case of chiral localized-electron states,
see Fig.~\ref{fig09}.
In the left panel of Fig.~\ref{fig09} 
we show the residual ground-state entropy $\frac{S(T=0,n,{\cal{N}})}{{\cal{N}}}$ versus electron density $\frac{n}{{\cal{N}}}$
for the Hubbard model on the double-tetrahedra chain as well as on the frustrated three-leg ladder with $t_2>2t_1$
(the data for both systems are identical).
If the hopping integral between neighboring sites along the triangular traps acquires a small pure imaginary component $ig$
(e.g., due to a magnetic field perpendicular to the triangular trap)
the degeneracy due to the chirality is lifted
and as a result the perturbed model exhibits a different dependence of the residual ground-state entropy 
on the electron density,
see the right panel of Fig.~\ref{fig09}.

Let us stress again that the considered lattices, although they have flat bands, do not show ferromagnetic ground-state order.
However, 
ferromagnetic ground states may appear if the flat band acquires small dispersion provided the Hubbard interaction is strong enough,\cite{dispersion-driven}
see Sec.~\ref{sec10} below.

\subsection{Sawtooth and kagome chains}
\label{sec082}

In this section we summarize some results of
Refs.~\refcite{fb-el1,fb-el3,fb-el4}.
The ground-state manifold discussed in the previous subsection belongs 
to lattices with only disconnected trapping cells.
As a result a trap may represent an isolated one-cell ferromagnetic cluster
where the associated $s=\frac{1}{2}$ spin  
can be flipped irrespective of the other cells.
Now we consider examples of Mielke's line graphs and Tasaki's decorated
lattices in one dimension,
where neighboring trapping cells share a common site, 
i.e., they are connected. 
We will consider here in some detail the sawtooth chain (one-dimensional
Tasaki lattice) and a kagome chain (linegraph of an unfrustrated two-leg ladder), see
Fig.~\ref{fig01}.
Again we will construct many-electron eigenstates 
and calculate their degeneracy.

The sawtooth chain consists of ${\cal{N}}=\frac{N}{2}$ cells (see Fig.~\ref{fig01}) 
and there are two branches of the one-electron energies:
\begin{eqnarray}
\varepsilon_{1,2}(\kappa)
=t_1\cos\kappa \mp\sqrt{t_1^2\cos^2\kappa + 2t_2^2(1+\cos\kappa)} +\mu.
\end{eqnarray}
For $t_2=\sqrt{2}t_1>0$ the lowest one-electron band is flat,
$\varepsilon_1(\kappa)=\varepsilon_1=-2t_1+\mu=-2t+\mu$.
The flat-band eigenstates may be taken as the ones localized in
${\sf{V}}$-shaped valleys,
i.e.,
$l_{2j}^\dagger\vert{\rm{vac}}\rangle=(c_{2j-1}^\dagger-\sqrt{2}c_{2j}^\dagger+c_{2j+1}^\dagger)\vert{\rm{vac}}\rangle$.

The kagome chain consists of ${\cal{N}}=\frac{N}{3}$ cells of diamond shape (see Fig.~\ref{fig01})
and there are three branches of the one-electron energies:
\begin{eqnarray}
\varepsilon_{1}(\kappa)=\varepsilon_1=-2t+\mu,
\;\;\;
\varepsilon_{2}(\kappa)=2t\cos\kappa +\mu,
\;\;\;
\varepsilon_{3}(\kappa)=2t(1+\cos\kappa) +\mu.
\end{eqnarray}
The localized states $l_{3j}^\dagger\vert{\rm{vac}}\rangle$ are created by the operators
$l_{3j}^\dagger=c^\dagger_{3j}-c^\dagger_{3j+1}-c^\dagger_{3j+2}+c^\dagger_{3j+3}$.
Note, however, that for even ${\cal{N}}$, in contrast to the case of the sawtooth chain,
a state with $\kappa=\pi$ from the dispersive band $\varepsilon_2(\kappa)$ has the flat-band energy.
This eigenstate has the form 
$\alpha^{\dagger}_{2,\pi}\vert{\rm{vac}}\rangle
=\frac{1}{\sqrt{2{\cal{N}}}}\sum_{j=0}^{{\cal{N}}-1}
(-1)^{j}
(c_{3j+1}^{\dagger}-c_{3j+2}^{\dagger})\vert{\rm{vac}}\rangle$.
Being in this state the electron is trapped on the two legs (two-leg state).
Thus we are faced with a set of ${\cal{N}}+1$ one-electron states having the energy $-2t+\mu$.
Note that a similar construction can be elaborated for another version of a kagome chain
with $\frac{N}{5}$ cells of hexagon shape,
see Ref.~\refcite{fb-el3}.

\begin{figure}[bt]
\centerline{\psfig{file=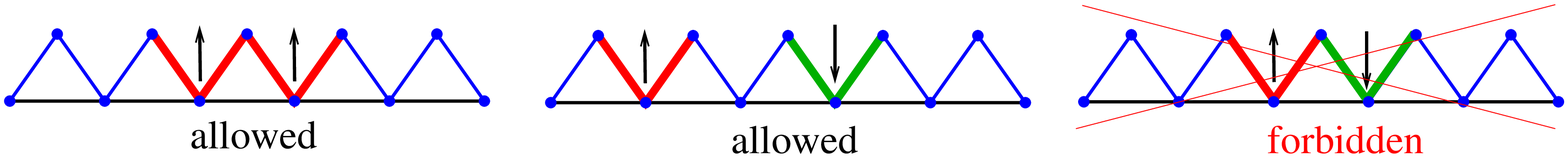,width=4.9in}}
\caption{(Color online)
Illustration of the occupation rules of the trapping cells for the Hubbard model
on the sawtooth chain.}
\label{fig_allow_forbidden}
\end{figure}

We pass to the many-electron states for Hubbard repulsion $U>0$ 
and begin with the sawtooth chain.\cite{fb-el1,fb-el3,fb-el4}
In analogy to the last subsection, Sec.~\ref{sec081},
we first consider those $n$-electron states where the electrons are located in
non-contiguous (i.e., disconnected) traps. They
are ground states in the $n$-electron subspace with energy
$n\varepsilon_1$.
As long as we consider the occupation of disconnected traps only, the spin of the electrons is
again irrelevant 
and we have a $2^n$-fold degeneracy of such a state, if we  fix the location of the trapped electrons
($n$ independent ferromagnetic clusters, each consists of one cell).
However, these states are not the only ground states in the $n$-electron subspace.
An example of another type of states is given by
$l_{2j_1,\uparrow}^\dagger\ldots l_{2(j_1+n-1),\uparrow}^\dagger\vert{\rm{vac}}\rangle$;
this state corresponds to $n$ electrons, all with spin $\uparrow$, sitting within $n$ contiguous valleys, 
i.e., it represents a fully polarized ferromagnetic cluster of $n$ connected
cells, cf. Fig.~\ref{fig_allow_forbidden} (left) for $n=2$.
If only  $\uparrow$-electrons are considered, the difference to the previously discussed
localized multi-magnon states becomes particularly evident, since for the magnon
states the occupation of neighboring trapping cells was not allowed.
Although the traps have common sites, the Hubbard repulsion is inactive, since the electrons have the same spin, 
i.e., due to the Pauli principle the simultaneous occupation of a common site is
ruled out.
Clearly, because of the SU(2) invariance of the Hubbard model this state is $(n+1)$-fold degenerate,
i.e., it is one component of a SU(2)-multiplet.
Let us emphasize, 
that the occupation of neighboring cells with electrons of different spin is
forbidden, i.e., such states are not the eigenstates,
cf. Fig.~\ref{fig_allow_forbidden} (right).
To find all ground states in the $n$-electron subspace
we must consider all possible splittings of $n$ into a sum $n=n_1+n_2+\ldots\,$, $n_i>0$ 
such that clusters of $n_i$ contiguous occupied valleys are separated. 
Furthermore, we have to include into the consideration the $(n_i+1)$-fold degeneracy of each
connected cluster
which gives an $[(n_1+1)(n_2+1)\ldots]$-fold degeneracy of a state with fixed position of the occupied traps.

Next we consider the total degeneracy of the ground-state manifold in the $n$-electron subspace
$g_{{\cal{N}}}(n)$.
There are two ways to calculate  $g_{{\cal{N}}}(n)$.
The first approach was presented in Ref.~\refcite{fb-el1}:
Interestingly, the localized-electron states can also be mapped onto a hard-dimer problem 
on an auxiliary simple chain. 
This mapping is similar (but not identical) and more sophisticated than the
mapping for localized-magnon states mentioned in Secs.~\ref{sec021}
and \ref{sec03}.
Thus, in contrast to
the mapping for Heisenberg spins each valley of the sawtooth
chain is assigned to {\bf two}
sites of the simple chain, i.e., the auxiliary simple chain has  $2{\cal{N}}$
sites.
Again, each state from the $n$-electron ground-state manifold corresponds to a spatial configuration of $n$ hard-dimers on the simple chain.
Only the easy case of maximum filling of $n=n_{\rm max}={\cal{N}}$ electrons (half filling of the flat band) has to be treated separately.
Finally we get\cite{fb-el1} for the degeneracies
\begin{eqnarray}
g_{\cal{N}}(n)={\cal{Z}}_{\rm{hd}}(n,2{\cal{N}})+({\cal{N}}-1)\delta_{n,{\cal{N}}},
\end{eqnarray}
where ${\cal{Z}}_{\rm{hd}}(n,2{\cal{N}})$ denotes the canonical partition function of $n$ hard dimers on the $2{\cal{N}}$-site simple chain.
Using the classical transfer-matrix approach we calculate the grand-canonical partition function
\begin{eqnarray}
\Xi_{\rm{hd}}(z,2{\cal{N}})=\sum_{n=0}^{\cal{N}} z^n{\cal{Z}}_{\rm{hd}}(n,2{\cal{N}})
=\lambda_+^{2{\cal{N}}} + \lambda_-^{2{\cal{N}}},
\nonumber\\
\lambda_{\pm}=\frac{1}{2}\pm\sqrt{\frac{1}{4}+z}, 
\; \; \;
 z=e^{\frac{\mu_0-\mu}{T}},
 \; \; \;
 \mu_0=2t.
\label{activity}
\end{eqnarray}
The canonical partition function can be determined through the relation
${\cal{Z}}_{\rm{hd}}(n,2{\cal{N}})=\left.\frac{1}{n!}\frac{d^n\Xi_{\rm{hd}}(z,2{\cal{N}})}{dz^n}\right\vert_{z=0}$.

An alternative way to compute  the contribution of the
localized states to the partition function,
${\cal{Z}}(n,{\cal{N}})={\cal{Z}}_{\rm{hd}}(n,2{\cal{N}})$, for  the initial
${\cal{N}}$-cell Hubbard chain was used in Refs.~\refcite{fb-el3,fb-el4}.
Let us illustrate this alternative
approach in some detail. 
Within this approach 
we 
take into account not only the geometrical degeneracy related to various possibilities of occupying $n$ cells 
of the ${\cal{N}}$-cell chain
but also the components of the SU(2)-multiplets.
Instead of the true component of the multiplet, 
one may choose, for counting purposes, a representative state,
e.g., the one where all spins $\uparrow$ are in the left of each $n$ contiguously occupied valleys and the spins $\downarrow$ are on the right side.
Then the counting problem becomes as follows:
We have to count the number of configurations ${\cal{Z}}(n,{\cal{N}})$ of three states in the trapping cells, 
namely, empty and occupied with $\sigma=\uparrow$ or $\sigma=\downarrow$,
subject to the constraint that no states $\uparrow$ is allowed to appear as the right neighbor of a $\downarrow$.
This problem can be solved with the help of a $3\times 3$ transfer matrix in the grand-canonical-ensemble
setup,
namely,
\begin{eqnarray}
\Xi(z,{\cal{N}})=\sum_{n=0}^{\cal{N}} z^n{\cal{Z}}(n,{\cal{N}})
={\rm{Tr}} {\bf {T}}^{{\cal{N}}}=\xi_+^{\cal{N}} + \xi_-^{\cal{N}} + \xi_0^{\cal{N}},
\nonumber\\
\xi_\pm =\frac{1}{2}+z\pm\sqrt{\frac{1}{4}+z},
\;\;\;
\xi_0 = 0.
\label{transfer_II}
\end{eqnarray}
We have used a transfer matrix
\begin{eqnarray}
{\bf{T}}
=
\left(
\begin{array}{ccc}
T(0,0)          & T(0,\uparrow)          & T(0,\downarrow) \\ 
T(\uparrow,0)   & T(\uparrow,\uparrow)   & T(\uparrow,\downarrow) \\ 
T(\downarrow,0) & T(\downarrow,\uparrow) & T(\downarrow,\downarrow) 
\end{array}
\right)
=
\left(
\begin{array}{ccc}
1 & 1 & 1 \\ 
z & z & z \\ 
z & 0 & z 
\end{array}
\right),
\label{transfer_IIa}
\end{eqnarray}
where $z$ denotes the activity defined in Eq.~(\ref{activity}).
Comparing Eqs.~(\ref{activity}) and (\ref{transfer_II}) we find 
that both expressions for $\Xi(z,{\cal{N}})=\Xi_{\rm{hd}}(z,2{\cal{N}})$ are
equivalent, since $\xi_{\pm}=\lambda_{\pm}^2$. 

Now let us consider the kagome chain for electron numbers $n>1$. 
There is a difference to the sawtooth chain, since in addition to the hard-dimer states
(relevant for both systems) there are additional leg states.
First we consider the case $n=2$.
In addition to the ${\cal{Z}}(2,{\cal{N}})$ hard-dimer states (electrons sit
in the diamond-shaped trapping cells only)
we can put one electron into one of ${\cal{N}}$ diamond cells and the other electron in a two-leg state. 
Taking into account the SU(2) symmetry we obtain $3{\cal{N}}$ extra states in the subspace with $n=2$ electrons
compared to the sawtooth chain.
Similarly, there are extra states for $n>2$. 
In particular, localized states exist up to maximum filling $n=n_{\rm max}={\cal{N}}+1$,
where all the ${\cal{N}}$ diamonds and the legs are occupied. 
Note, however that these extra states become irrelevant for the thermodynamics in the limit $N \to \infty$, see below.
The degeneracy of the ground states for the periodic kagome chain is given by\cite{fb-el3}
\begin{eqnarray}
g^{\rm{kagome}}_{\cal{N}}(n)
=
(1-\delta_{n,{\cal{N}}+1}) g^{\rm{sawtooth}}_{{\cal{N}}}(n)
+
(1-\delta_{n,0})\left[(n+1){\cal{C}}_{{\cal{N}}}^{n-1} + \delta_{n,2}\right].
\end{eqnarray}
Knowing the ground-state energies and degeneracies for $n=0,\ldots,n_{\max}$, $n_{\max}={\cal{N}}$ for the sawtooth chain or $n_{\max}={\cal{N}}+1$ for the kagome chain,
we can calculate the contribution of these states to the grand-canonical partition function
$\Xi(T,\mu,N)=\sum_{n=0}^{n_{\max}}g_{{\cal{N}}}(n)z^n$.
This contribution 
dominates the  thermodynamics
at low temperatures for a chemical potential around the value
$\mu_0=2t$.
Although $\Xi(T,\mu,N)$ is different for finite sawtooth and kagome chains, 
it approaches the same function in the thermodynamic limit\cite{fb-el3}
\begin{eqnarray}
\label{056}
\Xi(T,\mu,N)=\left(\frac{1}{2}+\sqrt{\frac{1}{4}+e^{\frac{2t-\mu}{T}}}\right)^{2{\cal{N}}}.
\end{eqnarray}
Thus the sawtooth and the kagome chains exhibit identical behavior for
$N\to \infty$  
with a difference only in the relation between $N$ and ${\cal{N}}$.

\begin{figure}[bt]
\centerline{\psfig{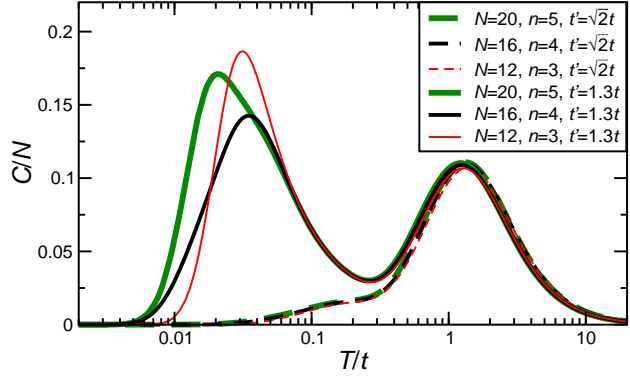}} 
\vspace*{8pt}
\caption{(Color online)
The Hubbard model on the sawtooth chain of $N=12,\,16,\,20$ sites: 
Specific heat per site for $n=\frac{{\cal{N}}}{2}$ and $U \to \infty$. 
The results for ideal flat-band geometry 
($t_2=\sqrt{2}t_1$, dashed curves)
are compared to the ones for slightly violated flat-band condition 
($t_2=1.3t_1$, solid curves).}
\label{fig10}
\end{figure}

\begin{figure}[bt]
\centerline{\psfig{file=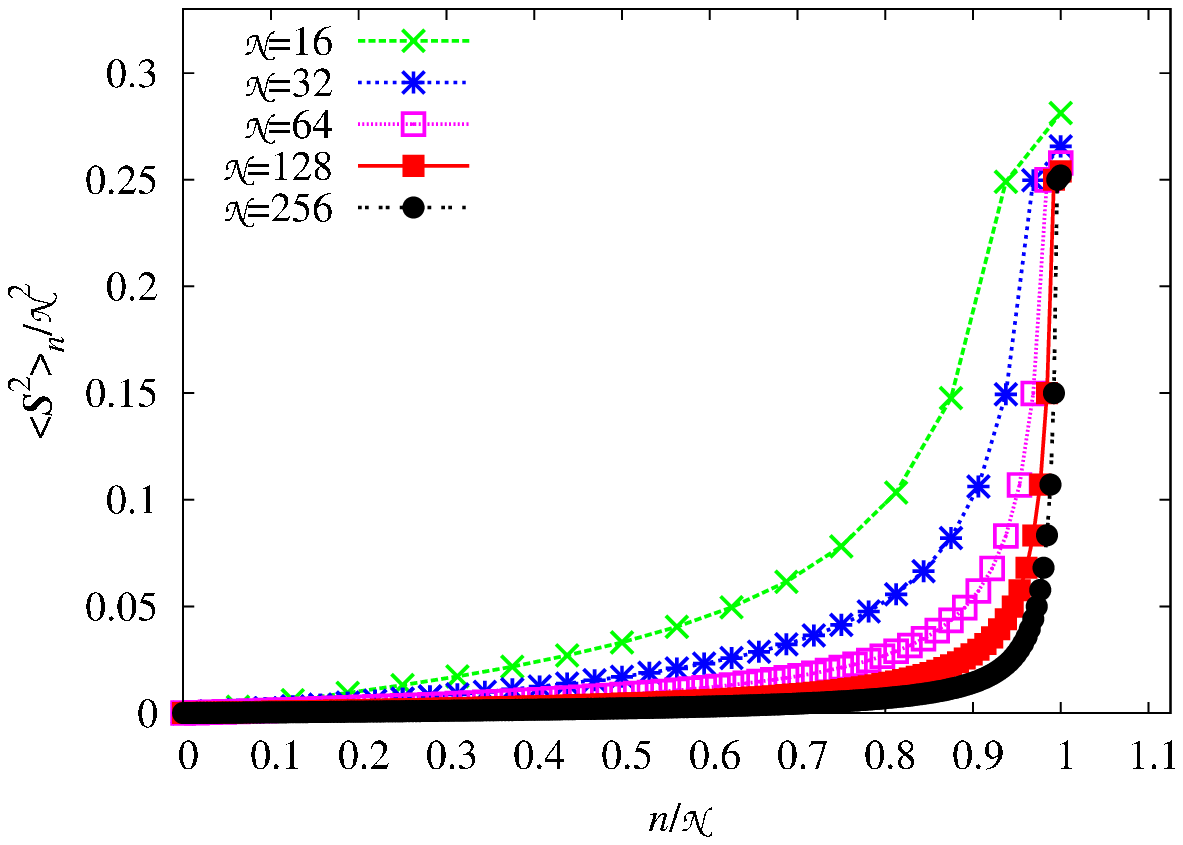,width=2.25in}\hspace{10mm}\psfig{file=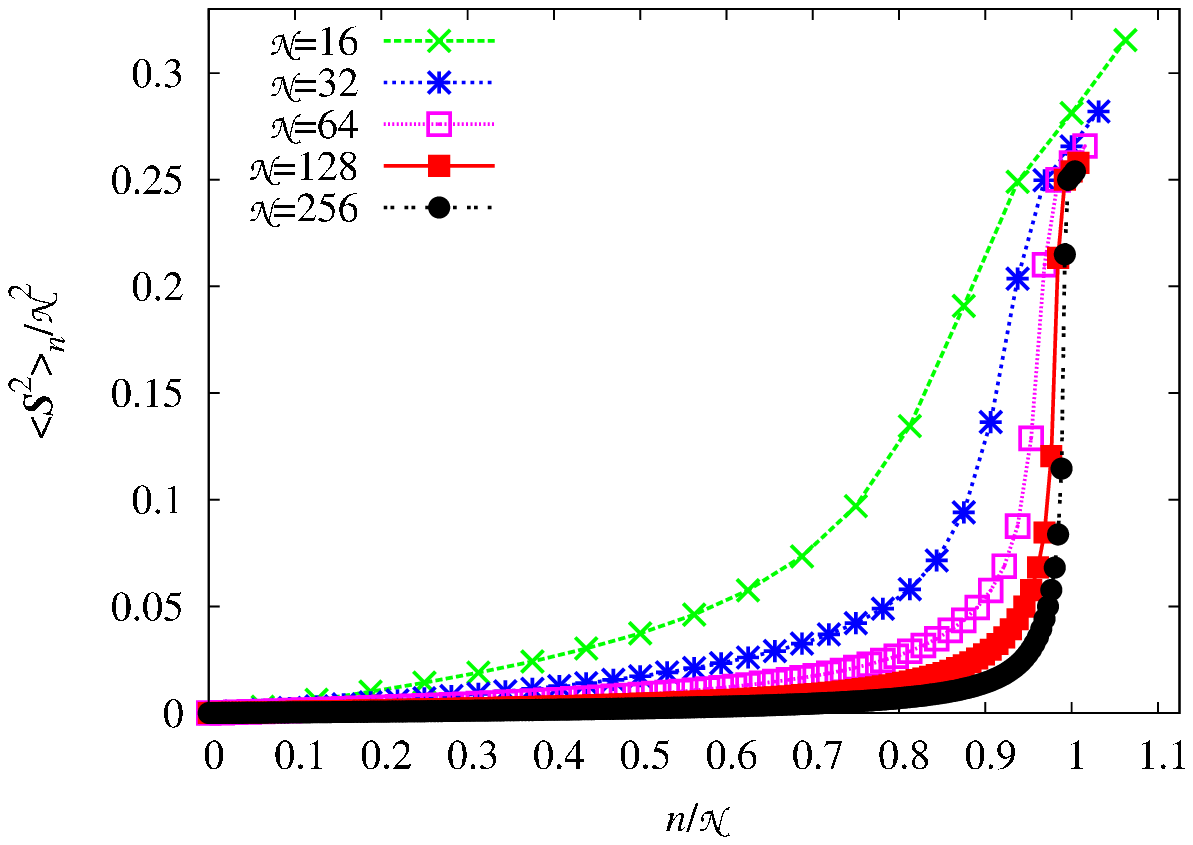,width=2.25in}}
\vspace*{8pt}
\caption{(Color online)
The sawtooth chain (left) and the kagome chain (right): 
The averaged ground-state magnetic moment per cell
$\frac{\langle{\bf{S}}^2\rangle_n}{{\cal{N}}^2}$ as a function the electron
density $\frac{n}{{\cal{N}}}$
for ${\cal{N}}=16,\,32,\,64,\,128,\,256$.}
\label{fig11}
\end{figure}

A detailed discussion of the thermodynamic properties and a comparison to exact-diagonalization data can be found in
Refs.~\refcite{fb-el1} and \refcite{fb-el3}.
Here we focus only on the low-temperature dependence of the specific heat,\cite{fb-el1,fb-el3,vj} see Fig.~\ref{fig10}, 
and on the ground-state magnetic moment, see Fig.~\ref{fig11}.
The specific heat in the grand-canonical ensemble shows an extra low temperature maximum if $\mu$ is around $\mu_0$.\cite{fb-el1,fb-el3}
The specific heat of the localized states in the canonical ensemble vanishes, 
since the localized-electron energies depend only on the number of electrons
[in fact, they constitute the (degenerate) ground-state manifold].
This is illustrated by exact-diagonalization results 
for the Hubbard sawtooth chain for
$n=\frac{{\cal{N}}}{2}=\frac{N}{4}$ in Fig.~\ref{fig10},
where obviously $\frac{C(T)}{N}$ (dashed curves) remains very small up to about $\frac{T}{t} \simeq 0.05$.   
However, a deviation form the flat-band condition, 
i.e., in case of $t_2 \ne \sqrt{2}t_1$, 
produces a splitting of the ground-state manifold 
and an additional low-temperature maximum of the specific heat emerges as demonstrated in Fig.~\ref{fig10} for $t_2= 1.3t_1$.

The study of the integrated number of states above the degenerate ground state in the subspaces with different $n<{\cal{N}}$ and $U>0$ shows
a noticeable density of states at rather low energies for small and even for rather large values of $U$.
One may conclude that these low-lying excitations originate from  states which belong to the ground-state manifold for $U=0$.\cite{fb-el1}
The question whether the density of low-energy excitations increases with increase of $N$ 
leading to a vanishing excitation gap for $N\to\infty$\footnote{The absence of the gap would imply quantitative corrections to the hard-dimer description (\ref{056}).} 
remains open and requires further studies.
For the sawtooth chain the separation of the ground-state manifold (for $t_2=\sqrt{2}t_1$) within the grand-canonical setup is controlled by the charge gap
$\Delta\mu=[E({\cal{N}}+1)-E({\cal{N}})]-[E({\cal{N}})-E({\cal{N}}-1)]$.
Exact-diagonalization calculations show\cite{fb-el1} that it opens linearly in $U$ and
reaches a saturation value as $U\to\infty$ showing no visible finite-size effects.

Next, we consider the ground-state magnetic properties of the sawtooth chain and the kagome chain.
Since both chains belong to the class of Mielke's or Tasaki's flat-band
ferromagnets, there is an exact proof of a fully polarized ferromagnetic
ground state for $n=n_{\max}$, see Refs.~\refcite{le1,le2,le3}.
Based on our discussion given above there is a simple geometrical
explanation:  For $n=n_{\max}$ all cells are occupied and we are faced with 
a fully polarized ferromagnetic connected cluster that includes {\bf all}
cells. Hence no independent spin flips of disconnected clusters are possible. 
Lowering $n < n_{\max}$ just the possibility of independent spin flips of
disconnected clusters leads to a decreasing averaged total magnetic moment.       
The mapping of the localized-electron states onto a classical transfer-matrix problem 
is the crucial step to calculate the averaged total magnetic moment.\cite{fb-el1,fb-el3,fb-el4}  
For that we consider the operator ${\bf S}^2$ and
perform the equal-weight average of ${\bf S}^2$
over all degenerate ground states for the given number of electrons $n$.
Then we consider the quantity
\begin{eqnarray}
0\le \frac{\langle{\bf{S}}^2\rangle_n}{{\cal{N}}^2}\le\frac{S_{\max}(S_{\max}+1)}{{\cal{N}}^2},
\;\;\;
S_{\max}=\frac{n}{2}.
\end{eqnarray}
It achieves its maximal value, 
if the ground state in the subspace with $n$ electrons is the saturated ferromagnetic state.
If it has a nonzero value less than the maximal value (nonsaturated
ferromagnetism) 
the ground-state manifold in the subspace with $n$ electrons contains a
substantial part of ferromagnetic ground states.
Again both transfer-matrix techniques (see above) are suitable to calculate
$\langle{\bf{S}}^2\rangle_n=3\langle{S^z}^2\rangle_n$. 
We illustrate here briefly the approach based on the  $3\times 3$
transfer-matrix technique, see Eqs.~(\ref{transfer_II}) and
(\ref{transfer_IIa}), to calculate this quantity.
The average over all degenerate ground states for a given number of electrons is 
$\langle{S^z}^2\rangle_n={\cal{N}}\sum_{j=0}^{{\cal{N}}-1}\langle S_0^zS_j^z\rangle_n$,
where $S_j^z$ is the $z$-component of the spin operator of the trap $j$.
Within the grand-canonical ensemble we may find 
$\langle S_0^zS_j^z\rangle_{\zeta}=\sum_{n=2}^{\cal{N}}{\zeta}^n{\cal{Z}}(n,{\cal{N}})\langle S_0^zS_j^z\rangle_n$ 
[note that here $\zeta=e^{\frac{\mu_0-\mu}{T}}$ is the activity
which was denoted by $z$ in Eq.~(\ref{activity})]
which yields the required canonical $zz$ correlation function $\langle S_0^zS_j^z\rangle_n$
via the relation
${\cal{Z}}(n,{\cal{N}})\langle S_0^zS_j^z\rangle_n=\left.\frac{1}{n!}\frac{d^n \langle S_0^zS_j^z\rangle_{\zeta}}{d{\zeta}^n}\right\vert_{{\zeta}=0}$.
Finally, within the $3\times 3$ transfer matrix technique $\langle S_0^zS_j^z\rangle_{\zeta}$
is obtained by
\begin{eqnarray}
\langle S_0^zS_j^z\rangle_{\zeta}
=
{\rm{Tr}}({\bf{S}}{\bf{T}}^j {\bf{S}} {\bf{T}}^{{\cal{N}}-j}),
\;\;\;
{\bf{S}}
=
\left(
\begin{array}{ccc}
0 & 0           & 0\\
0 & \frac{1}{2} & 0 \\
0 & 0           & -\frac{1}{2}
\end{array}
\right),
\label{S2_vs_n}
\end{eqnarray}
where ${\bf{T}}$ is the transfer matrix given in Eq.~(\ref{transfer_IIa}).

Based on Eq.~(\ref{S2_vs_n}) in Ref.~\refcite{fb-el3} the averaged magnetic moment
$\frac{\langle{\bf{S}}^2\rangle_n}{{\cal{N}}^2}$  was calculated up to
${\cal{N}}=256$, i.e.,  up to $N=512$ ($N=768$) for the sawtooth (kagome)
chain (cf. Fig.~\ref{fig11}). 
While for the sawtooth chain (left panel) 
only hard-dimer states exist,
for the kagome chain (right panel) also the ferromagnetic
states related to the existence of leg states have to be taken into account.
However, their contribution vanishes in the thermodynamic limit,
cf. also the discussion of thermodynamics.
Obviously, the averaged magnetic moment 
becomes smaller with increasing of ${\cal{N}}$
for any fixed $0<\frac{n}{{\cal{N}}}<1$.
In fact
$\frac{\langle{\bf{S}}^2\rangle_n}{{\cal{N}}^2}$ decreases linearly with
$\frac{1}{{\cal{N}}}$, cf. Ref.~\refcite{fb-el4} and also
Eq.~(\ref{1D_S2_ns_n_exact}) in Sec.~\ref{sec09}, where we discuss the
flat-band
ferromagnetism in one dimension    
using a percolation representation.
Finally, 
the region of ground-state ferromagnetism shrinks to one particular electron density $\frac{n}{{\cal{N}}}=1$ in the thermodynamic limit.
In the next section we will illustrate that the existence of ground-state ferromagnetism in a finite region 
of electron density $\frac{n}{{\cal{N}}}$ can be observed when going from dimension one to dimension two.

\section{Flat-Band Ferromagnetism as a Pauli-Correlated Percolation Problem}
\label{sec09}

A recently elaborated percolation representation of localized 
electrons in flat bands\cite{fb-el4} provides an important
step in understanding of flat-band ferromagnetism and the corresponding para--ferromagnetic transition.     
Some ideas on the percolation representation were introduced in the early papers by A.~Mielke and
H.~Tasaki\cite{le1,le2,le3} without presenting a detailed analysis.
Here we report on the percolation analogy published in Refs.~\refcite{fb-el3_4,fb-el4,mykola,fb-el5} in some
detail.
As discussed in the previous section the emergence of ground-state
ferromagnetism is related to ferromagnetic clusters of connected
cells occupied by electrons. Due to the Pauli principle the occupation of
neighboring  cells (sharing common sites) by electrons, 
and therefore the occupation of a cluster built by  connected cells, 
is possible only if such electrons form a symmetric (i.e., ferromagnetic) spin state.
As the density of
electrons increases, ferromagnetic clusters of increasing size appear. The
ferromagnetic transition corresponds to the emergence of a
ferromagnetic cluster of connected cells
containing a nonzero fraction of the electrons, i.e., we face a
percolation problem.    
The special features of this percolation problem are related to the
degeneracy, $m+1$, of a ferromagnetic cluster $C$ of $|C|$ cells ($|C|$
sites of the auxiliary lattice) containing
$m=|C|$ electrons. This
degeneracy 
gives differing weights to different clustering of electrons.
This unusual percolation problem is called Pauli-correlated percolation\cite{fb-el4} or Pauli
percolation.\cite{fb-el5}

Within the canonical ensemble (i.e.,
for fixed electron density $\frac{n}{{\cal{N}}}$)
the percolation problem deals with all possible geometric configurations $q$ of $n$ particles distributed over 
${\cal{N}}$ sites of the auxiliary lattice with 
a nontrivial weight of each configuration $q$
\begin{eqnarray}
\label{059}
W(q)=\prod_{i=1}^{M_q}\left(\vert C_i\vert+1\right).
\end{eqnarray}
Here the  weight $W(q)$ arises from the above mentioned spin degeneracy of a ferromagnetic cluster of size $\vert C_i\vert$ 
in the configuration $q$.
$M_q$ denotes the number of clusters in the configuration $q$, $\sum_{i=1}^{M_q}\vert C_i\vert=n$.
The expectation value of a quantity $A$ is given by the usual expression
$\langle A\rangle=\frac{1}{Z}\sum_q A(q)W(q)$,
where the canonical partition function is $Z=\sum_q W(q)$.
Here the sum over $q$ runs over all configurations for a given number of electrons $n$.
As mentioned already above,
the appearance of ferromagnetism is related  to the emergence of a cluster containing a nonzero fraction of electrons.
Thus it can be pinned down using standard tools of the percolation theory.\cite{percolation1,percolation2,percolation3}
To detect ferromagnetism  the relevant observable is the square of the total spin
${\bf{S}}^2$, see also the last section.
For $n$ particles distributed over the lattice with a particular geometric configuration $q$
it can be written as 
\begin{eqnarray}
\label{060}
{\bf{S}}_q^2
=\sum_{i=1}^{M_q} \frac{\vert C_i\vert}{2} \left(\frac{\vert C_i\vert}{2}+1\right)
=\sum_{l=1}^n{\cal{N}} n_q(l)\frac{l}{2}\left(\frac{l}{2}+1\right).
\end{eqnarray}
In the first form the contribution from each cluster $C_i$,
$i=1,\ldots,M_q$,
of the configuration $q$ is taken into account,
while the second form contains the normalized number of cluster of size $l$ in the configuration $q$ denoted by $n_q(l)$.
Such quantities [i.e., $n_q(l)$] are typical ones in percolation studies.
The crucial differences is owing to the nontrivial weight factor $W(q)$ 
(which reflects $m+1$ possible orientations of the total spin of a ferromagnetic cluster of
$m$ electrons)
since for the standard percolation all geometric configurations have the same weight and $W(q)=1$.

Instead of the canonical ensemble with fixed number of occupied sites $n$ (or electron density $\frac{n}{{\cal{N}}}$),
one may consider the grand-canonical ensemble letting each site of the lattice to be occupied with an a priori probability $p$ or left empty with an a priori probability $1-p$.
Eq.~(\ref{059}) now becomes
\begin{eqnarray}
\label{061}
W(q)=\prod_{i=1}^{M_q} e^{\mu \vert C_i\vert}\left(\vert C_i\vert+1\right)
=\prod_{i=1}^{M_q} \left(\frac{p}{1-p}\right)^{\vert C_i\vert}\left(\vert C_i\vert+1\right),
\end{eqnarray}
where $e^{\mu}=\frac{p}{1-p}$ is the fugacity which is used to tune the number of electrons.
Eq.~(\ref{061}) leads to the same formula for the expectation value,
$\langle A\rangle=\frac{1}{Z}\sum_q A(q)W(q)$,
and the grand-canonical partition function,
$Z=\sum_q W(q)$,
where, however, now the sum over $q$ runs over all configurations of $n=0,1,\ldots,{\cal{N}}$ particles (electrons).

Pauli percolation exhibits a number of characteristic features.
First, it corresponds to an effective (entropic) repulsion between particles
which, although is long-range, saturates, is range free and does not depend on the shape of the cluster, and is many-body.
Therefore it should not be astonishing 
that such quantities as the percolation threshold or the cluster-size distribution differ from the corresponding quantities for the standard percolation.

Pauli percolation is interesting in its own right
as a rare example of a discontinuous (first-order) percolation transition with qualitatively more local interactions than in all currently considered examples, 
see Ref.~\refcite{fb-el5} and references therein.
It has a simple representation as a particular classical two-color, or contagion, percolation problem.\cite{fb-el5}
Indeed Pauli percolation corresponds to the following variation of the standard one: 
(i) sites can come in two colors,
say,
green (uninfected) or red (infected),
(ii) each site of a lattice is occupied and colored either green or red  or left empty,
(iii) only configurations where every cluster contains no more than one red site are taken into account.
Since merging two clusters of size $m$ and $n$ reduces their overall weight 
from $(m + 1)(n + 1)$ to $(m + n + 1)$,  
it closely resembles the ``product rule'' in the famous explosive percolation,\cite{explosive_percolation} 
but, unlike the latter, it does not require the elevated degree of 
non-locality in the configuration-update process as the new site in the system gets occupied, see Ref.~\refcite{fb-el5}.

Although the initial studies of Pauli percolation were performed on one and two-dimensional lattices,
see below,
in the recent study concerning infinite dimensions\cite{fb-el5}
the analytical and numerical calculations refer to a regular random graph of $N$ sites.\cite{dorogovtsev}
Such graphs have a vanishing density of short cycles and mostly contain loops of size $\ln N$;
hence they are locally tree-like.
A rigorous analytical solution of Pauli percolation reported in Ref.~\refcite{fb-el5}
gives an explicit example of a rather rare discontinuous percolation transition in a random graph.
A sketch of this solution is as follows. 
First the Pauli percolation is considered on a Cayley tree\footnote{The distinction between a Bethe lattice and a Cayley tree 
is that the latter contains a root site with a coordination number $z-1$ while the bulk of a Bethe lattice is fully homogeneous\cite{fb-el5} (see also Ref.~\refcite{ostilli}). 
The standard percolation transition on the Bethe lattice is the continuous one with 
the percolation threshold $p_c=\frac{1}{z-1}$,
i.e., $p_c=\frac{1}{2}$ for $z=3$.\cite{essam,percolation2}}
(setting for concreteness the coordination number $z=3$) 
by solving the corresponding recursive relations in the two-color representation described above.
Examining the probability $P_\infty$ that the root site of an infinitely large tree is connected 
to its boundary one finds that it is zero (i.e., there is no percolation) until
$p<\frac{4}{5}$.\footnote{In the Ref.~\refcite{fb-el5} 
all the calculations are performed using $\tilde{p}$ as an a priori probability of site to be occupied and colored green or red;
it is connected to $p$ in Eq.~(\ref{061}) by the relations: 
$\tilde{p}=\frac{p}{p+1}$ and $p=\frac{\tilde{p}}{1-\tilde{p}}$.}
For $p\ge\frac{4}{5}$ the nonzero (i.e., percolating) solutions emerge.
Next, individual Cayley trees could be merged into the uniform Bethe lattice via the central site or bond addition. 
The value of $p$ at which the percolation solution in the Bethe lattice ${\cal{P}}_\infty\ne 0$
first 
emerges remains unchanged, $p=\frac{4}{5}$,
although the new percolation probability itself changes, ${\cal{P}}_{\infty}\ne P_{\infty}$.
The rigorous analysis in Ref.~\refcite{fb-el5} illustrates the discontinuous nature of the percolation transition. 
The value of $p$ at which it takes place is determined from the behavior 
of the central-site free energy.\cite{free-energy,cavity}
It has been found that ${\cal{P}}_{\infty}(p)$ at $p_c=0.823506\ldots$ has a discontinuous jump between zero and nonzero values. 

The study of Pauli percolation on a 3-regular random graph leads to an interesting conclusion for understanding ferromagnetism.
Namely, it implies that for the standard repulsive Hubbard model on the Tasaki-decorated 3-regular random graph 
the ferromagnetic ground state emerges as the electron density $\frac{n}{N}$ exceeds $0.208176\ldots$ 
(see the red line in Fig.~3 of Ref.~\refcite{fb-el5})
and its magnetization increases 
until it reaches a saturation value at $\frac{n}{N}=\frac{1}{3}$. 
Interestingly, since the Bethe lattice is an 
example of the expander graph it intrinsically possesses a gap in the excitation
spectra,\cite{thorpe_phonon_gap,laumann_goldstone_bethe} which in turn can protect the ferromagnetism in such structures.

The rest of this section on Pauli percolation contains 
a brief discussion of
the exact solution on a simple chain\cite{fb-el4} 
and
Monte Carlo simulations on a square lattice.\cite{fb-el4}

\subsection{Pauli percolation on a simple chain}
\label{sec091}

Alternatively to the approaches used in Sec.~\ref{sec082} for the
investigation of the
one-dimensional sawtooth-chain or kagome-chain Hubbard model we present here
the percolation approach to consider these models, see Refs.~\refcite{le3,fb-el4}.  
The treatment of the strongly correlated electron problem in one dimension
as a Pauli percolation problem is useful (i) as a preparation for the treatment of the two-dimensional problem, see Sec.~\ref{sec092}, 
(ii) to provide one-dimensional data to compare with two-dimensional
ones, and (iii) as comparison with standard percolation.
Moreover, it is valuable in its own right, since an exact solution of this unconventional percolation problem
is possible.

We analyze one-dimensional Pauli percolation using a transfer-matrix
method.\cite{fb-el4}
To illustrate the percolation approach and to provide
results for comparison we first consider the standard percolation in one dimension which is exactly
solvable.\cite{percolation2}
We work in the grand-canonical ensemble and seek for the grand-canonical partition function $Z=Z(z,{\cal{N}})$ of a percolating system
as a sum of probabilities of all possible random realizations.
$Z(z,{\cal{N}})$ can be written in terms of a $2\times 2$ transfer matrix ${\bf{T}}$
as 
\begin{eqnarray}
\label{trans-mat}
Z(z,{\cal{N}})={\rm{Tr}}{\bf{T}}^{\cal{N}},
\;\;\;
{\bf{T}}
=
\left(
\begin{array}{cc}
1 & 1 \\
z & z
\end{array}
\right).
\end{eqnarray}
The matrix element $T(n_i,n_{i+1})$ corresponds to the pair of neighboring sites $i$ and $i+1$ on a simple (periodic) chain 
and acquires the value 1 if the site $i$ is empty or $z$ if the site $i$ is occupied
independently of the occupation number of site $i+1$.
The $2^{\cal{N}}$ terms of ${\rm{Tr}}{\bf{T}}^{\cal{N}}$ correspond to all possible
realizations of the percolating system.
To determine the fugacity $z$ let us calculate the average occupation number of the site $i$:
\begin{eqnarray} 
\label{eq_ni}
\langle n_i\rangle=\frac{{\rm{Tr}}({\bf{T}}^{\cal{N}}{\bf{N}})}{Z},
\;\;\;
{\bf{N}}
=
\left(
\begin{array}{cc}
0 & 0 \\
0 & 1
\end{array}
\right).
\end{eqnarray}
The matrix elements of the matrix ${\bf{N}}{\bf{T}}$ acquire the value 0 if the site $i$ is empty or $z$ if the site $i$ is occupied
this way counting only those realizations (among 
$2^{\cal{N}}$ ones) of the percolating system
that have an occupied site $i$.
Diagonalizing the transfer matrix ${\bf{T}}$ 
one easily finds in the thermodynamic limit ${\cal{N}}\to\infty$ that $\langle n_i\rangle= \frac{z}{1+z}$.
On the other hand, the average site-occupation number should be equal to $p$, $\langle n_i\rangle= p$.
This yields the relation between $z$ and $p$:
$\frac{z}{1+z}=p$ or $z=\frac{p}{1-p}$.
 
We consider now the calculation of the average number of clusters of size $l$ 
(normalized by the lattice size ${\cal{N}}$) 
to be denoted by $n(l)$.
Simple arguments immediately yield:\cite{percolation2}
$n(l)=(1-p)^2p^l$.
To fix the cluster of length $l$, within the transfer-matrix method
 we start with an empty site,
then we have a string (cluster) of $l$ occupied sites,
and the last site of the string is followed by an empty one.
This has to be written in terms of matrices.
Let us introduce 
the matrix ${\bf{S}}$ with only one nonzero element 1 if the site $i$ is empty and the site $i+1$ is occupied,
the matrix ${\bf{C}}$ with only one nonzero element $z$ if both sites $i$ and $i+1$ are occupied,
and
the matrix ${\bf{F}}$ with only one nonzero element $z$ if the site $i$ is occupied and the site $i+1$ is empty.
To calculate $n(l)$ we have to replace  in
Eq.~(\ref{trans-mat}) the product of a sequence of $l+1$ ${\bf{T}}$-matrices by the product ${\bf{S}}{\bf{C}}^{l-1}{\bf{F}}$,
i.e.,
\begin{eqnarray}
\label{064}
n(l)=\frac{{\rm{Tr}}({\bf{T}}^{{\cal{N}}-l-1}{\bf{S}}{\bf{C}}^{l-1}{\bf{F}})}{Z},
\;\;\;
{\bf{S}}
=
\left(
\begin{array}{cc}
0 & 1 \\
0 & 0
\end{array}
\right),
\;\;\;
{\bf{C}}
=
\left(
\begin{array}{cc}
0 & 0 \\
0 & z
\end{array}
\right),
\;\;\;
{\bf{F}}
=
\left(
\begin{array}{cc}
0 & 0 \\
z & 0
\end{array}
\right).
\end{eqnarray}
Straightforward calculations in Eq.~(\ref{064}) in the thermodynamic limit ${\cal{N}}\to\infty$ reproduce the cited above formula $n(l)=(1-p)^2p^l$.
Similarly we get the pair (site-occupation) correlation function\cite{weinrib}
\begin{eqnarray} \label{065}
g(l)=\langle n_in_{i+l}\rangle-\langle n_i\rangle\langle n_{i+l}\rangle
=\frac{{\rm{Tr}}({\bf{T}}^{{\cal{N}}-l}{\bf{N}}{\bf{T}}^{l}{\bf{N}})}{Z}-p^2
=p(1-p)\delta_{l,0},
\end{eqnarray}
and the pair connectivity (the probability that two sites $n$ and $n+l$ are both occupied and belong to the same cluster)
\begin{eqnarray}
\Gamma(n,n+l)=\frac{{\rm{Tr}}({\bf{T}}^{{\cal{N}}-l}{\bf{N}}{\bf{C}}^{l})}{Z}
=p\cdot p^l,
\end{eqnarray}
which were obtained by more elementary means in textbooks on percolation.\cite{percolation2}

Advantages of the transfer-matrix approach become undoubted while applying it to the Pauli percolation.\cite{fb-el4} 
We have simply to introduce the transfer-matrix for the Pauli percolation in one dimension
\begin{eqnarray}
\label{067}
{\bf{T}}
=
\left(
\begin{array}{ccc}
T(0,0)          & T(0,\uparrow)          & T(0,\downarrow)         \\
T(\uparrow,0)   & T(\uparrow,\uparrow)   & T(\uparrow,\downarrow)  \\
T(\downarrow,0) & T(\downarrow,\uparrow) & T(\downarrow,\downarrow)
\end{array}
\right)
=
\left(
\begin{array}{ccc}
1 & 1 & 1\\
z & z & z\\
z & 0 & z
\end{array}
\right),
\end{eqnarray}
as well as the other matrices emerging in the transfer-matrix calculations
\begin{eqnarray}
{\bf{N}}
=
\left(
\begin{array}{ccc}
0 & 0 & 0\\
0 & 1 & 0\\
0 & 0 & 1
\end{array}
\right),
\;\;\;
{\bf{S}}
=
\left(
\begin{array}{ccc}
0 & 1 & 1\\
0 & 0 & 0\\
0 & 0 & 0
\end{array}
\right),
\;\;\;
{\bf{C}}
=
\left(
\begin{array}{ccc}
0 & 0 & 0\\
0 & z & z\\
0 & 0 & z
\end{array}
\right),
\;\;\;
{\bf{F}}
=
\left(
\begin{array}{ccc}
0 & 0 & 0\\
z & 0 & 0\\
z & 0 & 0
\end{array}
\right).
\end{eqnarray}
These are now $3\times 3$ matrices, cf. Sec.~\ref{sec082}, because of
the two spin values of an electron:
The site can be empty, occupied by the electron with spin $\uparrow$, or occupied by the electron with spin $\downarrow$.
The zero in the third row of the transfer matrix $\bf T$ in Eq.~(\ref{067})
appears, 
since the configuration $\downarrow_i\uparrow_{i+1}$ is forbidden.
After straightforward although lengthy calculations one gets in the thermodynamic limit
\begin{eqnarray}
\label{069}
p(z)=1+4z-\frac{\sqrt{1+4z}}{1+4z},
\;\;\;
z(p)=\frac{p(2-p)}{4(1-p)^2},
\nonumber\\
n(l)=\frac{4(1-p)^3}{(2-p)^2}(l+1)\alpha^l,
\;\;\;
\alpha=\frac{p}{2-p},
\nonumber\\
g(l)=-(1-p)^2\alpha^{2\vert l\vert},
\nonumber\\
\Gamma(n,n+l)=p\cdot \left(1+\frac{1-p}{2-p}l\right)\alpha^l.
\end{eqnarray}

\begin{figure}[bt]
\centerline{\psfig{file=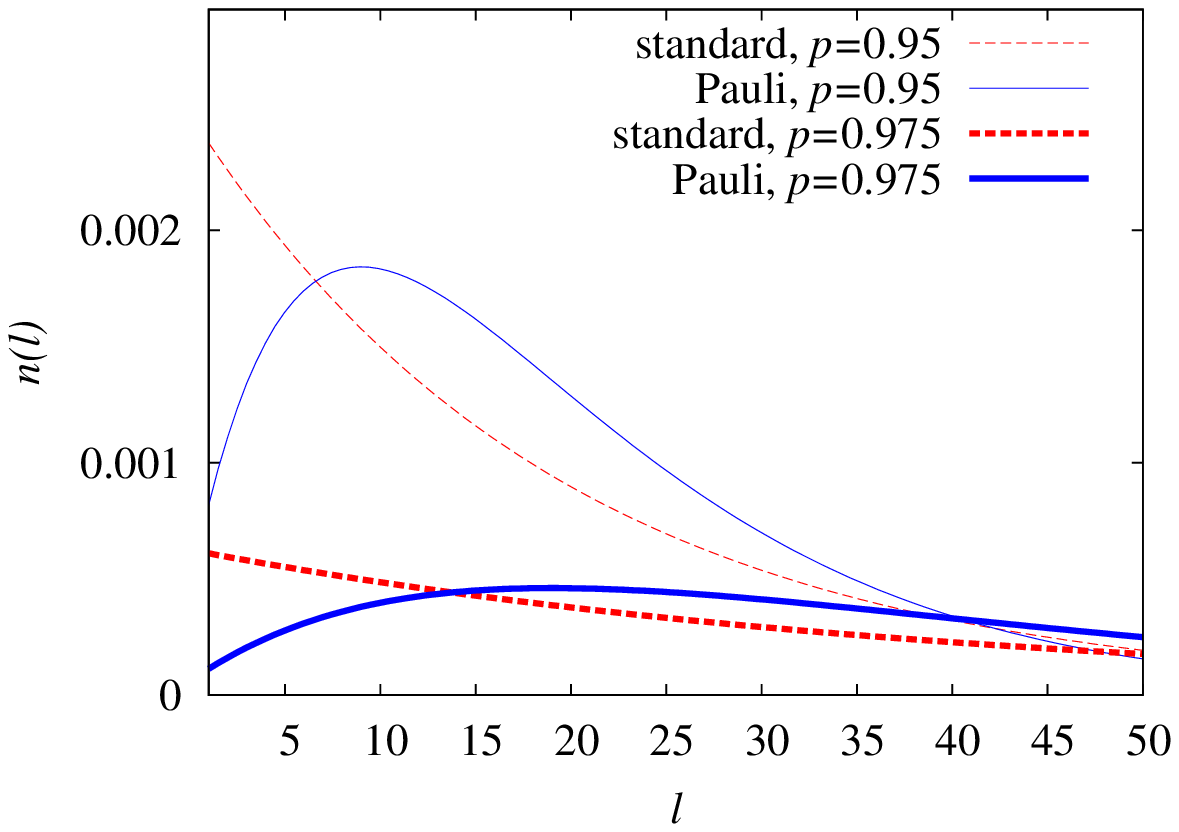,width=2.25in}\hspace{10mm}\psfig{file=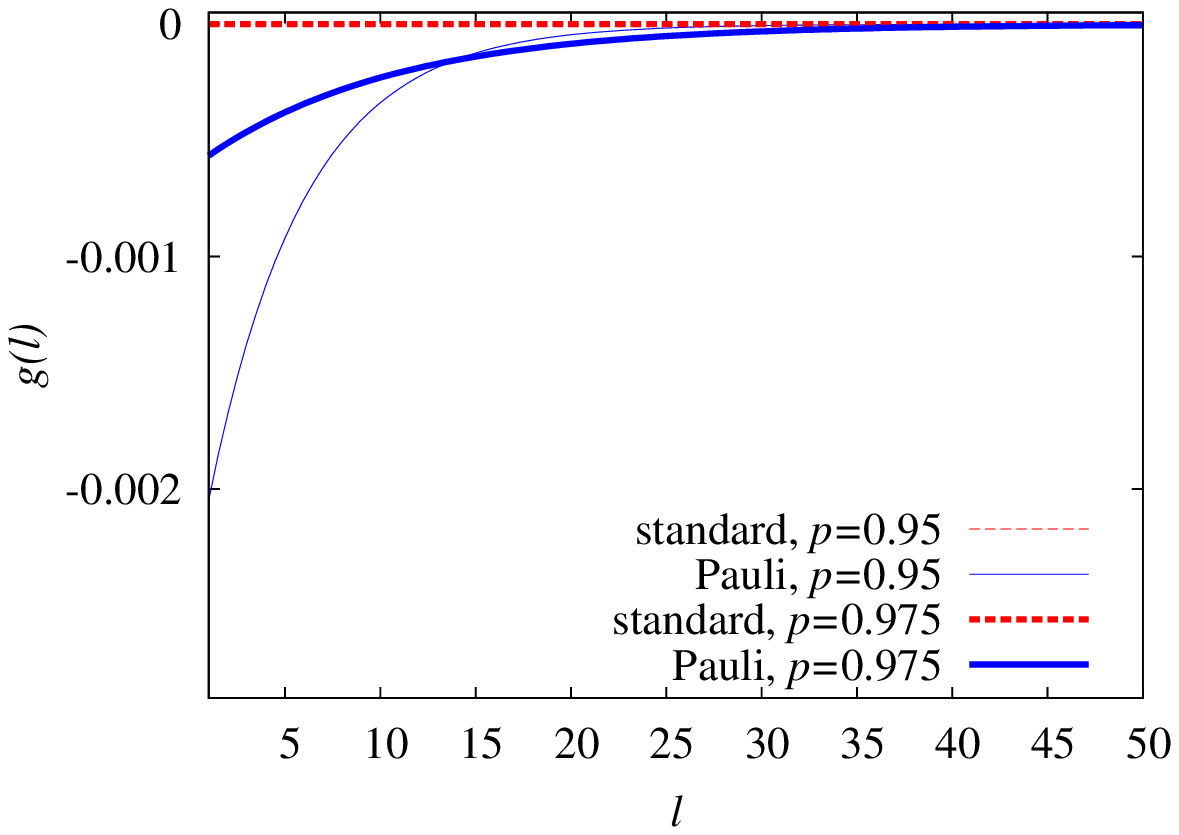,width=2.25in}}
\vspace*{8pt}
\caption{(Color online)
The average number of clusters  $n(l)$ 
(left) 
and the pair correlation function  $g(l)$ 
(right) 
for $p=0.95$ and $p=0.975$ 
for standard (dotted red lines) and Pauli (solid blue lines) percolation in one dimension, 
see Eqs.~(\ref{064}), (\ref{065}) and Eq.~(\ref{069}).}
\label{fig15_16}
\end{figure}

The Pauli percolation shows a number of striking differences from the standard
percolation. 
The most obvious  differences can be seen in the average number of
clusters of size $l$, $n(l)$,  and in the  pair correlation
function $g(l)$, see Fig.~\ref{fig15_16}. 
Although the percolation threshold remains unchanged, $p_c=1$,
the quantity $n(l)$ for Pauli percolation has a maximum $l^*>1$ for $p>0.8$ moving along $l^*\approx-(1+\frac{1}{\ln\alpha})$ for $p\to
1$.
For the standard percolation $n(l)$ decays monotonically with $l$.
The pair correlation function $g(l)$ decays with a correlation length $\xi=-\frac{1}{2\ln\alpha}$
that diverges as $(p_c-p)^{-1}$ when $p\to p_c=1$.
Since the pair correlation function $g(l)$ is negative, the interaction is repulsive, i.e., particle (electron) 
positions anticorrelate.
(For the standard percolation there are no nontrivial pair correlations.)

From the point of view of magnetism, Eqs.~(\ref{060}) and (\ref{069}) allow to calculate the macroscopic magnetic moment 
(in the thermodynamic limit)
\begin{eqnarray}
\langle {\bf{S}}^2\rangle
=\frac{3p(2-p)}{8(1-p)}{\cal{N}},
\;\;\;
p<1.
\label{1D_S2_ns_n_exact}
\end{eqnarray}
This formula provides an extension to ${\cal{N}}\to\infty$ of the finite-size results reported in Fig.~\ref{fig11}.
Clearly Eq.~(\ref{1D_S2_ns_n_exact}) implies that $\frac{\langle {\bf{S}}^2\rangle}{{\cal{N}}^2}\to 0$ if $0\le p<1$, 
as already discussed in Sec.~\ref{sec082}. Only at $p=1$ we have
$\lim_{{\cal{N}}\to\infty} \frac{\langle {\bf{S}}^2\rangle}{{\cal{N}}^2} >0.$

\subsection{Flat-band ferromagnetism and Pauli percolation in two dimensions}
\label{sec092}

To study flat-band ferromagnetism of Hubbard electrons in two dimensions we consider the Tasaki lattice (Fig.~\ref{fig02}) 
which is the two-dimensional counterpart of the sawtooth chain. 
The trapping cells, where the electrons can be localized, contain five sites, 
namely one site of the underlying square lattice [label $({\bf{m}},1)$ in Fig.~\ref{fig02}] 
and the four sites connected to site $({\bf{m}},1)$ by the hopping integral $t_2$ (red zigzag paths in Fig.~\ref{fig02}).   
Since every cell of the initial Tasaki lattice of $N$ sites contains precisely one site of the underlying square lattice,  
the corresponding lattice of the related Pauli percolation problem 
is just the underlying square lattice of ${\cal N}=\frac{N}{3}$ sites.

The two-dimensional percolation problem is more complicated and it can be studied numerically, only.\cite{percolation2}
However, having in mind that the initial model is a strongly correlated quantum many-body model 
the treatment of the relevant degrees of freedom by a classical percolation description 
provides a significant simplification of the initial problem.
In contrast to the standard percolation,
simple random sampling of geometrical configurations is not sufficient for averaging in the Pauli-percolation case,
since various geometrical configurations $q$ have different weights $W(q)$.
Going beyond standard numerical schemes,\cite{hoshen,newman}
to take into account efficiently various geometrical configurations $q$ according to their weights $W(q)$,
one has to implement importance sampling choosing samples according to the distribution of $W(q)$.
In the canonical ensemble for a fixed number of electrons,
a new configuration $q_2$ is generated from the given one $q_1$ 
by a random permutation of two sites in order to ensure the fixed number of electrons.
In the grand-canonical ensemble,
a site is simply chosen and if it is empty (occupied), a particle is proposed
to be inserted (removed).
The new configuration is accepted with the Metropolis probability $\min[1,\frac{W(q_2)}{W(q_1)}]$.
In addition, for the grand-canonical simulations exchange Monte Carlo steps have been employed.\cite{hukushima}
As a result, the numerical investigation of the Pauli percolation becomes more challenging,
since it suffers from critical slowing down
(that does not occur in standard-percolation simulations).
Labeling of clusters is done in two different ways:
(i) using a modified Newman-Ziff algorithm\cite{newman} which locally updates cluster labeling for fixed number of occupied sites
and
(ii) using the Hoshen-Kopelman algorithm\cite{hoshen} which makes a global update.
Central results of numerical simulations concern 
$\langle {\bf S}^2\rangle$ and 
the average number $n(l)$ of clusters of size $l$ normalized by the size of
the lattice ${\cal N}$.
For the numerical simulations  lattice sizes ${\cal N}={\cal L} \times  {\cal L}$ up to $
{\cal L}=270$ were used, where  periodic or open boundary conditions were imposed.

\begin{figure}[bt]
\centerline{\psfig{file=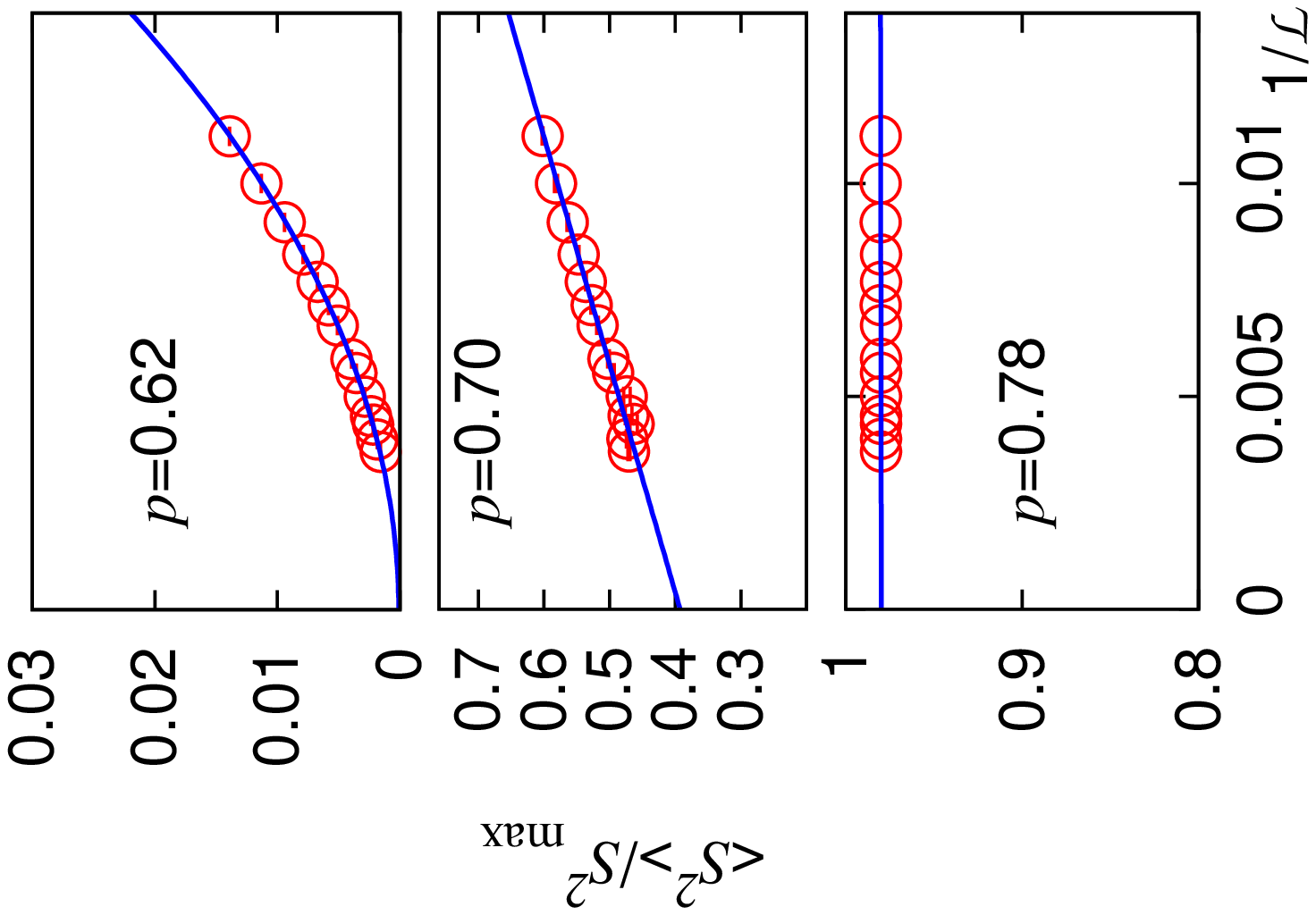,width=2.25in,angle=270}\hspace{10mm}\psfig{file=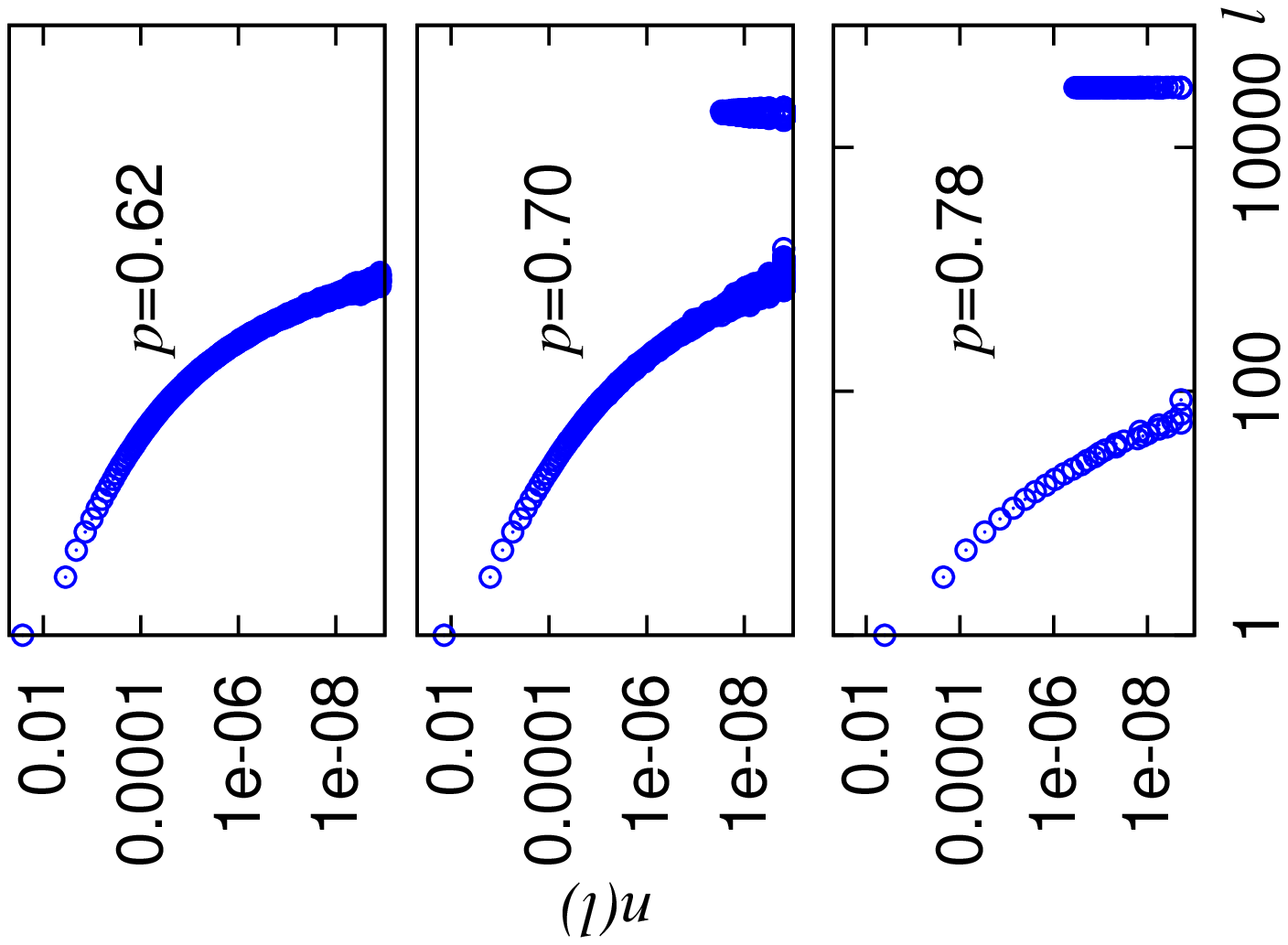,width=2.25in,angle=270}}
\vspace*{8pt}
\caption{(Color online)
The Hubbard model on the two-dimensional Tasaki lattice. 
Left: Finite-size scaling of the square of the magnetic moment 
$\frac{\langle {\bf S}^2\rangle}{{\bf S}^2_{\max}}$.
Right: Normalized number of clusters $n(l)$ for ${\cal L} = 200$.}
\label{fig12a}
\end{figure}

\begin{figure}[bt]
\centerline{\psfig{file=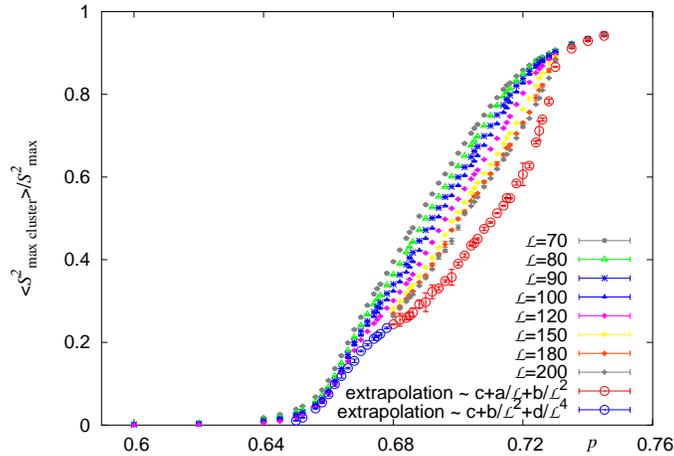,width=2.25in,angle=270}}
\vspace*{8pt}
\caption{(Color online)
The Hubbard model on the two-dimensional Tasaki lattice. 
Square of the magnetic moment of the largest cluster, 
$M^2=\frac{\langle {\bf S}^2_{\rm{maxcluster}}\rangle}{{\bf S}^2_{\max}}$,
versus electron density $p = \frac{n}{{\cal N}}$.}
\label{fig12b}
\end{figure}

The average square of the magnetic moment $\langle {\bf S}^2\rangle$ is the quantity which directly signalizes
the appearance of the ground-state ferromagnetism.
In canonical simulations,
one can obtain the square of the magnetic moment of the largest cluster $M^2=\frac{\langle {\bf S}^2_{\rm{maxcluster}}\rangle}{{\bf S}^2_{\max}}$, 
where ${\bf S}^2_{\max}=\frac{n}{2}(\frac{n}{2}+1)$,
or the square of the magnetic moment $\frac{\langle {\bf S}^2\rangle}{{\bf S}^2_{\max}}$
as a function of electron density $p=\frac{n}{{\cal{N}}}=3\frac{n}{N}$.
Furthermore,
one can perform a finite-size scaling of the square of the magnetic moment
to get $\lim_{{\cal N} \to \infty} M^2$,
see Fig.~\ref{fig12a} (left panel).
We find that the scaling behavior changes from
$M^2 = a + b{\cal L}^{-2} + \cdots$ for low electron densities
$p$ to $M^2 = a + b{\cal L}^{-1} + \cdots$ at higher $p$. 
While for $p=0.62$ the macroscopic moment scales to zero with system size,
for $p=0.70$ it approaches a nonzero value of  about $0.40$.  Further increasing of
the electron density, i.e., to $p=0.78$,
yields an almost size-independent magnetic moment of about $0.98$.
Additionally, the cluster-size distribution $n(l)$ indicates
the emergence of a large component for $p>0.62$, see right panel of
Fig.~\ref{fig12a}.
The average square of the magnetic moment $M^2$
is presented in Fig.~\ref{fig12b} 
for finite systems along with the extrapolated values.


\begin{figure}[bt]
\centerline{\psfig{file=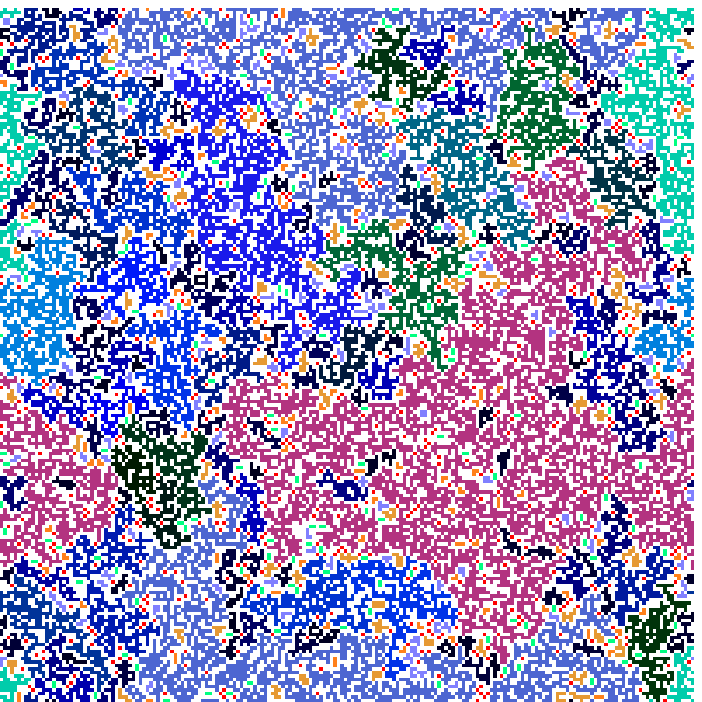,width=1.75in}\hspace{5mm}\psfig{file=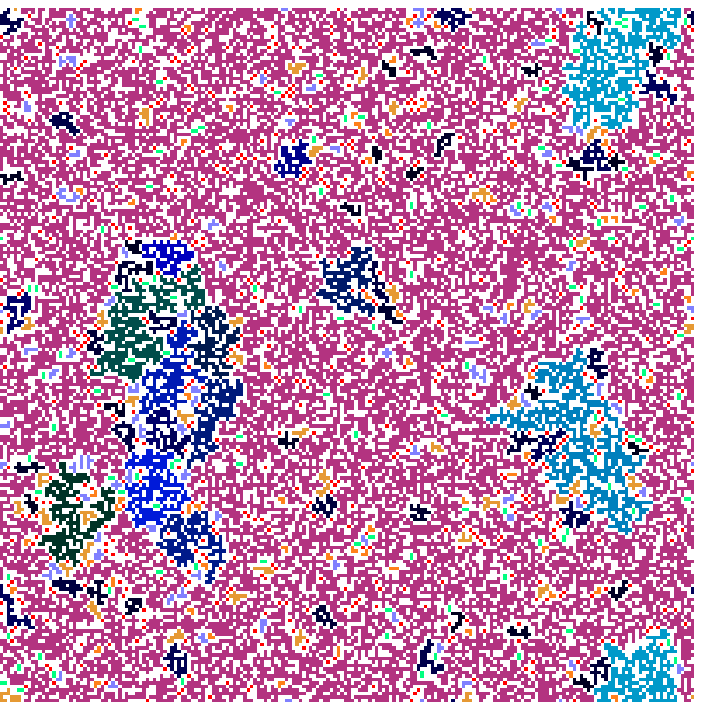,width=1.75in}}
\vspace{10mm}
\centerline{\psfig{file=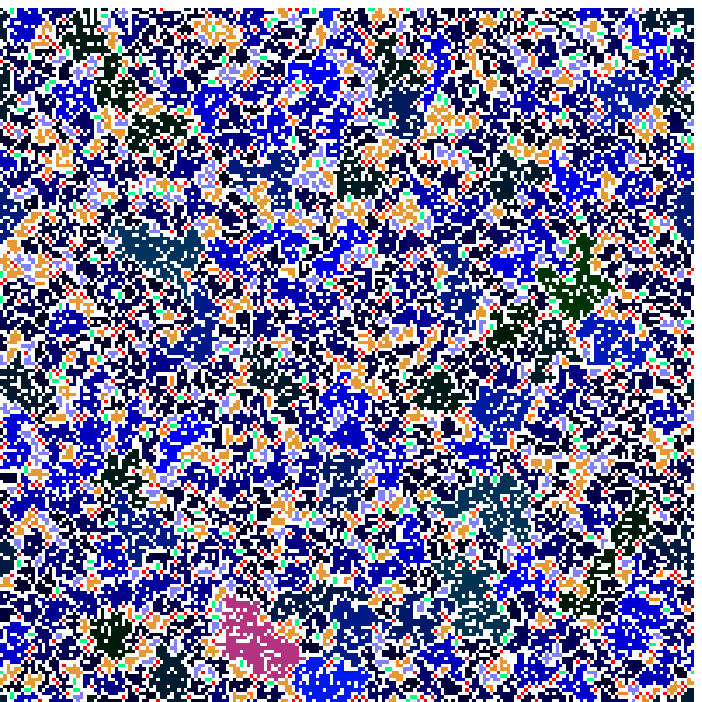,width=1.75in}\hspace{5mm}\psfig{file=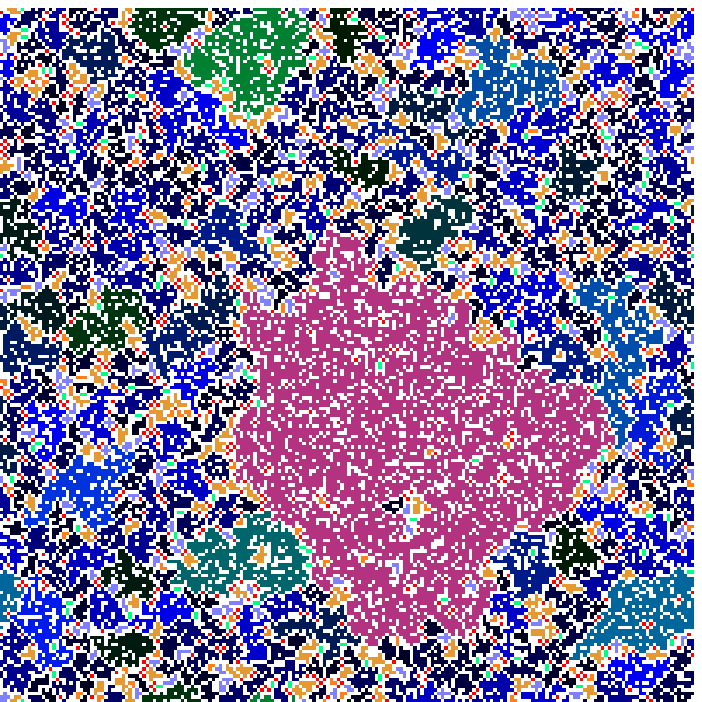,width=1.75in}}
\vspace{5mm}
\centerline{\psfig{file=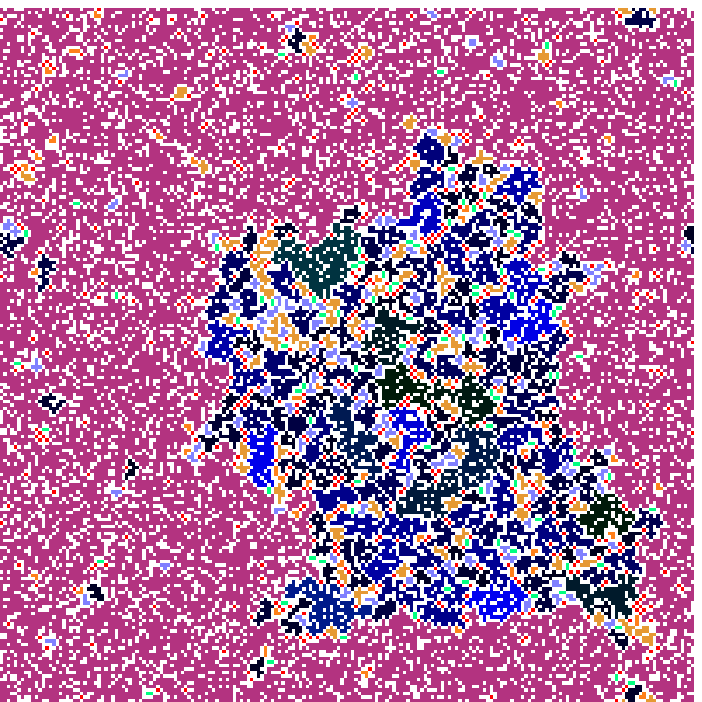,width=1.75in}\hspace{5mm}\psfig{file=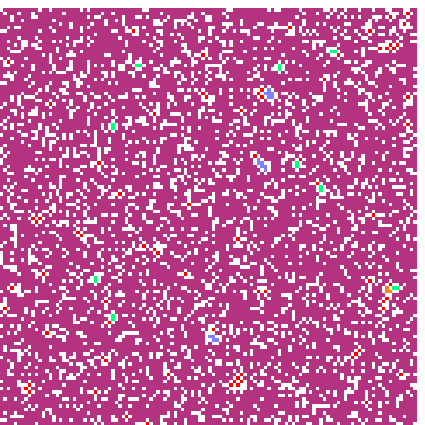,width=1.75in}}
\vspace*{8pt}
\caption{(Color online)
Snapshots of configurations for standard (upper row) and Pauli-correlated percolation 
(middle and bottom rows) on a square lattice of size ${\cal L} \times {\cal L}$, ${\cal L} =200$, 
for small deviations from the critical concentration $p_c$.
The pink color corresponds to the largest cluster for the given concentration. 
The concentrations (from left to right) are 
$p_1=0.574$ and $p_2=0.6$ for standard percolation, where $p_c=0.592746\ldots\,$, 
and 
$p_3=0.62$ (paramagnetic), $p_4=0.65$ (phase-separated), $p_5=0.7$ (phase-separated), and $p_6=0.78$ (ferromagnetic) for Pauli percolation.} 
\label{fig12}
\end{figure}

Altogether, from these findings one may conclude, that
the critical density for the Pauli percolation exceeds that of the standard counterpart $p_c=0.592746\ldots\,$, 
see, e.g., Ref.~\refcite{percolation1}.
For higher densities, there appears a regime of unsaturated ferromagnetism,
which is characterized by $0<\frac{\langle {\bf S}^2\rangle}{{\bf S}^2_{\max}}<1$.
Canonical simulations of the Pauli percolation exhibit some features arising in
a phase-coexistence regime.
The high-density phase appears to form first as a compact nonpercolating object with a macroscopic magnetic moment.
For higher densities, including $p=0.7$,
the ferromagnetic phase spans across the system.
Therefore, the phase-separated regime is drastically different from standard percolation.
Grand-canonical simulations exhibit a jump at a certain chemical potential between densities 
$(\frac{n}{{\cal{N}}})_-$ around 0.63(1) and $(\frac{n}{{\cal{N}}})_+\approx 0.75(2)$.
All findings are consistent with the visual analysis of typical snapshots of configurations for the standard and Pauli percolation,
see Fig.~\ref{fig12}.
For electron densities below the threshold the system tends to form many small clusters which increase the weight.
Slightly above the threshold the larger cluster is still suppressed by the existence of 
many small clusters which help to increase the weight of a configuration.

In summary,
the simulations of the Pauli percolation in two dimensions show the following.
Effective repulsive interaction leads to a breaking up of the clusters,
and thus to a first-order grand-canonical transition in two dimensions.
The critical density is higher than that of standard (site) percolation.
For the underlying two-dimensional Tasaki-Hubbard model these findings imply ground-state ferromagnetism
in a range of electron density $\frac{n}{N}$ from 0.21(1) to $\frac{1}{3}$.

\section{Dispersion-Driven Ferromagnetism in a Flat-Band Hubbard System}
\label{sec10}

Introducing dispersion (violation of the flat-band geometry) 
typically modifies the balance of interaction and kinetic energy and tends to destabilize
ferromagnetism in the Hubbard model.
However, it was demonstrated by several studies 
that the flat-band ferromagnetic ground states  are robust 
if the flat band becomes (slightly) dispersive and the Hubbard repulsion $U$ 
is larger than a threshold $U_c>0$, 
where $U_c$ depends on the degree of the deviation from the flat-band
geometry.\cite{flband-ferro-stability}
By contrast, recently it has been found in Ref.~\refcite{dispersion-driven}
that unexpectedly for the class of flat-band Hubbard systems without
ground-state ferromagnetism, see Sec.~\ref{sec081},
just the dispersion, i.e., the kinetic energy,
will open the route to ferromagnetic ground states.
This dispersion-driven ferromagnetism resembles the famous {\it
order-from-disorder} mechanism,\cite{villain,shender}
i.e., due to distortions
an ordered ground state is selected from the degenerate flat-band
ground-state manifold, where paramagnetic states
dominate.

To this end we investigate the frustrated diamond chain, where for 
ideal geometry the lowest one-electron band is flat.
However, the ground state even for full occupation of all trapping cells 
is nonmagnetic for arbitrary on-site repulsion $U$,
since the trapping cells for the frustrated diamond chain are isolated from each other 
and the resulting huge set of paramagnetic states prevails against the ferromagnetic eigenstate, 
cf. Sec.~\ref{sec081}.
Below we will illustrate how a dispersion of the former flat band leads to
ground-state ferromagnetism.

We consider the Hubbard model (\ref{044}) on a distorted frustrated diamond-chain lattice, see Fig.~\ref{fig01},
i.e., we assume $t_1\ne t_3$ but fix their average $t_1+t_3=2t$.
The electron density is fixed to $n = {\cal N}$, 
i.e., for ideal geometry all trapping cells are occupied and the flat band is half filled. 
For the sake of comparison we first give the bandwidth $w_d$ of the dispersive band above the flat band for  $t_1=t_3$: 
$w_d=\sqrt{\frac{t_2^2}{4}+8t^2}-\frac{t_2}{2}\approx 2\frac{(t_3+t_1)^2}{t_2}$ ($\frac{t}{t_2}\ll 1$).
The bandwidth $W_f$ of the former flat band due to distortion  $t_1\ne t_3$ is 
$W_f \approx 2\frac{(t_3-t_1)^2}{t_2}$.
We introduce the dimensionless parameter
$\Omega=\left\vert\frac{t_3-t_1}{t_3+t_1}\right\vert$.
Then $\Omega^2\approx\frac{W_f}{w_d}$ is a appropriate measure of the strength of deviation from the ideal flat-band geometry.

\begin{figure}[bt]
\centerline{\psfig{file=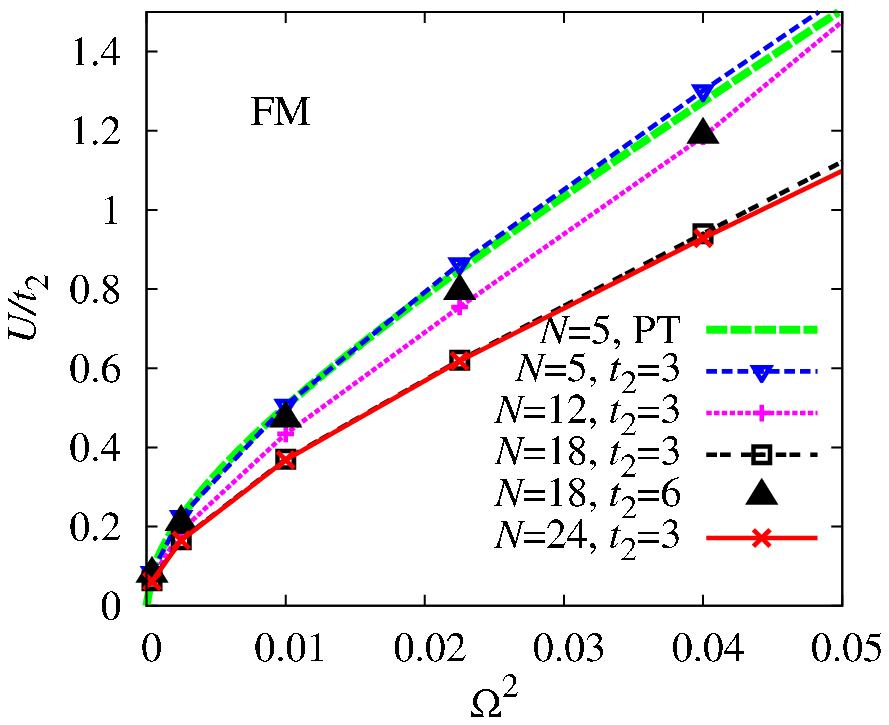,width=2.25in}\hspace{10mm}\psfig{file=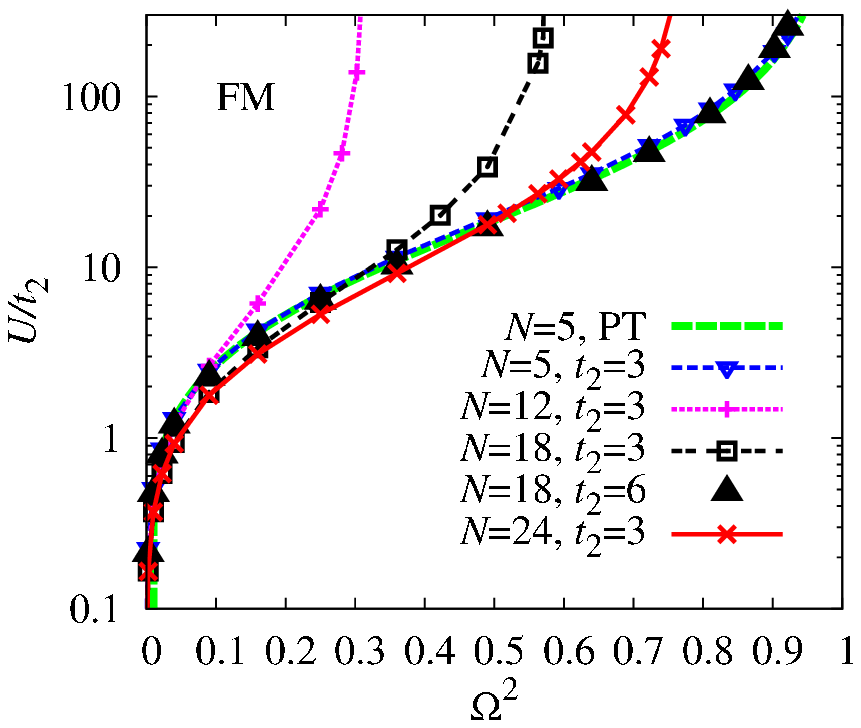,width=2.25in}}
\vspace*{8pt}
\caption{(Color online)
Phase diagram of the distorted Hubbard diamond chain with electron density $\frac{n}{N}=\frac{1}{3}$ (half-filled lowest band). 
Ferromagnetism (denoted by ``FM'') appears for on-site repulsions $U$ above a critical value $U_c$.
$U_c$ is shown as a function of the dimensionless bandwidth parameter
$\Omega^2$.
The various critical lines $U_c(\Omega^2)$ are obtained 
by fourth-order perturbation theory (symbol ``PT''), see Eq.~(\ref{077}), 
and 
by exact diagonalization for finite systems of 
$N= 5$ (open boundary conditions) 
and 
$N= 12,18,24$ (periodic boundary conditions) 
for hopping parameter sets with $t_3+t_1= 2$ and $t_2= 3$ and $t_2=6$.}
\label{fig12ttt}
\end{figure}

To determine the ground-state properties,  in
Ref.~\refcite{dispersion-driven}  exact diagonalization and fourth-order perturbation
theory are used. 
The main result, namely the ground-state phase diagram,  
is illustrated in Fig.~\ref{fig12ttt}.
For not too large deviations from ideal geometry controlled by $\Omega^2<1$ and for sufficiently large
on-site repulsion  $U>U_c$,
the ground state of the Hubbard diamond chain with half-filled lowest band is ferromagnetic
(the region denoted as ``FM'').

The Lanczos exact diagonalization for distorted chains were performed up to $N=30$ 
to determine the critical value $U_c$ above which the ground state is the ferromagnetic SU(2)-multiplet 
with $\langle {\bf S}^2 \rangle = \frac{n}{2}(\frac{n}{2}+1)$. 
Note that for the largest system $N=30$,
$U=\infty$ was assumed to find the maximal dispersion
$\Omega_c^2\approx 0.884$ above which ground-state ferromagnetism does not exist.
There is an obvious finite-size dependence of $U_c(\Omega^2)$,
but the general shape of the curve is the same for all $N$, cf.
Fig.~\ref{fig12ttt}.
Interestingly the region of ferromagnetism increases  with growing system size $N$.
That provides evidence that the dispersion-driven ferromagnetism does not  disappear for $N \to
\infty$.
The influence of the vertical hopping integral $t_2$ on the phase boundary
$U_c(\Omega^2)$
is visible from the corresponding curves for $N=18$. 
For small bandwidth $\Omega^2$ the region of ferromagnetism slightly shrinks with
increasing of $t_2$,
whereas $U_c$ decreases with increasing $t_2$ for $\Omega^2  \gtrsim 0.25$.

Complementary to  the   exact-diagonalization study the fourth-order
perturbation theory
was used.
All essential features of the perturbation approach are already 
provided by a 5-site cluster that contains two trapping cells
(the sites $m,1$, $m,2$, $m,3$, $m+1,1$, and $m+1,2$ connected by six bonds, see Fig.~\ref{fig01})
in the subspace of $n=2$ electrons.
The unperturbed Hamiltonian contains the $t_2$- and $U$-terms, the $t_1$-
and the  $t_3$-terms constitute the perturbation.
The  ground-state energy of the unperturbed Hamiltonian is $-2t_2$, it is fourfold
degenerate (one singlet, and threefold degenerate triplet).
The perturbation theory up to fourth order gives for the triplet state
\begin{eqnarray} \label{e_t}
E_t=-2t_2-\frac{(t_3-t_1)^2}{t_2}
-\frac{(t_3+t_1)^2(t_3-t_1)^2}{2t_2^3}+\frac{(t_3-t_1)^4}{t_2^3}
+\ldots,
\end{eqnarray}
i.e., $E_t$ is independent of $U$.
In contrast, 
the energy of the singlet state, $E_s$, depends on $U$.
The fourth-order result can be found in Ref.~\refcite{dispersion-driven}.
In the small-$U$ limit the dominating term in $E_s(U)$ is 
\begin{eqnarray} \label{e_s_1}
E_s(U \to 0)=
-\frac{2(t_3-t_1)^4}{Ut_2^2},
\end{eqnarray}
whereas in the limit $U\to\infty$ it is given by
\begin{eqnarray}  \label{e_s_2}
E_s(U \to \infty)=-2t_2-\frac{(t_3-t_1)^2}{t_2}
-\frac{(t_3+t_1)^2(t_3-t_1)^2}{4t_2^3}+\frac{3(t_3-t_1)^4}{4t_2^3}
+\ldots.
\end{eqnarray}
Obviously, we have $E_s<E_t$ for small $U$, but 
$E_s>E_t$ for large $U$, i.e.,
there is a critical value $U_c$ above which ferromagnetism emerges.
Moreover, Eqs.~(\ref{e_t}) -- (\ref{e_s_2}) show that the appearance of ground-state ferromagnetism for deviations from ideal geometry is a
result of fourth-order processes.
From $E_t=E_s(U_c)$ one gets a compact analytical formula 
\begin{eqnarray}
\label{077}
\frac{U_c}{t_2}
=
\frac{\sqrt{16+65\Omega^2}+9\Omega}{1-\Omega^2}\Omega.
\end{eqnarray}
From Fig.~\ref{fig12ttt} one can conclude,
the 
perturbation theory confirms the exact-diagonalization results for the existence of
ferromagnetism.
The quantitative agreement between perturbation theory and exact diagonalization  is the better the smaller
$\Omega^2$ and the larger $t_2$.
Higher-order perturbation theory naturally would enlarge the region of quantitative agreement
with  exact-diagonalization data.

In summary,
the exact-diagonalization data and the perturbation theory show
a dispersion-driven appearance of ground-state ferromagnetism of the Hubbard model on a frustrated lattice. 
For a half-filled lowest one-electron band,
the ground state for the ideal flat-band case has perfect charge order
[$n={\cal{N}}$ electrons are localized  in ${\cal{N}}$ trapping cells (vertical dimers) thus avoiding the on-site Hubbard repulsion]
but no spin oder
(spins of electrons in different cells are uncorrelated leading to $2^{{\cal{N}}}$-fold degeneracy of the ground state).
A small deviation  from the ideal flat-band geometry removes  the localization
of the electrons  
and they  
(i) either form a complicate many-electron ground state with a $U$-dependent energy which increases as $U$ increases
(when $U$ is small) 
(ii) or form a symmetric (i.e., ferromagnetic) spin multiplet ground state 
with a $U$-independent energy thus avoiding $U$ 
(when $U$ is large enough).
We mention finally that 
such a scenario is not restricted to the specific geometry of the frustrated diamond chain
and may be expected for other lattices with isolated cells discussed in Sec.~\ref{sec081}
such as the frustrated ladder or some two-dimensional lattices.\cite{patrick}

\section{Closely Related Studies on Electron Systems}
\label{sec11}

\subsection{Construction of exact ground states of the Hubbard model using positive semidefinite operator properties}
\label{sec111}

Alternatively to  Mielke's or Tasaki's description of the flat-band scenario  
and to the route to ferromagnetism described in the previous section,
Z.~Gul\'{a}csi, A.~Kampf, D.~Vollhardt and coworkers use a technique      
to construct exact ground states (including ferromagnetic ones) of the Hubbard model that is based  
on positive semidefinite operator properties.\cite{gulacsi1,gulacsi2}
In what follows we will illustrate some basic features of this method.
For more details on the method and more applications to strongly correlated
electron systems we refer the interested reader to Refs.~\refcite{gulacsi1,gulacsi2}.
Instead of dealing with an (arbitrary) Hamiltonian $H$ 
we may always consider the operator $O_{p}=H-E_0$, where $E_0$ is the ground-state energy,
which is a positive semidefinite operator with only non-negative eigenvalues.
Therefore we can start by an exact rewriting of the initial Hamiltonian in the form
\begin{eqnarray}
\label{078}
H=\sum_i O_{p,i}+C_{g,H},
\end{eqnarray}
where the c-number $C_{g,H}$ is a function of the Hamiltonian parameters and $O_{p,i}$ are positive semidefinite operators.
It is reasonable to introduce block operators $A_{n,\sigma}$  
which are composed of operators acting on the sites of a finite block $n$.
Then the operators $O_{p,i}$  contain types of
expressions  of the
form $A^\dagger_{n,\sigma}A_{n,\sigma}$ which preserve the positive
semidefinite form of $O_{p,i}$.
To get a valid transformation to the form $H=\sum_i O_{p,i}+C_{g,H}$ a specific 
relationship between the parameters of the initial Hamiltonian and
those contained in the
expressions of the  $O_{p,i}$ in terms of the block operators $A_{n,\sigma}$ (the so-called matching equations) must be
fulfilled.
The reason why Eq.~(\ref{078}) merits attention is that the deduction of the ground state $\vert{\rm{GS}}\rangle$
can be performed by constructing the most general state which satisfies the equation $O_p\vert{\rm{GS}}\rangle=0$.
Solving this equation opens a route for deducing exact results for nonintegrable systems
independent of dimensionality.
The construction of exact ground states is then performed in four steps:
(i) the transformation of the Hamiltonian in a positive semidefinite form,
(ii) the solution of the matching equations,
(iii) the construction of the ground states,
(iv) the proof of the uniqueness.
In a final step then the physical properties are deduced from the obtained ground state.

As an illustrative example we consider a diamond Hubbard chain with the Hamiltonian
\begin{eqnarray}
\label{079}
H=H_0+H_U,
\nonumber\\
H_0=\sum_{\sigma=\uparrow,\downarrow}\sum_{m} 
\left\{
\left[
te^{i\frac{\delta}{2}}
\left(
c_{m,1,\sigma}^\dagger c_{m-1,3,\sigma}+c_{m,3,\sigma}^\dagger c_{m,1,\sigma}
\right.
\right.
\right.
\nonumber\\
\left.
\left.
\left.
+c_{m,2,\sigma}^\dagger c_{m-1,3,\sigma}+c_{m,3,\sigma}^\dagger c_{m,2,\sigma}
\right)
\right.
\right.
\nonumber\\
\left.
\left.
+t_\perp c_{m,1,\sigma}^\dagger c_{m,2,\sigma}
+t_\parallel c_{m-1,3,\sigma}^\dagger c_{m,3,\sigma}
+{\rm{H.c.}}
\right]
\right.
\nonumber\\
\left.
+\epsilon\left(c_{m,1,\sigma}^\dagger c_{m,1,\sigma}+c_{m,2,\sigma}^\dagger c_{m,2,\sigma}\right)
\right\}, \quad
\nonumber\\
H_U=U\sum_{m} 
\left(
n_{m,1,\uparrow} n_{m,1,\downarrow}+n_{m,2,\uparrow} n_{m,2,\downarrow}+n_{m,3,\uparrow} n_{m,3,\downarrow}
\right).
\end{eqnarray}
For the case $t_\perp=t_\parallel=0$ and $\delta=\frac{\pi}{2}$
the one-electron spectrum consists of three flat bands (``Aharonov-Bohm cage'' limit).
The  exact ground states for electron densities $\frac{n}{N}=\frac{1}{3}$ and $\frac{n}{N}<\frac{1}{3}$ 
represent  then simply  an example of Mielke's and Tasaki's flat-band ferromagnetism,
see Ref.~\refcite{gulacsi1}.
This example, to some extent, can be understood without introducing the positive-semidefinite-operators technique:
The $O_p$ operator is simply a diagonalized $H_0$ 
and the ground states $\vert{\rm{GS}}\rangle$ are trivially constructed from the localized-Wannier-eigenstate operators.

One can obtain exact many-electron ground states also for other special cases
which do not belong to the scenario of Mielke's and Tasaki's flat-band
ferromagnetism.
We illustrate as an example  the case $t_\parallel>0$, $\epsilon=-t_\perp+\frac{2}{t_{\perp}}$, and
$\delta=\pi$; then the one-electron spectrum consists of a dispersive lowest band and two upper flat bands.
In order to rewrite $H_0$ as a positive semidefinite form
one defines the noncanonical fermionic
block operators
\begin{eqnarray}
\label{080}
A_{m,\sigma}
=
a_1c_{m-1,3,\sigma}+a_2c_{m,1,\sigma}+a_3c_{m,3,\sigma}+a_4c_{m,2,\sigma},
\end{eqnarray}
which fulfill $(A_{m,\sigma})^2=0$ and $\{A_{m_1,\sigma},A^\dagger_{m_2,\sigma}\}\ne\delta_{m_1,m_2}$.
Using Eq.~(\ref{080}) one gets the expression for $A^\dagger_{m,\sigma}A_{m,\sigma}$ and with
the requirement $-\sum_{m,\sigma} A^\dagger_{m,\sigma}A_{m,\sigma}=H_0$
one arrives at the matching conditions
$a_2^*a_1=a_3^*a_2=a_4^*a_3=a_1^*a_4=-te^{i\frac{\delta}{2}}$,
$a_2^*a_4=-t_\perp$,
$a_3^*a_1=-t_\parallel$,
$\vert a_1\vert^2+\vert a_3\vert^2=\epsilon+\vert a_2\vert^2=\epsilon+\vert a_4\vert^2$.
As a result, Eq.~(\ref{080}) becomes
\begin{eqnarray}
\label{081}
A_{m,\sigma}
=
\sqrt{t_\parallel}
\left[c_{m-1,3,\sigma}-c_{m,3,\sigma}
-2t_\perp e^{i\frac{\delta}{2}}\left(c_{m,2,\sigma}-c_{m,1,\sigma}\right)\right].
\end{eqnarray}
Consequently, the Hamiltonian has been transformed into the positive semidefinite form
\begin{eqnarray}
\label{079-}
H=\sum_{m,\sigma} A_{m,\sigma}A_{m,\sigma}^\dagger +U P+ E_{\rm{GS}},
\nonumber\\
P=\sum_m\left(n_{m,\uparrow}-1\right)\left(n_{m,\downarrow}-1\right),
\nonumber\\
E_{\rm{GS}}=\left(\epsilon+U+t_\perp\right)n - \left(3U+4t_\perp+\frac{1}{t_\perp}\right){\cal{N}}; 
\;\;\; 
{\cal{N}}=\frac{N}{3}.
\end{eqnarray}
For electron number $n=4{\cal{N}}$
the ground state is
\begin{eqnarray}
\vert{\rm{GS}}\rangle
=C
\left(\prod_m A^\dagger_{m,-\sigma}A^\dagger_{m,\sigma}\right)
F_{\sigma}^\dagger\vert{\rm{vac}}\rangle \; , 
\;
F_{\sigma}^\dagger
=
\prod_m  c^\dagger_{m,i_1,\sigma}c^\dagger_{m,i_2,\sigma},  
\end{eqnarray}
where $C$ is a normalization constant, $m$ runs over all ${\cal{N}}$ cells, and $i_1\ne i_2$ are two arbitrary sites in each cell.
The ground state can be also found for higher electron densities
$\frac{n}{N}>\frac{4}{3}$.\cite{gulacsi1}
The physical properties of these ground-state solutions correspond to a correlated half-metal.

Let us mention that the procedure described above is not restricted to the
case when only flat one-electron bands are present. 
For instance in Ref.~\refcite{gulacsi2} the authors provide  a solution for
a model, where all one-electron bands are dispersive.

\subsection{Flat bands and randomness}
\label{sec112}

In flat-band systems with perfect geometry the electrons are localized due
to destructive quantum interference.
On the other hand, the standard Anderson localization of electrons in a
dispersive band is caused by disorder.\cite{anderson-localization}
The  interplay between flat-band localization and disorder may lead to
unusual features.\cite{random1,random2,random3}
A spectacular effect is a ``inverse Anderson transition'' found in three-dimensional
flat-band systems, 
i.e., a disorder-driven delocalization of electrons.\cite{random1} 
To demonstrate this effect, 
the authors of Ref.~\refcite{random1} consider a system with only flat bands
-- a diamond lattice with fourfold-degenerated orbitals on each site.
The system is described by the standard tight-binding Hamiltonian and the
disorder is introduced via mutually independent random 
on-site energies.
The random variables 
are uniformly distributed over the range from $-\frac{W}{2}$ to $\frac{W}{2}$.
To study the transition of the one-electron states the authors examine the level statistics.
They found a disorder-induced localization-delocalization-localization transition
which in this specific model is characterized by the two critical values of the
disorder strength,
$W_{c1}\approx 13.5$ and $W_{c2}\approx 37.0$.
While at $W_{c2}$ the well-known Anderson transition takes place, 
the first transition at $W_{c1}< W_{c2}$ is an unknown new transition from localized states to extended states
driven by disorder. 
The localization for $W<W_{c1}$ is not due to the strength of disorder but to the
flat-band localization which survives under a certain strength of disorder.

A further analysis of tight-binding flat-band systems  in the presence of disorder
was reported in Ref.~\refcite{random2}.
The authors of Ref.~\refcite{random2} focus on the weak-disorder limit and consider the two-dimensional checkerboard lattice 
(also called the planar pyrochlore lattice or the square lattice with
crossings).
For the random on-site potential a Gaussian distribution is assumed, 
periodic systems of size $L \times L$, $34\le L\le 80$,
are examined numerically.
To study localization  they use the probability distribution
of the  spacing between adjacent eigenvalues:
Extended and localized phases are identified by Wigner-Dyson and Poisson distributions, respectively,
while the critical point corresponds to a distinct, universal distribution.
The numerical results for the probability distribution of
the spacing between adjacent levels 
is intermediate between those for Wigner-Dyson and Poisson distributions,
which indicates that the system is at a critical point.
Thus, weak disorder in a two-dimensional tight-binding model with a flat band gives rise to eigenstates that are
critical, i.e.,  they are neither Anderson localized nor
spatially extended.

In a series of recent papers\cite{random3} the interplay of
flat-band geometry and randomness   has
been studied for several lattices in detail.
As already mentioned in Sec.~\ref{sec021} the authors also give 
a scheme for generating flat-band lattices 
by adding Fano defects to an initial lattice with dispersive bands and
performing  appropriate transformations of the Fano lattice to a
corresponding flat-band
lattice, see Ref.~\refcite{random3}.

In what follows we will illustrate some results     
of Ref.~\refcite{random3}.
Let us consider as an example a tight-binding diamond chain (with zero
``frustrating'' vertical bond $t_2$, see Fig.~\ref{fig01}). The
corresponding  one-electron band energies are
$\left\{0,\pm 2 \sqrt{1+\cos\kappa}\right\}$. Note that $t_1=t_3=1$ and
$t_2=0$; then the flat band has zero energy.
We consider the eigenstate of the Hamiltonian of the form 
$\vert\Psi\rangle=\sum_{m}(a_mc_{m,a}^\dagger + b_m c_{m,b}^\dagger+ c_m c_{m,c}^\dagger)\vert{\rm{vac}}\rangle$.\footnote{We use 
here the notations (indices) of
Ref.~\refcite{random3}. To make the relationship to the notations (indices) used throughout
our paper
(see Fig.~\ref{fig01}) one has to  
use the relations $n,a\to n,2$,
$n,b\to n-1,3$,
and
$n,c\to n,1$.}
The basic equations to be discussed further are as follows:
\begin{eqnarray}
\label{sf}
i\dot{a}_n+\epsilon_{n,a}a_n &=& -\nabla^2b_{n+1},
\nonumber\\
i\dot{b}_n+\epsilon_{n,b}b_n &=& -\nabla^2(a_{n}+c_{n}),
\nonumber\\
i\dot{c}_n+\epsilon_{n,c}c_n &=&-\nabla^2b_{n+1}.
\end{eqnarray}
Here $\nabla^2f_n=f_n+f_{n-1}$ and $\epsilon_{n,\alpha}$, $\alpha=a,b,c$, is the random potential.
The tight-binding model described by Eq.~(\ref{sf})  contains intersecting dispersive and flat bands at zero energy.
The disorder is introduced as usual  
by uniformly distributed random on-site  energies $\epsilon_{n,\alpha}\in\left[-\frac{W}{2},\frac{W}{2}\right]$.

Various standard quantities such as the participation ratio, the compactness index, the localization length $\xi$ 
(asymptotic decay rate of the eigenmode tails) 
etc.
were calculated numerically by diagonalization of Eq.~(\ref{sf}).
The authors of Ref.~\refcite{random3} found different
behavior for the low-energy ($\vert E\vert <\frac{W}{2}$) and high-energy ($\vert E\vert >\frac{W}{2}$) regimes.
In particular, for the correlation length at the flat-band energy $E=0$ 
$\xi(W)\sim W^{-\gamma}$ with the unusual exponent $\gamma=1.30\pm0.01$ was found
in contrast to the exponent $\gamma=2$ for a dispersive band, i.e., the flat-band states are more delocalized.
Moving slightly away from the flat-band energy the anomalous exponent persists as long as $\vert E\vert\lesssim W$.
As $E$ increases further, there is a rapid crossover to the conventional $\gamma=2$
exponent.
Comparing the behavior of other measures of localization,
the authors conclude that the flat-band states display criticality in the weak-disorder regime.

As mentioned above, the authors of  Ref.~\refcite{random3} found a relation
between flat-band lattices and lattices with Fano defects.
Using the example of the frustrated two-leg ladder (see
Fig.~\ref{fig01}, note that in Ref.~\refcite{random3} the notation ``cross-stitch lattice'' is used instead of ``frustrated two-leg ladder'')
they discuss the  connection between the existence of flat-band states and
the appearance of Fano  resonances\cite{fano} for wave
propagation. This relationship between the representations  
allows to relate flat-band physics to features of  Fano resonances.\cite{random3}
For the frustrated two-leg ladder with $t_1=1$
the one-electron eigenstates  can be written as
$\vert\Psi\rangle=\sum_{m}(a_mc_{m,a}^\dagger + b_m
c_{m,b}^\dagger)\vert{\rm{vac}}\rangle$.\footnote{Again we use
here the notations (indices) of
Ref.~\refcite{random3}. To make the relationship to the notations (indices)
used throughout
our paper
(see Fig.~\ref{fig01}) one has to  
use the relations $n,a\to n,1$ and $n,b\to n,2$.}
To find the coefficients $a_m$ and $b_m$ one has to solve the equations
\begin{eqnarray}
Ea_n=\epsilon_{n,a}a_n-a_{n+1}-a_{n-1}-b_{n-1}-b_{n+1}-t_2b_n,
\nonumber\\
Eb_n=\epsilon_{n,b}b_n-a_{n+1}-a_{n-1}-b_{n-1}-b_{n+1}-t_2a_n.
\end{eqnarray}
In the absence of a potential $\epsilon_{n,b}=\epsilon_{n,b}=0$ the resulting
energy bands  read $E_{\rm{FB}}=t_2$, $E(\kappa)=-4\cos\kappa -t_2$.

By using the transformation
\begin{eqnarray}
p_n=\frac{1}{\sqrt{2}}(a_n+b_n),
\;\;\;
f_n=\frac{1}{\sqrt{2}}(a_n-b_n),
\nonumber\\
\epsilon_{n,+}=\frac{1}{2}(\epsilon_{n,a}+\epsilon_{n,b}),
\;\;\;
\epsilon_{n,-}=\frac{1}{2}(\epsilon_{n,a}-\epsilon_{n,b})
\end{eqnarray}
one obtains a new representation of the initial problem 
that corresponds to a lattice with dispersive degrees of freedom $p_n$ and side-coupled Fano states $f_n$\cite{fano}
\begin{eqnarray}
\label{det}
E p_n=(\epsilon_{n,+}-t_2)p_n+\epsilon_{n,-}f_n-2(p_{n+1}+p_{n-1}),
\nonumber\\
E f_n=(\epsilon_{n,+}+t_2)f_n+\epsilon_{n,-}p_n.
\end{eqnarray}
In Ref.~\refcite{random3} such a lattice with side-coupling defects is called  ``Fano
lattice''.
This procedure can be inverted:
Choose a dispersive chain, 
add a set of uncoupled Fano states $f_n$ with energies $E_{f_n}$,
assign locally each $f_n$ to a site with $p_n$, 
and perform local rotations (transformations) in the space $\{p_n,f_n\}$.
As a result, one obtains the original frustrated two-leg ladder lattice.
This construction can be generalized and it yields complex lattices having flat-band localized states.

The above described Fano-lattice representation of the frustrated two-leg ladder 
allows to discuss the impact of perturbations of the initial model from a
completely different viewpoint. 
We consider a site-dependent on-site potential $\epsilon_n$. 
For $\epsilon_{n,-}=\frac{1}{2}(\epsilon_{n,a}-\epsilon_{n,b})=0$ the Fano states will remain
decoupled, but their degeneracy
is lifted, since $\epsilon_{n,+}\ne 0$, see Eq.~(\ref{det}).
A particular case corresponds to uncorrelated random numbers $\epsilon_{n,a}=\epsilon_{n,b}=\epsilon_n$ 
with uniformly distributed random variables $\epsilon_{n}\in\left[-\frac{W}{2},\frac{W}{2}\right]$.
The Fano states stay decoupled, however, acquire an energy spread of the order of $W$ around $E_{\rm{FB}}=t_2$.
The dispersive lattice becomes Anderson localized with a localization length $\xi\sim W^{-2}$ for weak disorder $W\le 4$.
If the symmetry constraint is relaxed,
i.e., $\epsilon_{n,-}\ne 0$, the Fano states are locally coupled to the dispersive chain.
Because of the pure local coupling, they can be eliminated resulting in a equation for
the dispersive lattice with a Lorentzian (Cauchy) disorder\cite{ishii} if $\vert E-E_{\rm{FB}}\vert\le\frac{W}{2}$.
If $\vert E-E_{\rm{FB}}\vert\ge\frac{W}{2}$ the Lorentzian disorder does not appear and the correlation-length exponent
is $\gamma=2$.
These analytical arguments are in excellent agreement with numerical
findings.\cite{random3}
However, at $E_{\rm{FB}}=0$ the localization-length exponent is found to
be $\gamma=1.30\pm 0.01$ for the diamond
chain, by contrast to $\gamma=1$ as expected for the  Lorentzian   
distribution.  On the other hand, for other models, e.g., for the
frustrated two-leg ladder
(or ``cross-stitch lattice''), the
expected exponent $\gamma=1$ is found.
This unexpected finding for the diamond chain remains a puzzle to be
explained.

\subsection{Conductance through flat-band clusters}
\label{sec113}

The conductance through various nanosystems such as molecular devices or
nanowires has been extensively studied both theoretically and experimentally
in recent years.
The main nanotransport phenomena 
(e.g., Coulomb blockade, conductance quantization, resonant tunneling etc.) 
are meanwhile well studied and understood.
For nanoclusters with flat-band  geometry the existence of   
localized states may lead to distinct features,
i.e., localized states may manifest themselves in an intriguing way in the
conductance.\cite{lopes,transp_kagochains}

\begin{figure}[bt]
\centerline{\psfig{file=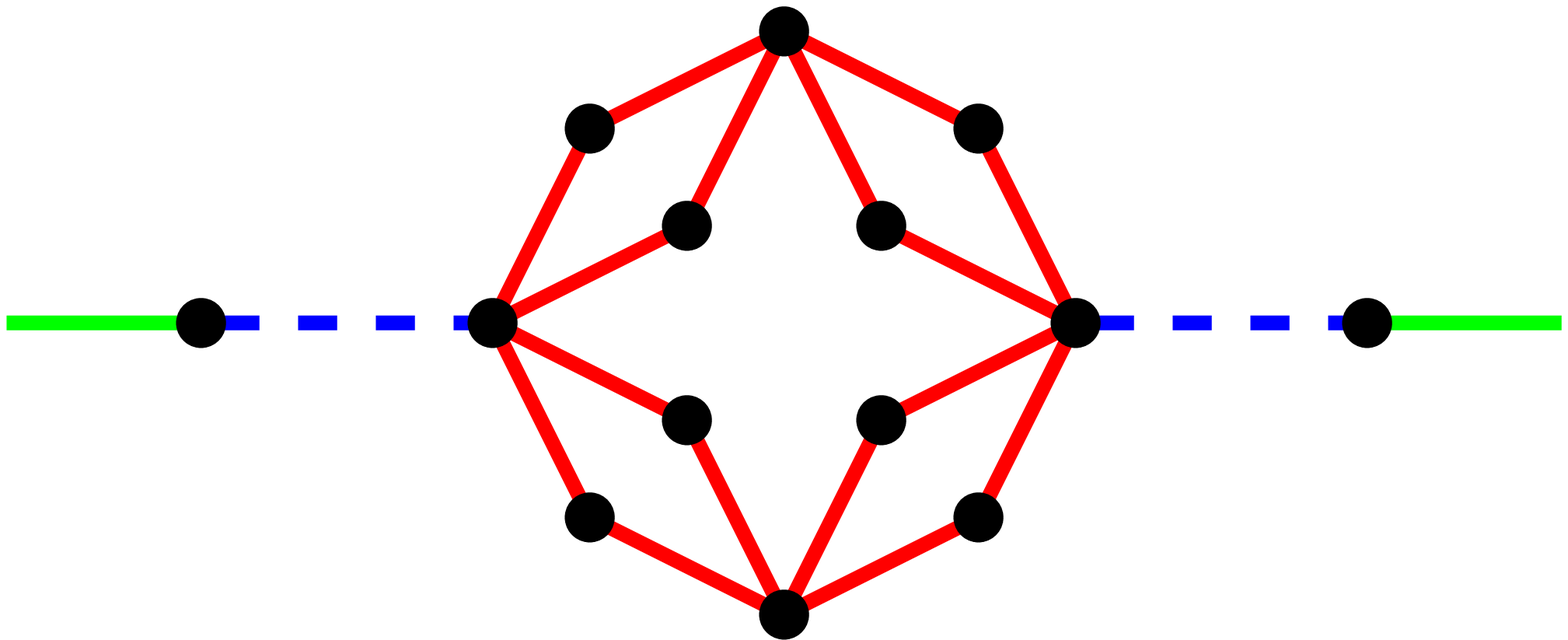,width=2.5in}\hspace{10mm}\psfig{file=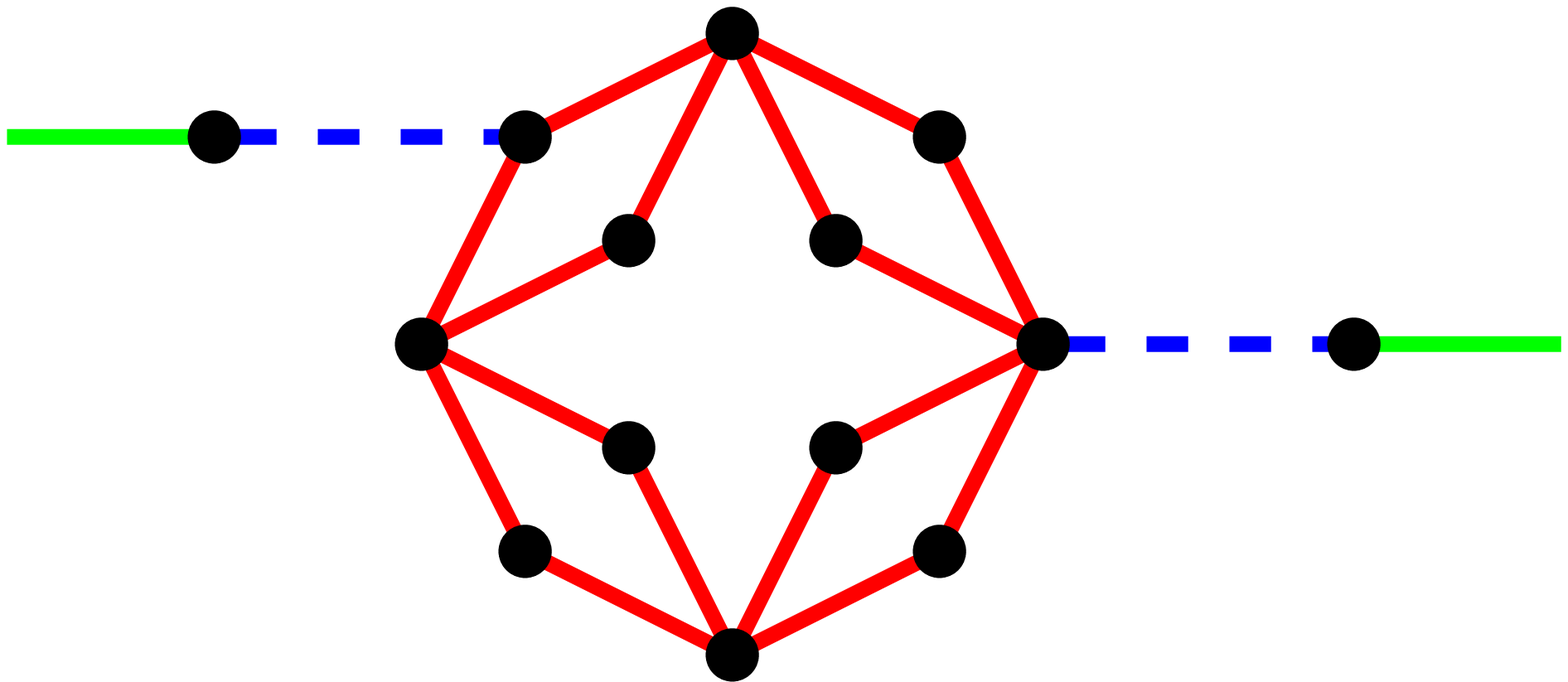,width=2.35in}}
\vspace*{8pt}
\caption{(Color online)
Periodic ${\cal{N}}$-cell diamond chain (${\cal{N}}=4$) connected to semi-infinite leads via hoping amplitudes $t_L$ and $t_R$ (dashed bonds).}
\label{fig13ttt}
\end{figure}

We follow Ref.~\refcite{lopes} and consider as an example a finite periodic
 diamond chain
(i.e., a diamond ring) without frustration (the vertical bond is zero, cf. Fig.~\ref{fig01})
which is connected to left and right semi-infinite leads, see Fig.~\ref{fig13ttt}.
The Hamiltonian of the two-terminal device is as follows:
\begin{eqnarray}
H&=&H_{\rm{leads}}+H_{\rm{cluster}}+H_{\rm{int}},
\nonumber\\
H_{\rm{leads}}
&=&
-t\sum_{j=-\infty}^0\left(\vert j-1\rangle\langle j\vert+{\rm{H.c.}}\right)
-t\sum_{j=N+1}^{\infty}\left(\vert j\rangle\langle j+1\vert+{\rm{H.c.}}\right),
\nonumber\\
H_{\rm{cluster}}
&=&
-t\sum_{m=1}^{\cal{N}}
\left[e^{i\frac{\Phi_o}{2{\cal{N}}}}\left(\vert m,3\rangle\langle m,1\vert + \vert m,1\rangle\langle m-1,3\vert\right)
\right.
\nonumber\\
&&\left.
+
e^{-i\frac{\Phi_i}{2{\cal{N}}}}\left(\vert m,2\rangle\langle m,3\vert + \vert m-1,3\rangle\langle m,2\vert\right)+{\rm{H.c.}}\right],
\nonumber\\
H_{\rm{int}}
&=&
-t_L\left(\vert 0\rangle\langle 1\vert+{\rm{H.c.}}\right)
-t_R\left(\vert N\rangle\langle N+1\vert+{\rm{H.c.}}\right).
\end{eqnarray}
The phase factors in $H_{\rm{cluster}}$ are related to fluxes, where the fluxes $\Phi$ and $\Phi_i$ are threading the plaquettes and the inner
ring,
respectively,
and $\Phi_o=\Phi_i+{\cal{N}}\Phi$. 
Furthermore, $\vert j\rangle=c_j^{\dagger}\vert {\rm{vac}}\rangle$ (respectively
$\vert m,\alpha\rangle=c_{m,\alpha}^{\dagger}\vert {\rm{vac}}\rangle$,
$\alpha=1,2,3$) is a state, where the fermion is located at the lattice site $j$
of the lead (at the lattice site $m,\alpha$ of the diamond ring).
The  pairs of indices
labeling the sites of the diamond ring enumerate the cells $m=1,\ldots,{\cal{N}}$ and the position
$\alpha=1,2,3$ of the site within a cell, 
see Fig.~\ref{fig01}.  
The connecting hopping amplitudes $t_L$ and $t_R$ are assumed
to be small compared to the hopping integrals $t$. 
The site 1 ($N$) stands for the left (right) contact site of the cluster,
cf. the left and right panel in Fig.~\ref{fig13ttt}.
The spin variable $\sigma=\uparrow,\downarrow$ is omitted, since it is irrelevant in this tight-binding
treatment, i.e., for $U=0$.

Within the Landauer picture,\cite{imry,datta} 
which treats the electric current as a consequence of the injection of a carrier at
a contact and its probability to reach the other end,
the conductance is determined by the formula
\begin{eqnarray}
G=G_0 \vert {\sf{t}}(E)\vert^2 ,
\end{eqnarray}
where $G_0=\frac{e^2}{\pi\hbar}$ is the conductance quantum 
and $\vert {\sf{t}}(E)\vert^2$ is the transmission probability of a propagating electron with the energy $E$.
For the two-terminal device the propagating electron comes from the left lead 
with the energy $E=-2t\cos \kappa$, $-\pi\le \kappa<\pi$.
Going along standard reasoning of quantum scattering theory, 
one seeks for a solution of the stationary Schr\"{o}dinger equation,
$H\vert\psi\rangle=E\vert\psi\rangle$,
where  $\vert\psi\rangle=\sum_n \psi_n\vert n\rangle$ ($n=j$ or
$n=m,\alpha$). In the left and right
lead the amplitudes have the form:
\begin{eqnarray} \label{eq77}
\psi_j
=
\left\{
\begin{array}{ll}
e^{i\kappa j}+ {\sf{r}} e^{-i\kappa j}, & j\le 0,\\
{\sf{t}}e^{i\kappa j},                   & N+1\le j,
\end{array}
\right.
\end{eqnarray}
i.e., it is a linear combination of the incoming and  the reflected waves in the left lead and the transmitted wave in the right lead.
There are $3{\cal{N}}+2$ equations for $3{\cal{N}}+2$ unknown quantities ${\sf{r}}$, ${\sf{t}}$ and
$\psi_i$, 
where $i=m,\alpha$ runs over the $3{\cal{N}}$ sites of the diamond ring.
Solving these equations we determine ${\sf{t}}$ which is, in fact, the required amplitude of the transmission probability,
${\sf{t}}={\sf{t}}(E)$.
[Note 
that instead of the standard approach to calculate ${\sf{t}}(E)$ via Eq.~(\ref{eq77}) 
the authors of Ref.~\refcite{lopes} use a Green's function representation of the scattering problem.\cite{enss}]

Let us briefly describe some interesting findings reported in Ref.~\refcite{lopes}.
We begin with the case of zero flux.
If the leads are connected to a site $m,3$ of the  diamond ring (left panel in Fig.~\ref{fig13ttt})
one gets an expected dependence of the conductance on the gate voltage $V_{\rm{gate}}$.
It exhibits three Breit-Wigner-shaped peaks at energies $-2t,\,0,\,2t$.
The latter values correspond to the eigenvalues of the cluster which follows from the band energies,
$\left\{0,\pm 2t\sqrt{1+\cos\kappa}\right\}$ (the flat band has zero energy),
after substitution $\kappa=-\pi,-\frac{\pi}{2},0,\frac{\pi}{2}$.
By contrast to this usual conductance behavior, the situation is changed, if the left lead is
connected to a
site $m,1$ (right panel in Fig.~\ref{fig13ttt}): Then the $E=0$--peak in the conductance profile is absent, i.e.,
there is no conductance, if $E$ corresponds to the flat-band value.
The authors of Ref.~\refcite{lopes} explain this as follows.
The hopping term to a site $m,1$ of the diamond ring
describes  a hopping to an itinerant state of the ring as well as to a localized state.
Since the localized state is decoupled from all other states of the ring, 
it leads only to a reflected wave back into the left lead.
For $E=0$ this reflected wave interferes destructively with the incident wave 
resulting in the absence of the zero-energy peak in the conductance.

In the presence of finite fluxes the behavior at $E=0$ may be
different in case of the connection of the 
left lead to a site $m,1$ and the right lead to site $m,2$, $m-1,2$ etc.  
Due to fluxes 
a gap between the flat-band states with zero energy and itinerant states opens.
The conductance behavior for $E$ in a small region around $E=0$ is again
unusual:
The conductance is still zero precisely at $E=0$, but in close vicinity on both sides of
$E=0$ two well-pronounced peaks emerge, i.e., the conductance exhibits a double peak symmetric to zero
energy.
This is unexpected, since the energies of the itinerant states of the ring are far from this
zero-energy peak region
and this is related to variable extensions of localized states in the presence of fluxes.
For further details we refer the interested reader to Ref.~\refcite{lopes}.

In summary, localized states in a flat-band tight-binding cluster generate a striking behavior in the two-terminal electronic conductance.
The positions of the contact sites are extremely important.
The localized states may act as a filter of the zero-energy conductance peak in absence of
a flux
or may lead to the appearance of a conductance double peak around zero
energy.
Although the tight-binding calculations described  above do not take into account the interelectron interactions,
one may expect that the peculiarities will survive when interactions are
present, 
see Refs.~\refcite{lopes,conductance_inter,hubbard_molecule,single_mol_magnet}.

\section{Experiments}
\label{sec12}

The existence of ferromagnetism in a basic model of strongly correlated
electrons, the one-band repulsive Hubbard model, requires a particular
pattern of hopping integrals, i.e., a special lattice topology.  
The  search for this kind of ferromagnetism is an attractive goal for many researches
and it is an ambitious task to find its traces in specific materials.
In particular, flat-band ferromagnetism  opens the window to ferromagnets consisting entirely of
nonmagnetic elements such as organic molecules.   
There are various directions of ongoing experimental search for flat-band
ferromagnetism.

The realization of flat-band ferromagnetism in quantum dot arrays and in quantum atomic wires formed on solid surfaces 
was discussed in Ref.~\refcite{tamura-quantum-dots-wires}. 
While in real materials a lattice distortion may destabilize the ferromagnetism when the flat band is half filled,
artificial materials such as dot lattices do not have such disadvantages.
H. Tamura et al.\cite{tamura-quantum-dots-wires} demonstrated that flat-band ferromagnetism should be observable 
at reasonable temperatures in semiconductor dot arrays using existing fabrication technology.
They performed calculations for GaAs, InAs, and Si dot arrays of Lieb- and
kagome-lattice structures 
of typical interdot spacing of 5, 10, 50, and 100 nm and dot diameter two times smaller than the interdot spacing.
The designed structures, however, may require an ultra-low-temperature technique to detect flat-band ferromagnetism,
see the paper by K. Shiraishi et al. in Ref.~\refcite{tamura-quantum-dots-wires}.

Other examples for flat-band ferromagnets are related to polymers,\cite{flat-band-experiment-polymers}
where the search for purely organic ferromagnets is known as a challenging target. 
R. Arita et al.\cite{flat-band-experiment-polymers} found 
that a chain of five-membered rings (polyaminotriazole) should be ferromagnetic with an appropriate doping.
Furthermore,
the authors demonstrated that the predicted flat-band ferromagnetism of a single chain may occur as a bulk
property,
i.e., in a three-dimensional crystal of the polymer.

Another experimental realization of flat-band ferromagnetism has been discussed recently in Ref.~\refcite{fb-nature}. 
Specific organic molecules 
[tetracyano-p-quinodimethane (TCNQ) molecules] 
deposited on graphene epitaxially grown on Ru(0001) 
acquire a charge from the substrate and develop a magnetic moment which survives when the molecules form a monolayer. 
The authors of Ref.~\refcite{fb-nature} claim that the intermolecular bands are almost flat and half-filled 
and that the TCNQ/graphene/Ru(0001) system might be a realization of a flat-band Hubbard ferromagnet.

Just recently, the electronic structure of silicene has been examined and flat bands have been found.\cite{hatsugai}
The authors claim that if we can shift the chemical potential close to the flat band
(e.g., by chemical doping)
flat-band ferromagnetism may be expected.

Although, so far no unambiguous example of an experimental realization of a
flat-band ferromagnet was found this field of experimental research is very
promising not only for fundamental physics but also for applied physics
(having, e.g., in mind the possible realization of organic ferromagnets). 
Within this context it seems to be important that (small) deviations from the ideal
flat-band geometry are not necessarily contraindicative for the realization
of flat-band ferromagnetism.

\section{Summary and Outlook}
\label{sec13}

In this review we offer a unifying view on  spin and electron systems possessing flat bands.
The existence of flat bands implies similarities in mathematical description,
although the statistics of particles and the interactions lead to differences in properties.
A study of  strongly correlated flat-band systems leads to some surprises.
First of all, and contrary to what one may expect,
it appears that in spite of complicated (often frustrated) lattices which are usually required to get flat bands,
the theoretical description may be simpler and it involves specific methods of classical statistical mechanics.
Second, there are a number of intriguing properties of  strongly correlated flat-band systems 
with important potential applications.
For example, flat-band ferromagnetism offers a route to ferromagnets without magnetic atoms.
Other promising examples are the application for magnetic cooling using the magnetocaloric effect in frustrated quantum antiferromagnets 
or for the control of the charge transport through flat-band nanodevices.

We mention again that flat-band physics is a much broader field than it
was discussed in this
review;
examples are topological flat-band models,\cite{bergholtz}
superconductivity/superfluidity  in flat-band
systems,\cite{bcs_flat,nematic}
flat dispersion of phonons,\cite{phonons} flat bands in an extended Falicov-Kimball
model,\cite{falikov-kimball}
flat-band-like states in fractal networks,\cite{nandy}
``flat-band'' (i.e., localized-state) effects 	
in magnetic molecules\cite{fr_molecules,fr_molecules-review}
or nonequilibrium flat-band phenomena.\cite{biondi}
From a general point of view it might be interesting to examine systematically other flat-band systems
and to find more traces of flat-band physics in experiments.  
We hope this review will encourage further exploration of strongly correlated flat-band systems along the described lines.

\section*{Acknowledgements}

We acknowledge useful discussions and related collaborations with
S. Capponi,  
D. V. Dmitriev, 
S.-L. Drechsler,
A. Honecker, 
V. Ya. Krivnov, 
T. Krokhmalskii,
O. Krupnitska,
A. M. L\"{a}uchli,
R. Moessner,
S. Nishimoto,
and 
K. Shtengel.
We specially thank 
D. V. Dmitriev,
S. Flach,
Z. Gul\'{a}csi,
V. Ya. Krivnov,
A. A. Lopes,
and
H. Tanaka
for valuable comments on the manuscript.
We thank 
A. Andrusyk
and
T. Verkholyak
for critical reading of the manuscript.

The present study was supported by the DFG (project RI615/21-1).
O.~D. acknowledges financial support of the Abdus Salam International Centre for Theoretical Physics (Trieste) 
through the Senior Associate award.
O.~D. and J.~R. are grateful to the MPIPKS (Dresden) for the kind hospitality in October-December of 2013.
O.~D. is grateful to the University of Magdeburg for the kind hospitality in April-June and October-December of 2014.
O.~D. acknowledges kind hospitality of the organizers 
of the 567. WE-Heraeus-Seminar on ``Integrable Lattice Models and Quantum Field Theories''
(Bad Honnef, Germany, June 28 - July 2, 2014)
and
of the Conference on Field Theory Methods in Low-Dimensional Strongly Correlated Quantum Systems
(ICTP, Trieste, Italy, August 25-29, 2014).


\section*{References}

\begin{thebibliography}{999}

\bibitem{fqh-flat-review}
S. A. Parameswaran, R. Roy and S. L. Sondhi,
{\it C. R. Physique} {\bf 14}, 816 (2013).

\bibitem{bergholtz}
E. J. Bergholtz and Z. Liu,
{\it Int. J. Mod. Phys.} {\bf B27}, 1330017 (2013).

\bibitem{liu-liu-wu}
Z. Liu, F. Liu and Y.-S. Wu,
{\it Chinese Phys.} {\bf B23}, 077308 (2014).

\bibitem{kagome_flat_1}
C. Zeng and V. Elser, 
{\it Phys. Rev.} {\bf B42}, 8436 (1990).

\bibitem{kagome_flat_2}
H. Asakawa and M. Suzuki,
{\it Physica} {\bf A205}, 687 (1994).

\bibitem{Ramirez} 
A. P. Ramirez, 
{\it Annu. Rev. Mater. Sci.} {\bf 24}, 453 (1994).

\bibitem{Greedan}
J. E. Greedan,
{\it J. Mater. Chem.} {\bf 11}, 37 (2001).

\bibitem{lm1}
J. Schnack, H.-J. Schmidt, J. Richter and J. Schulenburg, 
{\it Eur. Phys. J.} {\bf B24}, 475 (2001);
J. Schulenburg, A. Honecker, J. Schnack, J. Richter and H.-J. Schmidt, 
{\it Phys. Rev. Lett.} {\bf 88}, 167207 (2002);
J. Richter, J. Schulenburg, A. Honecker, J. Schnack and H.-J. Schmidt,
{\it J. Phys.: Condens. Matter} {\bf 16}, S779 (2004).

\bibitem{sp-Peierls}
J. Richter, O. Derzhko and J. Schulenburg,
{\it Phys. Rev. Lett.} {\bf 93}, 107206 (2004);
O. Derzhko, J. Richter and J. Schulenburg,
{\it physica status solidi (b)} {\bf 242}, 3189 (2005);
O. Derzhko and J. Richter,
{\it Phys. Rev.} {\bf B72}, 094437 (2005).

\bibitem{prb2004}
O. Derzhko and J. Richter,
{\it Phys. Rev.}  {\bf B70}, 104415 (2004).

\bibitem{mike-andreas}
M. E. Zhitomirsky and A. Honecker,
{\it J. Stat. Mech.: Theor. Exp.} P07012 (2004).

\bibitem{mike1}
M. E. Zhitomirsky and H. Tsunetsugu,
{\it Phys. Rev.} {\bf B70}, 100403(R) (2004);
M. E. Zhitomirsky and H. Tsunetsugu,
{\it Prog. Theor. Phys. Suppl.} 160, 361 (2005);
M. E. Zhitomirsky and H. Tsunetsugu,
{\it Phys. Rev.} {\bf B75}, 224416 (2007).

\bibitem{epjb2006}
O. Derzhko and J. Richter,
{\it Eur. Phys. J.} {\bf B52}, 23 (2006);
J. Richter, O. Derzhko and T. Krokhmalskii,
{\it Phys. Rev.} {\bf B74}, 144430 (2006).

\bibitem{cmp2005}
A. Honecker and J. Richter,
{\it Condensed Matter Physics (L'viv)} {\bf 8}, 813 (2005);
J. Richter, O. Derzhko and A. Honecker,
{\it Int. J. Mod. Phys.} {\bf B22}, 4418 (2008);
O. Derzhko, J. Richter and A. Honecker,
{\it J. Phys.: Conf. Ser.} {\bf 145}, 012059 (2009).

\bibitem{le1}
A. Mielke, 
{\it J. Phys.} {\bf  A24}, L73 (1991);
A. Mielke, 
{\it J. Phys.} {\bf  A24}, 3311 (1991);
A. Mielke, 
{\it J. Phys.} {\bf  A25}, 4335 (1992);
A. Mielke, 
{\it Phys. Lett.} {\bf  A174}, 443 (1993).

\bibitem{le2}
H. Tasaki,
{\it Phys. Rev. Lett.} {\bf 69}, 1608 (1992).

\bibitem{le3}
A. Mielke and H. Tasaki, 
{\it Commun. Math. Phys.} {\bf 158}, 341 (1993).

\bibitem{flband-ferro-stability}
H. Tasaki, 
{\it Phys. Rev. Lett.} {\bf 73}, 1158 (1994);
H. Tasaki, 
{\it Phys. Rev. Lett.} {\bf 75}, 4678 (1995);
H. Tasaki, 
{\it J. Stat. Phys.} {\bf 84}, 535 (1996);
H. Tasaki, 
{\it Commun. Math. Phys.} {\bf 242}, 445 (2003);
see also
K. Kusakabe and H. Aoki,
{\it Phys. Rev. Lett.} {\bf 72}, 144 (1994);
K. Kusakabe and H. Aoki,
{\it Physica} {\bf  B194-196}, 215 (1994);
H. Aoki,
{\it Int. J. Mod. Phys.} {\bf  B17}, 4953 (2003).

\bibitem{fb-el1}
O. Derzhko, A. Honecker and J. Richter, 
{\it Phys. Rev.} {\bf B76}, 220402(R) (2007);
A. Honecker, O. Derzhko and J. Richter,
{\it Physica} {\bf B404}, 3316 (2009).

\bibitem{fb-el2}
O. Derzhko, A. Honecker and J. Richter, 
{\it Phys. Rev.} {\bf B79}, 054403 (2009).

\bibitem{fb-el3}
O. Derzhko, J. Richter, A. Honecker, M. Maksymenko and R. Moessner, 
{\it Phys. Rev.} {\bf B81}, 014421 (2010);
O. Derzhko, M. Maksymenko, J. Richter, A. Honecker and R. Moessner, 
{\it Acta Physica Polonica} {\bf A118}, 736 (2010).

\bibitem{fb-el3_4}
M. Maksymenko, A. Honecker, R. Moessner, J. Richter and O. Derzhko, 
``The flat-band ferromagnetic transition as a Pauli-correlated percolation transition'',
preprint of the Institute for Condensed Matter Physics of the National Academy of Sciences of Ukraine (L'viv), 
ICMP-11-14E (2011).

\bibitem{fb-el4}
M. Maksymenko, A. Honecker, R. Moessner, J. Richter and O. Derzhko, 
{\it Phys. Rev. Lett.} {\bf 109}, 096404 (2012). 

\bibitem{mykola} 
M. Maksymenko, 
Ph.D. Thesis, Institute for Condensed Matter Physics, National Academy of Sciences of Ukraine, L'viv, 2012.

\bibitem{fb-el5}
M. Maksymenko, R. Moessner and K. Shtengel,
``Reversible first-order transition in Pauli percolation'',
arXiv:1401.6172.

\bibitem{lnp}
C. Lhuillier and G. Misguich, 
in {\it High Magnetic Fields: Applications in Condensed Matter Physics and Spectroscopy, Lecture Notes in Physics Vol. 595}, 
ed. C. Berthier, L. P. L\'{e}vy and G. Martinez (Springer-Verlag, Berlin, 2002), pp.~161-190; 
G. Misguich and C. Lhuillier, 
in {\it Frustrated Spin Systems}, ed. H. T. Diep (World Scientific, Singapore, 2013), pp.~235-319.

\bibitem{lnp645}
J. Richter, J. Schulenburg and A. Honecker,
in {\it Quantum Magnetism, Lecture Notes in Physics Vol. 645}, 
ed. U. Schollw\"{o}ck, J. Richter, D. J. J. Farnell and R. F. Bishop (Springer-Verlag, Berlin, 2004), pp.~85-153.

\bibitem{lacroix}
{\it Introduction to Frustrated Magnetism. Materials, Experiments, Theory},
ed. C. Lacroix, P. Mendels and F. Mila
(Springer-Verlag, Berlin, Heidelberg, 2011).


\bibitem{origin_FM}
M. C. Gutzwiller, 
{\it Phys. Rev. Lett.} {\bf 10}, 59 (1963);
J. Hubbard, 
{\it Proc. Roy. Soc. London} {\bf A276}, 238 (1963);
J. Kanamori, 
{\it Prog. Theor. Phys.} {\bf 30}, 275 (1963).

\bibitem{hubbard_nature}
Editorial,
{\it Nature Physics} {\bf 9}, 523 (2013).

\bibitem{tasaki_rev1}
H. Tasaki,
{\it J. Phys.: Condens. Matter} {\bf 10}, 4353 (1998). 

\bibitem{tasaki_rev2}
H. Tasaki,
{\it Prog. Theor. Phys.} {\bf 99}, 489 (1998). 

\bibitem{tasaki_rev3}
H. Tasaki,
{\it Eur. Phys. J.} {\bf B64}, 365 (2008).

\bibitem{review1}
J. Richter,
{\it Fizika Nizkikh Temperatur (Kharkiv)} {\bf 31}, 918 (2005)
[{\it Low Temperature Physics} {\bf 31}, 695 (2005)].

\bibitem{review2}
O. Derzhko, J. Richter, A. Honecker and H.-J. Schmidt,
{\it Fizika Nizkikh Temperatur (Kharkiv)} {\bf 33}, 982 (2007) 
[{\it Low Temperature Physics} {\bf 33}, 745 (2007)].

\bibitem{review3}
J. Richter and O. Derzhko, 
in {\it Condensed Matter Physics in the Prime of the 21st Century. Phenomena, Materials, Ideas, Methods}, 
ed. J. J\c{e}drzejewski (World Scientific, Singapore, 2008), pp.~237-270.

\bibitem{ising_degrees}
O. Derzhko, T. Krokhmalskii and J. Richter,
{\it Phys. Rev.} {\bf B82}, 214412 (2010);
O. Derzhko, T. Krokhmalskii and J. Richter,
{\it Teoret. Mat. Fiz.} {\bf 168}, 441 (2011)
[{\it Theoretical and Mathematical Physics} {\bf 168}, 1236 (2011)].

\bibitem{chiral1}
M. Maksymenko, O. Derzhko and J. Richter,
{\it Acta Physica Polonica} {\bf A119}, 860 (2011);
M. Maksymenko, O. Derzhko and J. Richter,
{\it Eur. Phys. J.} {\bf B84}, 397 (2011).

\bibitem{chiral2}
A. Mielke,
{\it Eur. Phys. J.} {\bf B85}, 184 (2012).

\bibitem{deviations1}
O. Derzhko, J. Richter and O. Krupnitska,
{\it Condensed Matter Physics (L'viv)} {\bf 15}, 43702 (2012).

\bibitem{deviations2}
O. Derzhko, J. Richter, O. Krupnitska and T. Krokhmalskii,
{\it Phys. Rev.} {\bf B88}, 094426 (2013);
O. Derzhko, J. Richter, O. Krupnitska and T. Krokhmalskii,
{\it Fizika Nizkikh Temperatur (Kharkiv)} {\bf 40}, 662 (2014) 
[{\it Low Temperature Physics} {\bf 40}, 513 (2014)].

\bibitem{huber-altman}
S. D. Huber and E. Altman,
{\it Phys. Rev.} {\bf B82}, 184502 (2010).

\bibitem{nishimoto}
S. Nishimoto, N. Shibata and C. Hotta, 
{\it Nat. Commun.} {\bf 4}, 2287 (2013). 

\bibitem{capponi}
S. Capponi, O. Derzhko, A. Honecker, A. M. L\"{a}uchli and J.~Richter,
{\it Phys. Rev.} {\bf B88}, 144416 (2013).

\bibitem{Sakai} 
H. Nakano and T. Sakai, 
{\it J. Phys. Soc. Jpn.} {\bf 79}, 053707 (2010);
T. Sakai and H. Nakano, 
{\it Phys. Rev.} {\bf B83}, 100405(R) (2011);
T. Sakai and H. Nakano, 
{\it J. Phys.: Conf. Ser.} {\bf 320}, 012016 (2011);
T. Sakai and H. Nakano, 
{\it physica status solidi (b)} {\bf 250}, 579 (2013).

\bibitem{dmitriev_krivnov}
V. Ya. Krivnov, D. V. Dmitriev, S. Nishimoto, S.-L. Drechsler and J. Richter,
{\it Phys. Rev.} {\bf B90}, 014441 (2014).

\bibitem{kikuchi}
H. Kikuchi, Y. Fujii, M. Chiba, S. Mitsudo, T. Idehara, T. Tonegawa,
K. Okamoto, T. Sakai, T. Kuwai and H. Ohta,
{\it Phys. Rev. Lett.} {\bf 94}, 227201 (2005);
H. Kikuchi, Y. Fujii, M. Chiba, S. Mitsudo, T. Idehara, T. Tonegawa,
K. Okamoto, T. Sakai, T. Kuwai, K. Kindo, A. Matsuo,
W. Higemoto, K. Nishiyama, M. Horvati\'{c} and C. Bertheir,
{\it Prog. Theor. Phys. Suppl.} 159, 1 (2005).

\bibitem{kikuchi_2}
H. Kikuchi, Y. Fujii, M. Chiba, S. Mitsudo, T. Idehara, T. Tonegawa,
K. Okamoto, T. Sakai, T. Kuwai and H. Ohta,
{\it Phys. Rev. Lett.} {\bf 97}, 089702 (2006).

\bibitem{rule}
K. C. Rule, A. U. B. Wolter, S. S\"{u}llow, D. A. Tennant, A. Br\"{u}hl, S. K\"{o}hler, B. Wolf, M. Lang and J. Schreuer,
{\it Phys. Rev. Lett.} {\bf 100}, 117202 (2008).

\bibitem{azurite-parameters}
H. Jeschke, I. Opahle, H. Kandpal, R. Valenti, H. Das, T. Saha-Dasgupta, O. Janson, H. Rosner, A. Br\"{u}hl,
B. Wolf, M. Lang, J. Richter, S. Hu, X. Wang, R. Peters, T. Pruschke and A. Honecker, 
{\it Phys. Rev. Lett.} {\bf 106}, 217201 (2011).

\bibitem{effective_xy2}
A. Honecker, S. Hu, R. Peters and J. Richter,
{\it J. Phys.: Condens. Matter} {\bf 23}, 164211 (2011).

\bibitem{azurite2014}
P. T. Cong, B. Wolf, R. S. Manna, U. Tutsch, M. de Souza, A. Br\"{u}hl and M. Lang,
{\it Phys. Rev.} {\bf B89}, 174427 (2014). 

\bibitem{tanaka}
H. Tanaka, N. Kurita, M. Okada, E. Kunihiro, Y. Shirata, K. Fujii, H. Uekusa, A. Matsuo, K. Kindo and H. Nojiri,
{\it J. Phys. Soc. Jpn.} {\bf 83}, 103701 (2014). 

\bibitem{dispersion-driven}
O. Derzhko and J. Richter,
{\it Phys. Rev.} {\bf B90}, 045152 (2014).

\bibitem{gulacsi1}
Z. Gul\'{a}csi, A. Kampf and D. Vollhardt,
{\it Phys. Rev. Lett.} {\bf 99}, 026404 (2007);
Z. Gul\'{a}csi, A. Kampf and D. Vollhardt,
{\it Prog. Theor. Phys. Suppl.} 176, 1 (2008).

\bibitem{gulacsi2}
Z. Gul\'{a}csi, A. Kampf and D. Vollhardt,
{\it Phys. Rev. Lett.} {\bf 105}, 266403 (2010);
Z. Gul\'{a}csi,
{\it Int. J. Mod. Phys.} {\bf B27}, 1330009 (2013);
M. Gul\'{a}csi, G. Kov\'{a}cs and Z. Gul\'{a}csi,
{\it Phil. Mag. Lett.} {\bf 94}, 269 (2014);
M. Gul\'{a}csi, G. Kov\'{a}cs and Z. Gul\'{a}csi,
{\it Europhys. Lett.} {\bf 107}, 57005 (2014).

\bibitem{random1}
M. Goda, S. Nishino and H. Matsuda,
{\it Phys. Rev. Lett.} {\bf 96}, 126401 (2006); 
S. Nishino, H. Matsuda and M. Goda,
{\it J. Phys. Soc. Jpn.} {\bf 76}, 024709 (2007).

\bibitem{random2}
J. T. Chalker, T. S. Pickles and P. Shukla,
{\it Phys. Rev.} {\bf B82}, 104209 (2010).

\bibitem{random3}
D. Leykam, S. Flach, O. Bahat-Treidel and A. S. Desyatnikov,
{\it Phys. Rev.} {\bf B88}, 224203 (2013);
S. Flach, D. Leykam, J. D. Bodyfelt, P. Matthies and A. S. Desyatnikov,
{\it Europhys. Lett.} {\bf 105}, 30001 (2014) and {\bf 106}, 19901 (2014);
J. D. Bodyfelt,  D. Leykam, C. Danieli, X. Yu and S. Flach,
{\it Phys. Rev. Lett.} {\bf 113}, 236403 (2014);
C. Danieli, J. D. Bodyfelt and S. Flach,
``Flatband engineering of mobility edges'',
arXiv:1502.06690.

\bibitem{lopes}
A. A. Lopes, A. A. Z. Ant\'{o}nio and R. G. Dias,
{\it Phys. Rev.} {\bf B89}, 235418 (2014);
A. A. Lopes and R. G. Dias,
``Simple approach for the two-terminal conductance through interacting clusters'',
arXiv:1407.5922;
see also
A. A. Lopes and R. G. Dias,
{\it Phys. Rev.} {\bf B84}, 085124 (2011).

\bibitem{tamura-quantum-dots-wires}
H.~Tamura, K.~Shiraishi, T.~Kimura and H.~Takayanagi, 
{\it Phys. Rev.}  {\bf B65}, 085324 (2002);
T.~Kimura, H.~Tamura, K.~Shiraishi and H.~Takayanagi,
{\it Phys. Rev.} {\bf B65}, 081307 (2002);
K.~Shiraishi, H.~Tamura and H.~Takayanagi,
{\it Physica} {\bf E24}, 107 (2004);
M.~Ichimura, K.~Kusakabe, S.~Watanabe and T.~Onogi, 
{\it Phys. Rev.} {\bf B58}, 9595 (1998); 
H.~Ishii, T.~Nakayama and J.-i.~Inoue, 
{\it Phys. Rev.} {\bf B69}, 085325 (2004).

\bibitem{nishino-3d-flatband}
S.~Nishino, M.~Goda and K.~Kusakabe, 
{\it J. Phys. Soc. Jpn.} {\bf 72}, 2015 (2003); 
S.~Nishino and M.~Goda, 
{\it J. Phys. Soc. Jpn.} {\bf 74}, 393 (2005).

\bibitem{flat-band-experiment-polymers}
R.~Arita, Y.~Suwa, K.~Kuroki and H.~Aoki, 
{\it Phys. Rev. Lett.} {\bf 88}, 127202 (2002); 
R.~Arita, Y.~Suwa, K.~Kuroki and H.~Aoki, 
{\it Phys. Rev.}  {\bf B68}, 140403(R) (2003);
Y.~Suwa, R.~Arita, K.~Kuroki and H.~Aoki, 
{\it Phys. Rev.}  {\bf B68}, 174419 (2003);
H.~Aoki,
{\it Appl. Surf. Sci.} {\bf 237}, 2 (2004).

\bibitem{lin-grapheneribbon}
H.-H.~Lin, T.~Hikihara, H.-T.~Jeng, B.-L.~Huang, C.-Y.~Mou and X.~Hu, 
{\it Phys. Rev.} {\bf B79}, 035405 (2009).

\bibitem{fb-nature}
M. Garnica, D. Stradi, S. Barja, F. Calleja, C. Diaz, M. Alcami, N. Martin, A.L. V\'{a}zquez de Parga, F. Martin and R. Miranda,
{\it Nat. Phys.} {\bf 9}, 368 (2013).

\bibitem{balents}
C. Wu, D. Bergman, L. Balents and S. Das Sarma,
{\it Phys. Rev. Lett.} {\bf 99}, 070401 (2007);
D. L. Bergman, C. Wu and L. Balents, 
{\it Phys. Rev.} {\bf B78}, 125104 (2008);
M. Zhang, H.-h. Huang, C. Zhang and C. Wu,
{\it Phys. Rev.} {\bf A83}, 023615 (2011).

\bibitem{finland1}
V. Apaja, M. Hyrk\"{a}s and M. Manninen,
{\it Phys. Rev.} {\bf A82}, 041402(R) (2010).

\bibitem{finland2}
M. Hyrk\"{a}s, V. Apaja and M. Manninen,
{\it Phys. Rev.} {\bf A87}, 023614 (2013).

\bibitem{noda}
K. Noda, K. Inaba and M. Yamashita,
``Flat-band ferromagnetism in the multilayer Lieb optical lattice'',
arXiv:1410.2166.

\bibitem{mukherjee}
S. Mukherjee, A. Spracklen, D. Choudhury, N. Goldman, P. \"{O}hberg, E. Andersson and R. R. Thomson,
``Observation of a localized flat-band state in a photonic Lieb lattice'',
arXiv:1412.6342.

\bibitem{kai}
M. Powalski, K. Coester, R. Moessner and K. P. Schmidt,
{\it Phys. Rev.} {\bf B87}, 054404 (2013);
K. Coester, W. Malitz, S. Fey and K. P. Schmidt,
{\it Phys. Rev.} {\bf B88}, 184402 (2013).

\bibitem{sawtooth-gen}
T. Nakamura and K. Kubo, 
{\it Phys. Rev.} {\bf B53}, 6393 (1996);
D. Sen, B. S. Shastry, R. E. Walstedt and R. Cava, 
{\it Phys. Rev.} {\bf B53}, 6401 (1996).

\bibitem{kagome-chains-gen}
Ch. Waldtmann, H. Kreutzmann, U. Schollw\"{o}ck, K. Maisinger and H.-U. Everts, 
{\it Phys. Rev.} {\bf B62}, 9472 (2000);
P. Azaria, C. Hooley, P. Lecheminant and A. M. Tsvelik,
{\it Phys. Rev. Lett.} {\bf 81}, 1694 (1998).

\bibitem{diamond}
K.~Takano, K.~Kubo and H.~Sakamoto,
{\it J. Phys.: Condens. Matter} {\bf 8}, 6405 (1996);
H.~Niggemann, G.~Uimin and J.~Zittartz, 
{\it J. Phys.: Condens. Matter} {\bf 9}, 9031 (1997);
H.~Niggemann, G.~Uimin and J.~Zittartz, 
{\it J. Phys.: Condens. Matter} {\bf 10}, 5217 (1998);
K.~Okamoto, T.~Tonegawa and M.~Kaburagi, 
{\it J. Phys.: Condens. Matter} {\bf 15}, 5979 (2003);
L.~\v{C}anov\'{a}, J.~Stre\v{c}ka and M.~Ja\v{s}\v{c}ur, 
{\it J. Phys.: Condens. Matter} {\bf 18}, 4967 (2006).

\bibitem{ladders-gen}
M. P. Gelfand,
{\it Phys. Rev.} {\bf B43}, 8644 (1991); 
F. Mila, 
{\it Eur. Phys. J.} {\bf B6}, 201 (1998); 
A. Honecker, F. Mila and M. Troyer, 
{\it Eur. Phys. J.} {\bf B15}, 227 (2000).

\bibitem{double_tetrahedra}
M. Mambrini, J. Tr\'{e}bosc and F. Mila, 
{\it Phys. Rev.} {\bf B59}, 13806 (1999);
O. Rojas and F. C. Alcaraz, 
{\it Phys. Rev.} {\bf B67}, 174401 (2003);
C. D. Batista and B. S. Shastry, 
{\it Phys. Rev. Lett.} {\bf 91}, 116401 (2003);
D. Antonosyan, S. Bellucci and V. Ohanyan, 
{\it Phys. Rev.} {\bf B79}, 014432 (2009).

\bibitem{fr_three}
V. Subrahmanyam, 
{\it Phys. Rev.} {\bf B50}, 16109 (1994);
A. L\"{u}scher, R. M. Noack, G. Misguich, V. N. Kotov and F. Mila, 
{\it Phys. Rev.} {\bf B70}, 060405(R) (2004);
G. Seeber, P. K\"{o}gerler, B. M. Kariuki and L. Cronin,
{\it Chem. Commun.}, 1580 (2004);
J. Schnack, H. Nojiri, P. K\"{o}gerler, G. J. T. Cooper and L. Cronin, 
{\it Phys. Rev.} {\bf B70}, 174420 (2004);
J.-B. Fouet, A. L\"{a}uchli, S. Pilgram, R. M. Noack and F. Mila, 
{\it Phys. Rev.} {\bf B73}, 014409 (2006);
N. B. Ivanov, J. Schnack, R. Schnalle, J. Richter, P. K\"{o}gerler, G. N. Newton, L. Cronin, Y. Oshima and H. Nojiri, 
{\it Phys. Rev. Lett.} {\bf 105}, 037206 (2010);
T. Sakai, M. Sato, K. Okamoto, K. Okunishi and C. Itoi,
{\it J. Phys.: Condens. Matter} {\bf 22}, 403201 (2010);
M. Lajko, P. Sindzingre and K. Penc, 
{\it Phys. Rev. Lett.} {\bf 108}, 017205 (2012).

\bibitem{kagome-gen}
P. Lecheminant, B. Bernu, C. Lhuillier, L. Pierre and P. Sindzingre, 
{\it Phys. Rev.} {\bf B56}, 2521 (1997); 
Ch. Waldtmann, H.-U. Everts, B. Bernu, C. Lhuillier, P. Sindzingre, P. Lecheminant and L. Pierre, 
{\it Eur. Phys. J.} {\bf B2}, 501 (1998); 
F. Mila, 
{\it Phys. Rev. Lett.} {\bf 81}, 2356 (1998).

\bibitem{kagome_new}
R. R. P. Singh and D. A. Huse, 
{\it Phys. Rev.} {\bf B76}, 180407(R) (2007);
S. Yan, D. A. Huse and S. R. White,
{\it Science} {\bf 332},  1173 (2011); 
S. Depenbrock, I. P. McCulloch and U. Schollw\"ock,
{\it Phys. Rev. Lett.} {\bf 109}, 067201 (2012);
H. Nakano and T. Sakai,
{\it J. Phys. Soc. Jpn.} {\bf 80}, 053704 (2011);
A. M. L\"{a}uchli, J. Sudan and E. S. S\o rensen, 
{\it Phys. Rev.} {\bf B83}, 212401 (2011); 
O. G\"otze, D. J. J. Farnell, R. F. Bishop, P. H. Y. Li and J. Richter, 
{\it Phys. Rev.} {\bf B84}, 224428 (2011);
Z. Y. Xie, J. Chen, J. F. Yu, X. Kong, B. Normand and T. Xiang,
{\it Phys. Rev.} {\bf X4}, 011025 (2014);
A. L. Chernyshev and M. E. Zhitomirsky,
{\it  Phys. Rev. Lett.} {\bf 113}, 237202 (2014);
O. G\"{o}tze and J. Richter, 
{\it Phys. Rev.} {\bf B91}, 104402 (2015).

\bibitem{sqkag}
R.~Siddharthan and A.~Georges,
{\it Phys. Rev.} {\bf B65}, 014417 (2002);
P.~Tomczak and J.~Richter, 
{\it J. Phys.} {\bf A36}, 5399 (2003);
J.~Richter, J.~Schulenburg, P.~Tomczak and D.~Schmalfu\ss, 
{\it Condensed Matter Physics (L'viv)} {\bf 12}, 507 (2009);
H. Nakano and T. Sakai,
{\it J. Phys. Soc. Jpn.} {\bf 82}, 083709 (2013).

\bibitem{ioannis}
I. Rousochatzakis, R. Moessner and J. van den Brink, 
{\it Phys. Rev.} {\bf B88}, 195109 (2013). 

\bibitem{bilayers-gen}
H.-Q.~Lin and J.~L.~Shen,
{\it J. Phys. Soc. Jpn.} {\bf 69}, 878 (2000);
P.~Chen, C.-Y.~Lai and M.-F.~Yang,
{\it Phys. Rev.} {\bf B81}, 020409(R) (2010);
A.~F.~Albuquerque, N.~Laflorencie, J.-D.~Picon and F.~Mila,
{\it Phys. Rev.} {\bf B83}, 174421 (2011);
Y.~Murakami, T.~Oka and H.~Aoki,
{\it Phys. Rev.} {\bf B88}, 224404 (2013).

\bibitem{dipla}
N. B. Ivanov and J. Richter, 
{\it Phys. Lett.} {\bf A232}, 308 (1997); 
J. Richter, N. B. Ivanov and J. Schulenburg, 
{\it J. Phys.: Condens. Matter} {\bf 10}, 3635 (1998); 
A. Koga, K. Okunishi and N. Kawakami, 
{\it Phys. Rev.} {\bf B62}, 5558 (2000); 
J. Schulenburg and J. Richter, 
{\it Phys. Rev.} {\bf B65}, 054420 (2002); 
A. Koga and N. Kawakami, 
{\it Phys. Rev.} {\bf B65}, 214415 (2002).

\bibitem{star}
J. Richter, J. Schulenburg, A. Honecker and D. Schmalfu\ss, 
{\it Phys. Rev.} {\bf B70}, 174454 (2004),
T.-P. Choy and Y. B. Kim,
{\it Phys. Rev.} {\bf B80}, 064404 (2009);
B.-J. Yang, A. Paramekanti and Y. B. Kim,
{\it Phys. Rev.} {\bf  B81}, 134418 (2010).

\bibitem{checkerboard-gen}
S. E. Palmer and J. T. Chalker, 
{\it Phys. Rev.} {\bf B64}, 094412 (2001); 
W. Brenig and A. Honecker, 
{\it Phys. Rev.} {\bf B65}, 140407 (2002); 
J.-B. Fouet, M. Mambrini, P. Sindzingre and C. Lhuillier, 
{\it Phys. Rev.} {\bf B67}, 054411 (2003).

\bibitem{pyrochlore-gen}
B. Canals and C. Lacroix, 
{\it Phys. Rev. Lett.} {\bf 80}, 2933 (1998).

\bibitem{strecka_lattice}
J. Stre\v{c}ka, L. \v{C}anov\'{a}, M. Ja\v{s}\v{c}ur and M. Hagiwara,
{\it Phys. Rev.} {\bf B78}, 024427 (2008);
D.-X. Yao, Y. L. Loh, E. W. Carlson and M. Ma,
{\it Phys. Rev.} {\bf B78}, 024428 (2008);
J. \v{C}is\'{a}rov\'{a} and J. Stre\v{c}ka, 
{\it Phys. Rev.} {\bf B87}, 024421 (2013);
J. \v{C}is\'{a}rov\'{a}, F. Michaud, F. Mila and J. Stre\v{c}ka,
{\it Phys. Rev.} {\bf B87}, 054419 (2013).

\bibitem{fr_molecules}
J. Richter, R. Schmidt and J. Schnack,
{\it  J. Magn. Magn. Mat.} {\bf 295}, 164 (2005); 
J.~Schnack,  R.~Schmidt and J.~Richter,
{\it Phys. Rev.} {\bf B76}, 054413 (2007);
I. Rousochatzakis, A. M. L\"{a}uchli and F. Mila
{\it Phys. Rev.} {\bf B77}, 094420 (2008);
J. Schnack and R. Schnalle,
{\it Polyhedron} {\bf 28}, 1620 (2009).

\bibitem{fr_molecules-review}
J. Schnack, 
{\it Dalton Trans.} {\bf 39}, 4677 (2010).

\bibitem{ab_cage}
J. Vidal, R. Mosseri and B. Dou\c{c}ot,
{\it Phys. Rev. Lett.} {\bf 81}, 5888 (1998);
J. Vidal, B. Dou\c{c}ot, R. Mosseri and P. Butaud, 
{\it Phys. Rev. Lett.} {\bf 85}, 3906 (2000);
J. Vidal, P. Butaud, B. Dou\c{c}ot and R. Mosseri, 
{\it Phys. Rev.} {\bf B64}, 155306 (2001);
see also
B. Dou\c{c}ot and J. Vidal,
{\it Phys. Rev. Lett.} {\bf 88}, 227005 (2002);
B. Dou\c{c}ot, L. B. Ioffe and J. Vidal,
{\it Phys. Rev.} {\bf B69}, 214501 (2004).

\bibitem{dice_lattice}
G. M\"{o}ller and N. R. Cooper, 
{\it Phys. Rev. Lett.} {\bf 108}, 045306 (2012).

\bibitem{fano}
U.~Fano, 
{\it Phys. Rev.} {\bf 124}, 1866 (1961).

\bibitem{fano-r}
A. E. Miroshnichenko, S. Flach and Y. S. Kivshar,
{\it Rev. Mod. Phys.} {\bf 82}, 2257 (2010).

\bibitem{zohar}
Z. Nussinov and J. van den Brink,
{\it Rev. Mod. Phys.} {\bf 87}, 1 (2015).

\bibitem{Schmidt}
H.-J. Schmidt,
{\it J. Phys.} {\bf A35}, 6545 (2002). 

\bibitem{linear_in}
H.-J. Schmidt, J. Richter and R. Moessner,
{\it J. Phys.} {\bf A39}, 10673 (2006). 

\bibitem{baxter}
R. J. Baxter,
{\it Exactly Solved Models in Statistical Mechanics} (Academic Press, London, 1982).

\bibitem{sq_lat_afm_in_a_field}
E. M\"{u}ller-Hartmann and J. Zittartz, 
{\it Z. Phys.} {\bf B27}, 261 (1977);
X. N. Wu and F. Y. Wu, 
{\it Phys. Lett.} {\bf A144}, 123 (1990);
X.-Z. Wang and J. S. Kim, 
{\it Phys. Rev. Lett.} {\bf 78}, 413 (1997);
S. J. Penney, V. K. Cumyn and D. D. Betts, 
{\it Physica} {\bf A330}, 507 (2003).

\bibitem{str-coup-appr}
J.-B. Fouet, F. Mila, D. Clarke, H. Youk, O. Tchernyshyov, P. Fendley and R. M. Noack, 
{\it Phys. Rev.} {\bf B73}, 214405 (2006);
J.-D. Picon, A. F. Albuquerque, K. P. Schmidt, N. Laflorencie, M. Troyer and F. Mila, 
{\it Phys. Rev.} {\bf B78}, 184418 (2008). 

\bibitem{aa}
A. Honecker and A. L\"{a}uchli, 
{\it Phys. Rev.} {\bf B63}, 174407 (2001).

\bibitem{klein}
D. J. Klein, 
{\it J. Chem. Phys.} {\bf 61}, 786 (1974).

\bibitem{fulde}
P. Fulde,
{\it Electron Correlations in Molecules and Solids} (Springer-Verlag, Berlin, Heidelberg, 1993).

\bibitem{essler}
F. H. L. Essler, H. Frahm, F. G\"{o}hmann, A. Kl\"{u}mper and V. E. Korepin,
{\it The One-Dimensional Hubbard Model} (Cambridge University Press, Cambridge, UK, 2005).

\bibitem{bkt}
V. L. Berezinskii, 
{\it Zh. Eksp. Teor. Fiz.} {\bf 59}, 907 (1970)
[{\it Sov. Phys. JETP} {\bf 32}, 493 (1971)]; 
J. M. Kosterlitz and D. J. Thouless, 
{\it J. Phys.} {\bf C6}, 1181 (1973).

\bibitem{bkt_tc}
R. Gupta, J. DeLapp, G. G. Batrouni, G. C. Fox, C. F. Baillie and J. Apostolakis, 
{\it Phys. Rev. Lett.} {\bf 61}, 1996 (1988); 
R. Gupta and C. F. Baillie, 
{\it Phys. Rev.} {\bf B45}, 2883 (1992);
M. Hasenbusch and K. Pinn, 
{\it J. Phys.} {\bf A30}, 63 (1997); 
M. Hasenbusch, 
{\it J. Phys.} {\bf A38}, 5869 (2005);
Y. Komura and Y. Okabe, 
{\it J. Phys. Soc. Jpn.} {\bf 81}, 113001 (2012);
Y.-D. Hsieh, Y.-J. Kao and A. W. Sandvik, 
{\it J. Stat. Mech.} P09001 (2013).

\bibitem{bkt_q}
H.-Q. Ding and M. S. Makivi\'{c}, 
{\it Phys. Rev.} {\bf B42}, 6827 (1990); 
H.-Q. Ding, 
{\it Phys. Rev.} {\bf B45}, 230 (1992);
K. Harada and N. Kawashima, 
{\it Phys. Rev.} {\bf B55}, R11949 (1997); 
K. Harada and N. Kawashima,
{\it J. Phys. Soc. Jpn.} {\bf 67}, 2768 (1998);
J. Carrasquilla and M. Rigol, 
{\it Phys. Rev.} {\bf A86}, 043629 (2012); 
G. Ceccarelli, J. Nespolo, A. Pelissetto and E. Vicari, 
{\it Phys. Rev.} {\bf B88}, 024517 (2013).

\bibitem{bkt_xxz}
F. H\'{e}bert, G. G. Batrouni, R. T. Scalettar, G. Schmid, M. Troyer and A. Dorneich, 
{\it Phys. Rev.} {\bf B65}, 014513 (2001); 
G. Schmid, S. Todo, M. Troyer and A. Dorneich, 
{\it Phys. Rev. Lett.} {\bf 88}, 167208 (2002).

\bibitem{tutsch}
U. Tutsch, B. Wolf, S. Wessel, L. Postulka, Y. Tsui, H. O. Jeschke, I. Opahle, T. Saha-Dasgupta, R. Valenti, 
A. Br\"{u}hl, K. Removi\'{c}-Langer, T. Kretz, H.-W. Lerner, M. Wagner and M. Lang,
{\it Nat. Commun.} {\bf 5}, 5169 (2014).

\bibitem{huber_exp1}
J. Ruostekoski,
{\it Phys. Rev. Lett.} {\bf 103}, 080406 (2009);
B. Damski, H. Fehrmann, H.-U. Everts, M. Baranov, L. Santos and M. Lewenstein,
{\it Phys. Rev.} {\bf A72}, 053612 (2005). 

\bibitem{huber_exp2}
K. Awaga, T. Okuno, A. Yamaguchi, M. Hasegawa, T. Inabe, Y. Maruyama and N. Wada,
{\it Phys. Rev.} {\bf B49}, 3975 (1994). 

\bibitem{girvin}
S. M. Girvin and T. Jach, 
{\it Phys. Rev.} {\bf B29}, 5617 (1984). 

\bibitem{phillips}
L. G. Phillips, G. De Chiara, P. \"{O}hberg and M. Valiente, 
{\it Phys. Rev.} {\bf B91}, 054103 (2015). 

\bibitem{tovmasyan}
M. Tovmasyan, E. P. L. van Nieuwenburg and S. D. Huber,
{\it Phys. Rev.} {\bf B88}, 220510(R) (2013). 

\bibitem{hosho}
S. Takayoshi, H. Katsura, N. Watanabe and H. Aoki,
{\it Phys. Rev.} {\bf A88}, 063613 (2013).

\bibitem{Hida2001} 
K. Hida, 
{\it J. Phys. Soc. Jpn.} {\bf 70}, 3673 (2001).

\bibitem{Richter2004} 
A. Honecker, J. Schulenburg and J. Richter,
{\it J. Phys.: Condens. Matter} {\bf 16}, S749 (2004).

\bibitem{earlier}
D. C. Cabra, M. D. Grynberg, P. C. W. Holdsworth, A. Honecker, P. Pujol, J. Richter, D. Schmalfu{\ss} and J. Schulenburg,
{\it Phys. Rev.} {\bf B71}, 144420 (2005);
A. Honecker, D. C. Cabra, M. D. Grynberg, P. C. W. Holdsworth, P. Pujol, J. Richter, D. Schmalfu{\ss} and J. Schulenburg,
{\it Physica} {\bf B359}, 1391 (2005).

\bibitem{Okamoto2011} 
Y. Okamoto, M. Tokunaga, H. Yoshida, A. Matsuo, K. Kindo and Z. Hiroi,
{\it Phys. Rev.} {\bf B83}, 180407(R) (2011).

\bibitem{inagaki}
Y. Inagaki, Y. Narumi, K. Kindo, H. Kikuchi, T. Kamikawa, T. Kunimoto, S. Okubo, H. Ohta, T. Saito, M. Azuma, M. Takano, H. Nojiri, M. Kaburagi and T. Tonegawa,
{\it J. Phys. Soc. Jpn.} {\bf 74}, 2831 (2005).

\bibitem{hamada}
T. Hamada, J.-i. Kane, S.-i. Nakagawa and Y. Natsume,
{\it J. Phys. Soc. Jpn.} {\bf 57}, 1891 (1988).

\bibitem{FM_crit}
W. Selke, 
{\it Z. Phys.} {\bf B27}, 81 (1977);
M. Yamada  and M. Takahashi, 
{\it J. Phys. Soc. Jpn.} {\bf 55}, 2024 (1986);
M. H\"{a}rtel, J. Richter, D. Ihle and  S.-L. Drechsler, 
{\it Phys. Rev.} {\bf B78}, 174412 (2008);
J. Sirker, V. Y. Krivnov, D. V. Dmitriev, A. Herzog, O. Janson, S. Nishimoto, S.-L. Drechsler and J. Richter,
{\it Phys. Rev.} {\bf B84}, 144403 (2011).

\bibitem{anisotropy_jmmm}
J. Richter, O. Krupnitska, T. Krokhmalskii and O. Derzhko,
{\it J. Magn. Magn. Mat.} {\bf 379}, 39 (2015).

\bibitem{vj}
V. Derzhko and J. J\c{e}drzejewski,
``How to recognize a nearly-flat-band ferromagnet by means of thermodynamic measurements?'',
arXiv:1004.2786.

\bibitem{percolation1}
D. Stauffer,
{\it Physics Reports} {\bf 54}, 1 (1979).

\bibitem{percolation2}
D. Stauffer and A. Aharony,
{\it Introduction to Percolation Theory} (Taylor \& Francis, 1994). 

\bibitem{percolation3}
M. B. Isichenko,
{\it Rev. Mod. Phys.}  {\bf 64}, 961 (1992). 

\bibitem{explosive_percolation}
D. Achlioptas, R. M. D'Souza and J. Spencer,
{\it Science} {\bf 323}, 1453 (2009);
Y. S. Cho, S. Hwang, H. J. Herrmann and B. Kahng,
{\it Science} {\bf 339}, 1185 (2013).

\bibitem{dorogovtsev}
S. N. Dorogovtsev, 
{\it Lectures on Complex Networks} (Oxford University Press, Oxford, 2010).

\bibitem{ostilli}
M. Ostilli, 
{\it Physica} {\bf{A391}}, 3417 (2012).

\bibitem{essam}
M. E. Fisher and J. W. Essam,
{\it J. Math. Phys.} {\bf 2}, 609 (1961).

\bibitem{cavity}
M. M\'{e}zard and G. Parisi,
{\it Eur. Phys. J.} {\bf B20}, 217 (2001);
C. Laumann, A. Scardicchio and S. L. Sondhi,
{\it Phys. Rev.} {\bf B78}, 134424 (2008);
O. Rivoire,
{\it J. Stat. Mech.} P07004 (2005);
B. Hsu, C. Laumann, A. L\"{a}uchli, R. Moessner and S. Sondhi,
{\it Phys. Rev.} {\bf A87}, 062334 (2013).

\bibitem{free-energy}
J. Barr\'{e} and B. Gon\c{c}alves,
{\it Physica} {\bf A386}, 212 (2007).

\bibitem{thorpe_phonon_gap}
M. F. Thorpe, 
{\it NATO Advanced Study Institute Series} {\bf B78}, 85 (1982).

\bibitem{laumann_goldstone_bethe}
C. R. Laumann, S. A. Parameswaran and S. L. Sondhi, 
{\it Phys. Rev.} {\bf B80}, 144415 (2009).

\bibitem{weinrib}
A. Weinrib,
{\it Phys. Rev.} {\bf B29}, 387 (1984).

\bibitem{hoshen}
J. Hoshen and R. Kopelman,
{\it Phys. Rev.} {\bf B14}, 3438 (1976).

\bibitem{newman}
M. E. J. Newman and R. M. Ziff,
{\it Phys. Rev.} {\bf E64}, 016706 (2001).

\bibitem{hukushima}
K. Hukushima and K. Nemoto,
{\it J. Phys. Soc. Jpn.} {\bf 65}, 1604 (1996).

\bibitem{villain} 
J. Villain, R. Bidaux, J. P. Carton and R. Conte,
{\it J. Phys.} {\bf 41}, 1263 (1980).

\bibitem{shender} 
E. F. Shender,
{\it Zh. Eksp. Teor. Fiz.} {\bf 83}, 326 (1982)
[{\it Sov. Phys. JETP} {\bf 56}, 178 (1982)].

\bibitem{patrick}
P. M\"{u}ller, J. Richter and O. Derzhko,
in preparation.

\bibitem{anderson-localization}
P. W. Anderson,
{\it Phys. Rev.} {\bf 109}, 1492 (1958);
see also 
B. Kramer and A. MacKinnon,
{\it Rep. Prog. Phys.} {\bf 56} 1469 (1993)
and references therein.

\bibitem{ishii}
K. Ishii,
{\it Prog. Theor. Phys. Suppl.} 53, 77 (1973).

\bibitem{transp_kagochains}
H. Ishii and T. Nakayama,
{\it Phys. Rev.} {\bf B73}, 235311 (2006);
M. Dey, S. K. Maiti and S. N. Karmakar,
{\it J. Appl. Phys.} {\bf 110}, 094306 (2011).

\bibitem{imry}
Y. Imry and R. Landauer,
{\it Rev. Mod. Phys.} {\bf 71}, S306 (1999).

\bibitem{datta}
S. Datta,
{\it Electronic Transport in Mesoscopic Systems} (Cambridge University Press, Cambridge UK, 1995).

\bibitem{enss}
T. Enss, V. Meden, S. Andergassen, X. Barnab\'{e}-Th\'{e}riault, W. Metzner and K. Sch\"{o}nhammer,
{\it Phys. Rev.} {\bf B71}, 155401 (2005).

\bibitem{conductance_inter}
Y. Meir and N. S. Wingreen,
{\it Phys. Rev. Lett.} {\bf 68}, 2512 (1992).

\bibitem{hubbard_molecule}
H. Ishida and A. Liebsch,
{\it Phys. Rev.} {\bf B86}, 205115 (2012).

\bibitem{single_mol_magnet}
P.-B. Niu, Y.-Y. Zhang, Q. Wang and Y.-H. Nie,
{\it Phys. Lett.} {\bf A376}, 1481 (2012).

\bibitem{hatsugai}
Y. Hatsugai, K. Shiraishi and H. Aoki,
``Flat bands in Weaire-Thorpe model and silicene'',
arXiv:1410.7885.

\bibitem{bcs_flat}
S. Miyahara, S. Kusuta and N. Furukawa,
{\it Physica} {\bf C460-462}, 1145 (2007);
V. I. Iglovikov, F. H\'{e}bert, B. Gr\'{e}maud, G. G. Batrouni and R. T. Scalettar,
{\it Phys. Rev.} {\bf B90}, 094506 (2014).

\bibitem{nematic}
G. Zhu, J. Koch and I. Martin,
``Nematic superfluidity and Wigner crystallization of bosons in frustrated kagome lattices'',
arXiv:1411.0043.

\bibitem{phonons}
M. Goda, R. Tanaka and S. Nishino,
{\it physica status solidi (c)} {\bf 1}, 3043 (2004).

\bibitem{falikov-kimball}
M. Udagawa, H. Ishizuka and Y. Motome,
{\it Phys. Rev. Lett.} {\bf 104}, 226405 (2010);
H. Ishizuka, M. Udagawa and Y. Motome,
{\it Phys. Rev.} {\bf B83}, 125101 (2011).

\bibitem{nandy}
A. Nandy, B. Pal and A. Chakrabarti,
{\it J. Phys.: Condens. Matter} {\bf 27}, 125501 (2015).

\bibitem{biondi}
M. Biondi, E. P. L. van Niewenburg, G. Blatter, S. D. Huber and S. Schmidt,
``Incompressible polaritons in a flat band'',
arXiv:1502.07854.

\end{thebibliography}
\end{document}